\documentclass[12pt]{scrartcl}%
\usepackage{amsmath}
\usepackage{amsfonts}
\usepackage{amssymb}
\usepackage{graphicx}%
\providecommand{\U}[1]{\protect\rule{.1in}{.1in}}
\newtheorem{theorem}{Theorem}

\newtheorem{conclusion}[theorem]{Conclusion}

\newtheorem{remark}[theorem]{Remark}

\begin{document}

\DeclareGraphicsExtensions{.eps,.pdf,.jpg,.png}

\LARGE\bf\sffamily 
\begin{center}
Life in the Rindler Reference Frame:\\ Does an Uniformly Accelerated Charge\\
Radiates? Is there a Bell\ `Paradox'?\\ Is Unruh Effect Real?
\end{center}

\smallskip

\normalsize
\large

\begin{center}
Waldyr A. Rodrigues Jr.{\normalsize\footnote{walrod@ime.unicamp.br}} 
and Jayme Vaz Jr.{\normalsize\footnote{vaz@ime.unicamp.br}}
\end{center}

\normalsize\rm\sffamily

\begin{center}
Departamento de Matem\'atica Aplicada - IMECC\\
Universidade Estadual de Campinas \\
13083-859 Campinas, SP, Brazil
\end{center}

\bigskip

\small\rm\rmfamily

\begin{center}
\begin{minipage}[c]{12cm}
\centerline{\bf\sffamily Abstract}
The determination of the electromagnetic field generated by a charge in
hyperbolic motion is a classical problem for which the majority view is that
the Li\'{e}nard-Wiechert solution which implies that the charge radiates) is
the correct one. However we analyze in this paper a less known solution due to
Turakulov that differs from the Li\'{e}nard-Wiechert one and which according
to him does not radiate. We prove his conclusion to be wrong. We analyze the
implications of both solutions concerning the validity of the Equivalence
Principle. We analyze also two other issues related to hyperbolic motion,\ the
so-called Bell's \textquotedblleft paradox\textquotedblright\ which is as yet
source of misunderstandings and the Unruh effect, which according to its
standard derivation in the majority of the texts, is a correct prediction of
quantum field theory. We recall that the standard derivation of the Unruh
effect does not resist any tentative of any rigorous mathematical
investigation, in particular the one based in the algebraic approach to field
theory which we also recall. These results make us to align with some
researchers that also conclude that the Unruh effect does not exist.
\end{minipage}
\end{center}

\normalsize

\tableofcontents

\section{Introduction}

There are some problems in Relativity theory that are continuously source OF
controversies, among them we discuss in this paper: (a) the problem of
determining if an uniformly accelerated charge does or does not radiate
\footnote{This problem is important concerning one of the formulations of the
Equivalence principle.}; (b) the so-called Bell's paradox and; (c) the Unruh
effect.\footnote{We call the reader's attention that the references quoted in
this paper are far from complete, so we apologize for papers not quoted.}.

In order to obtain some light on the controversies we discuss in details in
Section 2 the concept of (right and left) Rindler reference frames, Rindler
observers and a chart naturally adapted to a given Rindler frame. These
concepts are distinct and thus represented by different mathematical objects
and having this in mind is a necessary condition to avoid misunderstandings,
both OF mathematical as well as of physical nature.

In Section 3 we analyze Bell's \textquotedblleft paradox\textquotedblright%
\ that even having a trivial solution seems to not been understood for some
people even recently for it is confused with another distinct problem which if
one does not pay the required attention seems analogous to the one formulated
by Bell.

In Section 4 we discuss at length the problem of the electromagnetic field
generated by a charge in hyperbolic motion. First we present the classical
Li\'{e}nard-Wiechert solution, which implies that an observer at rest in an
inertial reference frame observes that the charge radiates. Next we analyze
(accepting that the Li\'{e}nard-Wiechert solution is correct) if an observer
comoving with the charge detects or no radiation. We argue with details that
contrary to some views it is possible for a real observer living in a real
laboratory\footnote{This, of course, means that the laboratory (whatever its
mathematical model) \cite{buver} must have finite spatial dimensions as
determined by the observer at any instant of its propertime.} in hyperbolic
motion to detect that the charge is radiating. Our conclusion is based
(following \cite{parrott1}) on a careful analysis of different concepts of
\emph{energy} that are used in the literature, the one defined in an inertial
reference frame and the other in the Rindler frame. In particular, we discuss
in details the error in Pauli's argument.

But now we ask: is it necessary to accept the Li\'{e}nard-Wiechert solutions
as the true one describing the electromagnetic field generated by a charge in
hyperbolic motion? To answer that question we analyzed the Turakulov
\cite{tura1} solution to this problem, which consisting in solving the wave
equation for the electromagnetic potential in a special systems of coordinates
where the equation gets separable. We have verified that Turakulov solution
(which differs form the Li\'{e}nard-Wiechert one) is correct (in particular,
by using the Mathematica software). Turakulov claims that in his solution the
charge does not radiate. However, we prove that his claim is wrong, i.e., we
show that as in the case of the Li\'{e}nard-Wiechert solution an observer
comoving with the charge can detect that is is emitting radiation.

In Section 5 we discuss, taking into account that it seems a strong result the
fact that a charge at rest in the Schwarzschild spacetime does not radiate
\cite{parrott1}, what the results of Section 4 implies for the validity or not
of one of the forms in which Equivalence principle is presented in many texts.

Section 6 is dedicated to the Unruh effect. We first recall the standard
presentation (emphasizing each one of the hypothesis used in its derivation)
of the supposed fact that Rindler observers are living in a thermal bath with
a Planck spectrum with temperature proportional to its local proper
acceleration and thus such radiation may\emph{ excite} detectors on
board.\ Existence of the Unruh radiation and Rindler particles seems to be the
majority view. However, we emphasize that rigorous mathematical analysis of
standard procedure (which is claimed to predict the Unruh effect) done by
several authors shows clearly that such a procedure contain several
inconsistencies. These rigorous analysis show that the Unruh effect does no
exist, although it may be proved that detectors in hyperbolic notion can get
excited, although the energy for that process comes form the source
accelerating the detector and it is not (as some claims) due to fluctuations
of the Minkowski vacuum. We recall in Appendix B a (necessarily resumed)
introduction to the algebraic approach to quantum theory as applied to the
Unruh effect in order to show how much we can trust each one of the
suppositions used in the standard derivation of the Unruh effect. Detailed
references are given at the \ appropriate places.

Section 7 presents our conclusions and in Appendix A we present our
conventions and some necessary definitions of the concepts of reference
frames, observers, instantaneous observers and naturally adapted charts to a
given reference frame.

\section{Rindler Reference Frame}

A proper understanding of almost any problem in Relativity theory requires
that we know (besides the basics of differential geometry\footnote{Basics of
differential geomety may be found in \cite{choquet,dodson,frankel,nakahara}.
Necessary concepts concerning Lorentzian manifods may be found in
\cite{oneil,sawu}.}) exactly the meaning and the precise mathematical
representation of the concepts of: (a) references frames and their
classification; (b) a naturally adapted chart to a given reference frame; (c)
observers and (d) instntaneous observers. The main results necessary for the
understanding of the present paper and some other definitions are briefly
recalled in Appendix A\footnote{More details may be found in
\cite{rc2016,gr2012}.}. Essential is to have in mind that most of the possible
reference frames used in Relativity theory are t\emph{heoretical instruments},
i.e., they are not physically realizable as a material systems. This is
particularly the case of the right and left Rindler reference frames and
respective observers that we introduce next.%

%TCIMACRO{\FRAME{ftbpFU}{6.4714in}{4.4391in}{0pt}{\Qcb{Some integral lines of
%the right $\QTR{bs}{R}$ and left $\QTR{bs}{L}$ Rindler reference frames}}%
%{}{Figure 1}{\special{ language "Scientific Word";  type "GRAPHIC";
%maintain-aspect-ratio TRUE;  display "USEDEF";  valid_file "T";
%width 6.4714in;  height 4.4391in;  depth 0pt;  original-width 7.2065in;
%original-height 4.9346in;  cropleft "0";  croptop "1";  cropright "1";
%cropbottom "0";  tempfilename 'OK8WNW0F.bmp';tempfile-properties "XPR";}} }%
%BeginExpansion
\begin{figure}[ptb]%
\centering
\includegraphics[width=13cm]{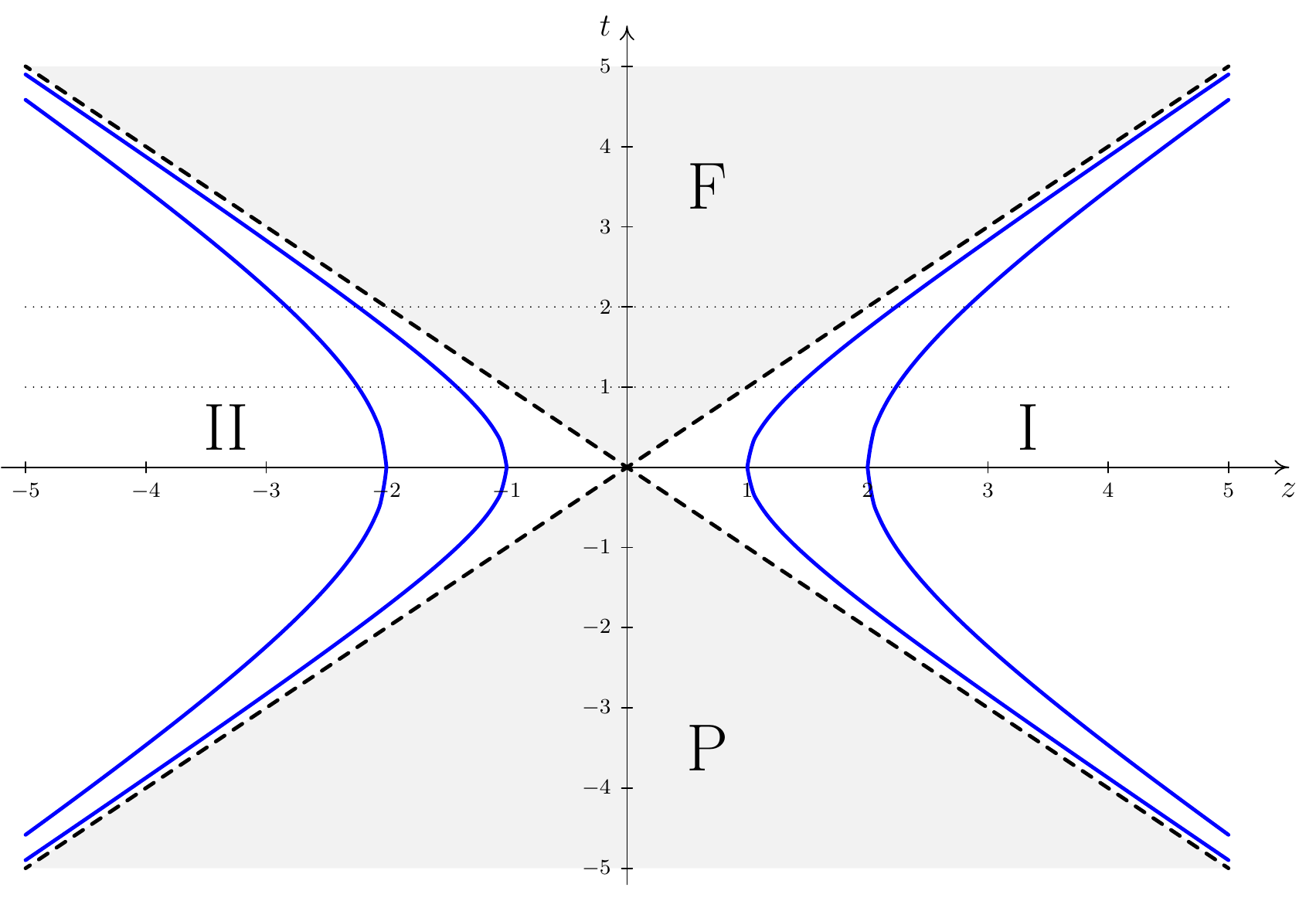}
\caption{Some integral lines of the right $\boldsymbol{R}$ and left
$\boldsymbol{L}$ Rindler reference frames}%
\end{figure}

Let $\sigma:I\rightarrow M$, $s\mapsto\sigma(s)$ a timelike curve in $M$
describing the motion of an accelerated observer (or an accelerated particle)
where $s$ is the proper time along $\sigma$. The coordinates of $\sigma$ in
ELP gauge (see Appendix A) are%
\begin{equation}
x_{\sigma}^{\mu}(s)=\boldsymbol{x}^{\mu}\circ\sigma~(s)\label{1}%
\end{equation}
and for motion along the $x^{3}=z$ axis it is
\begin{equation}
(x_{\sigma}^{o})^{2}-(x_{\sigma}^{3})^{2}=-\frac{1}{\mathrm{a}_{\sigma}^{2}%
},\label{2}%
\end{equation}
where $a_{\sigma}$ is a real constant for each curve $\sigma$. In Figure 1 we
can see two curves $\sigma$ and $\sigma^{\prime}$ for which $\frac
{1}{\mathrm{a}_{\sigma}}=1$ and $\frac{1}{\mathrm{a}_{\sigma^{\prime}}}=2$. To
understand the meaning of the parameter $\mathrm{a}_{\sigma}$ in Eq.(\ref{2})
we write%
\begin{equation}
x_{\sigma}^{0}(s)=\frac{1}{\mathrm{a}_{\sigma}}\sinh(\mathrm{a}_{\sigma
}s),~~~x_{\sigma}^{3}(s)=\frac{1}{\mathrm{a}_{\sigma}}\cosh(\mathrm{a}%
_{\sigma}s).\label{3}%
\end{equation}

The\ unit velocity vector of the observer is
\begin{equation}
\boldsymbol{v}_{\sigma}(s)=\sigma_{\ast}(s):=v^{\mu}(s)\frac{\partial
}{\partial x^{\mu}}=\cosh(\mathrm{a}_{\sigma}s)\frac{\partial}{\partial
t}+\sinh(\mathrm{a}_{\sigma}s)\frac{\partial}{\partial\mathrm{z}}.\nonumber
\end{equation}

Now, the acceleration of $\sigma$ is%
\begin{equation}
\boldsymbol{a}_{\sigma}=\frac{d}{ds}\sigma_{\ast}(s)=\mathrm{a}_{\sigma
}\left.  \left(  \sinh(\mathrm{a}_{\sigma}s)\frac{\partial}{\partial t}%
+\cosh(a_{\sigma}s)\frac{\partial}{\partial z}\right)  \right\vert _{\sigma}
\label{6}%
\end{equation}
and of course, $\boldsymbol{a}_{\sigma}\cdot\boldsymbol{v}_{\sigma}=0$ and
$\boldsymbol{a}_{\sigma}\cdot\boldsymbol{a}_{\sigma}=-\mathrm{a}_{\sigma}^{2}$.

\subsection{Rindler Coordinates}

Introduce first the regions I$,$II, F and P$\ $of Minkowski spacetime%
\begin{equation}
\mathcal{I=\{}(t,x,y,z)~~|~-\infty<t<\infty,-\infty<x<\infty,,-\infty
<y<\infty,0<z<\infty\},\label{7}%
\end{equation}
and two coordinate functions $(\boldsymbol{x}^{0},\boldsymbol{x}%
^{1},\boldsymbol{x}^{2},\boldsymbol{x}^{3})$ and $(\boldsymbol{x}^{\prime0}$%
,$\boldsymbol{x}^{\prime}{}^{1}$,$ \boldsymbol{x}^{\prime}{}^{2}$%
,$\boldsymbol{x}^{\prime}{}^{3})$ covering such regions. For $\mathfrak{e}\in M$ it
is $\{\boldsymbol{x}^{0}(\mathfrak{e})=x^{0}=t, \boldsymbol{x}^{1}%
(\mathfrak{e})=x, \boldsymbol{x}^{2}(\mathfrak{e})=y, \boldsymbol{x}%
^{3}(\mathfrak{e})=z\}$ and \{$\boldsymbol{x}^{\prime0}(\mathfrak{e})= 
\textrm{t},  \boldsymbol{x}^{\prime1}(\mathfrak{e})=\mathrm{x}%
=x, \boldsymbol{x}^{\prime2}(\mathfrak{e})=\mathrm{y}=y, \boldsymbol{x}%
^{\prime}{}^{3}(\mathfrak{e})=  \textrm{ z}\}$ with\footnote{Of course the
coordiantes $(t,x,y,z)$ cover all $M$ but the coordinates $(\mathrm{t,x,y,z})$
do not cover all $M$, they are singular at the origin.}
\begin{gather}
\mathrm{z}=\pm\sqrt{z^{2}-t^{2}},~~~\mathrm{t}=\tanh^{-1}\left(  \frac{t}%
{z}\right)  ,~~~~\left\vert \emph{z}\right\vert \geq\left\vert t\right\vert
,\nonumber\\
x^{0}=t=\mathrm{z}\sinh\mathrm{t},~~~x^{3}=z=\mathrm{z}\cosh\mathrm{t}%
~~~~\text{ in region \textrm{I},}\nonumber\\
x^{0}=t=-\mathrm{z}\sinh\mathrm{t},~~~x^{3}=z=-\mathrm{z}\cosh\mathrm{t~~~~}%
\text{ in region \textrm{II}}\label{8}%
\end{gather}
and%

\begin{align}
\mathrm{z}  &  =\pm\sqrt{t^{2}-z^{2}},~~~\mathrm{t}=\tanh^{-1}\left(  \frac
{z}{t}\right)  ,~~~~\left\vert t\right\vert \geq\left\vert z\right\vert
,\nonumber\\
x^{0}  &  =t=\mathrm{z}\cosh\mathrm{t},~~~x^{3}=z=\mathrm{z}\sinh
\mathrm{t~~~~}\text{ in region F,}\nonumber\\
x^{0}  &  =t=-\mathrm{z}\cosh\mathrm{t},~~~x^{3}=z=-\mathrm{z}\sinh
\mathrm{t~~~~}\text{ in region P.} \label{8a}%
\end{align}
The right Rindler \emph{reference frame} $\boldsymbol{R}\in\sec T$\textrm{I}
has support\ in region I and is defined by
\begin{gather}
\boldsymbol{R}=\frac{z}{\sqrt{z^{2}-t^{2}}}\frac{\partial}{\partial t}%
+\frac{t}{\sqrt{z^{2}-t^{2}}}\frac{\partial}{\partial z}=\frac{1}{\mathrm{z}%
}\frac{\partial}{\partial\mathrm{t}},\nonumber\\
z>0;~~~\left\vert z\right\vert \geq t. \label{9}%
\end{gather}
The left\ reference Rindler frame $\boldsymbol{L}\in\sec T$\textrm{II is
defined by }%
\begin{gather}
\boldsymbol{L}=\frac{z}{\sqrt{z^{2}-t^{2}}}\frac{\partial}{\partial t}%
+\frac{t}{\sqrt{z^{2}-t^{2}}}\frac{\partial}{\partial z}=\frac{1}{\mathrm{z}%
}\frac{\partial}{\partial\mathrm{t}},\nonumber\\
z<0:~~\left\vert z\right\vert \geq t. \label{9a}%
\end{gather}

Then, we see that in \textrm{I} $\mathcal{\subset}M$, $($\textrm{$t$}%
$,x^{1},x^{2},\mathrm{z})$ \ as defined in Eq.(\ref{8}) are a \emph{naturally
adapted coordinate system} to $\boldsymbol{R}$ [($nacs|\boldsymbol{R}$)] and
$\boldsymbol{L}$ [($nacs|\boldsymbol{L}$)]. With $D$ being the Levi-Civita
connection of $\boldsymbol{g}$, the acceleration vector field associated to
$\boldsymbol{R}$ is%

\begin{equation}
\boldsymbol{a}=D_{\boldsymbol{R}}\boldsymbol{R}=\frac{1}{\mathrm{z}}%
\frac{\partial}{\partial\mathrm{z}}. \label{10}%
\end{equation}

Also,%
\begin{equation}
\boldsymbol{a}_{\sigma}=\frac{d}{ds}\sigma_{\ast}(s)=\mathrm{a}_{\sigma
}\left.  \frac{\partial}{\partial\mathrm{z}}\right\vert _{\sigma} \label{11}%
\end{equation}
i.e., $\boldsymbol{a}_{\sigma}=\left.  D_{\boldsymbol{R}}\boldsymbol{R}%
\right\vert _{\sigma}=\left.  \frac{1}{\mathrm{z}}\frac{\partial}%
{\partial\mathrm{z}}\right\vert _{\sigma}=\left.  \mathrm{a}_{\sigma}%
\frac{\partial}{\partial\mathrm{z}}\right\vert _{\sigma}$. Moreover, recall
that since $\sigma$ is clearly an integral line of the vector field
$\boldsymbol{R}$, it is $\boldsymbol{v}_{\sigma}=\left.  \boldsymbol{R}%
\right\vert _{\sigma}.$

\begin{remark}
Note that in Eq.\emph{(\ref{9})} \emph{(}respectively \emph{Eq.(\ref{9a})) it
is necessary to impose }$z>0$ \emph{(respectively, }$z<0$\emph{) }this being
the reason for having defined the right and left Rindler reference frames.
\end{remark}

\subsection{Decomposition of $DR$}

Recall that the Minkowski metric field $\boldsymbol{g}=\eta_{\mu\nu}dx^{\mu
}\otimes dx^{\nu}$reads in Rindler coordinates (in region $\mathrm{I}$)
\begin{gather}
\boldsymbol{g}=g_{\mu\nu}dx^{\mu}\otimes dx^{v}=\mathrm{z}^{2}d\mathrm{t}%
\otimes d\mathrm{t}-d\mathrm{x}\otimes d\mathrm{x}-d\mathrm{y}\otimes
d\mathrm{y}-d\mathrm{z}\otimes d\mathrm{z}\nonumber\\
=\eta_{\mathbf{ab}}\boldsymbol{\gamma}^{\mathbf{a}}\otimes\boldsymbol{\gamma
}^{\mathbf{b}} \label{12a}%
\end{gather}
where $\{\boldsymbol{\gamma}^{0},\boldsymbol{\gamma}^{0},\boldsymbol{\gamma
}^{2},\boldsymbol{\gamma}^{3}\}=\{\mathrm{z}d$\textrm{$t$}$,d$\textrm{$x$%
}$,d\mathrm{y},d\mathrm{z}\}$ is an orthonormal coframe for $T^{\ast
}\mathrm{I}$ which is dual to the orthonormal frame $\{e_{0},e_{1},e_{2}%
,e_{3}\}=\{\boldsymbol{R=}\frac{1}{\mathrm{z}}\frac{\partial}{\partial
\mathrm{t}},\frac{\partial}{\partial\mathrm{x}},\frac{\partial}{\partial
\mathrm{y}},\frac{\partial}{\partial\mathrm{z}}\}$ for $T\mathrm{I}$. We write%
\begin{equation}
D_{\frac{\partial}{\partial\mathrm{x}^{\nu}}}d\mathrm{x}^{\mu}=-\Gamma
_{\mathbf{\cdot}\nu\iota}^{\mu\mathbf{\cdot\cdot}}d\mathrm{x}^{\iota
},~~~~~D_{e_{\mathbf{b}}}\boldsymbol{\gamma}^{a}=-\mathbf{\Gamma
}_{\mathbf{\cdot bc}}^{\mathbf{a\cdot\cdot}}\boldsymbol{\gamma}^{c}
\label{13aa}%
\end{equation}
and keep in mind that it is $\mathbf{\Gamma}_{\mathbf{\cdot b\cdot}%
}^{\mathbf{a\cdot c}}=-\mathbf{\Gamma}_{\mathbf{\cdot b\cdot}}^{\mathbf{c\cdot
a}}$ (and of course, $\Gamma_{\mathbf{\cdot}\nu\iota}^{\mu\mathbf{\cdot\cdot}%
}=\Gamma_{\mathbf{\cdot}\iota\nu}^{\mu\mathbf{\cdot\cdot}}$)

Define the $1$-form field (physically equivalent to $\boldsymbol{R}$)
\begin{equation}
R=\boldsymbol{g}(\boldsymbol{R},~)=R_{\mu}dx^{\mu}=\mathrm{z}d\mathrm{x}%
^{0}=\boldsymbol{\gamma}^{0}.\label{13}%
\end{equation}
Then, as well known\footnote{Se, e.g., \cite{rc2016}.} $DR$ has the invariant
decomposition%
\begin{equation}
DR=a\boldsymbol{\otimes}R\boldsymbol{+}\mathbf{\omega}_{R}+\varkappa
\boldsymbol{+}\frac{1}{3}\mathfrak{E}h\boldsymbol{,}\label{14}%
\end{equation}
with
\begin{gather}
a:=\boldsymbol{g}(\boldsymbol{a},~),\nonumber\\
\mathbf{\omega}_{R}\boldsymbol{:}=\omega_{\mu\nu}dx^{\mu}\otimes dx^{v}%
=\frac{1}{2}\left(  R_{\sigma;\tau}-R_{\tau;\sigma}\right)  h_{\mu}^{\sigma
}h_{\nu}^{\tau}dx^{\mu}\otimes dx^{v}\nonumber\\
\varkappa\boldsymbol{:}=\varkappa_{\mu\nu}dx^{\mu}\otimes dx^{v}=\left[
\frac{1}{2}\left(  R_{\sigma;\tau}+R_{\tau;\sigma}\right)  h_{\mu}^{\sigma
}h_{\nu}^{\tau}-\frac{1}{3}\mathfrak{E}h_{\sigma\tau}h_{\mu}^{\sigma}h_{\nu
}^{\tau}\right]  dx^{\mu}\otimes dx^{v}\nonumber\\
\mathfrak{E:}=\operatorname{div}\boldsymbol{R}=R_{;\mu}^{\mu}=\delta
R\nonumber\\
h:=\left(  g_{\mu\nu}-R_{\mu}R_{\nu}\right)  dx^{\mu}\otimes dx^{v}\label{15}%
\end{gather}
where $a\boldsymbol{,}\omega,\varkappa$ and $\mathfrak{E}$ are respectively
the (form) \emph{acceleration}, the \emph{rotation tensor} (or vortex) of $R$,
$\varkappa$ is the shear tensor of $R$ and $\mathfrak{E}$ is the
\emph{expansion ratio} of\ $R$.

Now, $d\boldsymbol{\gamma}^{0}=d\mathrm{z}\wedge d$\textrm{$x$}$^{0}=\frac
{1}{\mathrm{z}}\boldsymbol{\gamma}^{3}\wedge\boldsymbol{\gamma}^{0}$ and thus
$\boldsymbol{\gamma}^{0}\wedge d\boldsymbol{\gamma}^{0}=0$ which implies that
$\mathbf{\omega}_{R}=0$. See Appendix A and details in \cite{rc2016}

This means that the Rindler reference frame $\boldsymbol{R}$ is locally
synchronizable, but since $R$ is not an exact differential $\boldsymbol{R}$ is
\emph{not }proper time synchronizable, something that is obvious once we look
at Figure 1 and see that for each time $t>0$ of the inertial reference frame
$\boldsymbol{I}=\partial/\partial t$ the Rindler observers following paths
$\sigma$ and $\sigma^{\prime}$ (which have of course, different proper
accelerations) have also different speeds, so their clocks (according to an
inertial observer) tic-tac at different ratios.

\subsection{\label{dist}Constant Proper Distance Between $\sigma$ and
$\sigma^{\prime}$}

We can easily verify using the orthonormal coframe introduced above that since
$d\boldsymbol{\gamma}^{\mathbf{i}}=0$, $\mathbf{i}=1,2,3$ it is
$\mathbf{\Gamma}_{\mathbf{ab}}^{\mathbf{i}}=\mathbf{\Gamma}_{\mathbf{ba}%
}^{\mathbf{i}}$ for $\mathbf{i}=1,2,3$ and $\mathbf{a,b}=0,1,2,3$ and also
from the form of $d\boldsymbol{\gamma}^{0}$ we realize that \ $\mathbf{\Gamma
}_{\cdot\mathbf{00}}^{\mathbf{0\cdot\cdot}}=\mathbf{\Gamma}_{\cdot
\mathbf{0\cdot}}^{\mathbf{0\cdot0}}=-\mathbf{\Gamma}_{\cdot\mathbf{0\cdot}%
}^{\mathbf{0\cdot0}}=0$. Thus,
\begin{equation}
\mathfrak{E}=\delta R\boldsymbol{=}-\boldsymbol{\gamma}^{\mathbf{a}}\lrcorner
D_{\boldsymbol{e}_{\mathbf{a}}}(\boldsymbol{\gamma}^{\mathbf{0}}%
)=\mathbf{\Gamma}_{\cdot\mathbf{ab}}^{\mathbf{0\cdot\cdot}}\boldsymbol{\gamma
}^{\mathbf{a}}\lrcorner\boldsymbol{\gamma}^{\mathbf{b}}=\eta^{\mathbf{ab}%
}\mathbf{\Gamma}_{\cdot\mathbf{ab}}^{\mathbf{0\cdot\cdot}}=-\mathbf{\Gamma
}_{\cdot\mathbf{a\cdot}}^{\mathbf{a\cdot0}}=\mathbf{\Gamma}_{\cdot\mathbf{a0}%
}^{\mathbf{a\cdot\cdot}}=0 \label{16a}%
\end{equation}
and we realize that each observer following an integral line of
$\boldsymbol{R}$, say $\sigma_{1}$ will maintain a \emph{constant}
\emph{proper distance} to any of its neighbor observers which are following a
different integral line of $\boldsymbol{R}$.\ 

Of course, proper distance between an observer following $\sigma$ and another
one following $\sigma^{\prime}$ is operationally obtained\ in the following
way: Using Rindler coordinates at an event, say $\mathfrak{e}_{1}%
=(0,0,0,\mathrm{z}_{1})$ the observer following $\sigma$\emph{ }send a light
signal to $\sigma^{\prime}$\ (in the direction $\boldsymbol{e}_{3}$) which
arrives at the $\sigma^{\prime}$ worldline at\ the event \ $\mathfrak{e}%
_{2}=(\mathrm{t}_{2},0,0,\mathrm{z}_{1}+\ell)$ where it is immediately
reflected back to $\sigma$ arriving at event $\mathfrak{e}_{3}=(\mathrm{t}%
_{3},0,0,\mathrm{z}_{1})$. So, the total coordinate time for the two way trip
of the light signal is $t_{3}$ and immediately we get (from the null geodesic
equation followed by the light signal)
\begin{align}
\mathrm{t}_{2} &  =\ln\left(  1+\frac{\ell}{\mathrm{z}_{1}}\right)
,\nonumber\\
\mathrm{t}_{3}-\mathrm{t}_{2} &  =\ln\left(  1+\frac{\ell}{\mathrm{z}_{1}%
}\right)  \label{16aa}%
\end{align}
and thus
\begin{equation}
\mathrm{t}_{3}=2\ln\left(  1+\frac{\ell}{\mathrm{z}_{1}}\right)  .\label{16A}%
\end{equation}
Now, the observer at $\sigma$ evaluates the total proper time for the total
trip of the signal, it is $\mathrm{z}_{1}t_{3}$. The \emph{proper distance} is
by definition%
\begin{equation}
d_{\sigma\sigma^{\prime}}:=\frac{1}{2}\mathrm{z}_{1}\mathrm{t}_{3}%
=\mathrm{z}_{1}\ln\left(  1+\frac{\ell}{\mathrm{z}_{1}}\right)  .\label{17A}%
\end{equation}
Eq.(\ref{17A}) shows that proper distance and coordinate distance are
different in a Rindler reference frame.

\begin{remark}
A look at \emph{Figure 1} shows immediately that inertial observers in
$\boldsymbol{I}=\partial/\partial t$ will find that the distance between
$\sigma$ and $\sigma^{\prime}$ is shortening with the passage of $t$ time. It
is opportune to take into account that despite the fact that the Rindler
coordinate times for the going and return paths are equal (the coordinate time
being equal to proper time in $\sigma$\emph{) }measured by the inertial
observers are different and indeed as it is intuitive the return path is
realized in a shorter inertial time.
\end{remark}

\begin{remark}
Of course, if $\boldsymbol{R=}\frac{1}{\mathrm{z}}\partial/\partial
$\textrm{$t$} is physically realizable by a rocket with the constraint that,
e.g., $\mathrm{z}_{1}\leq\mathrm{z}\leq(\mathrm{z}_{1}+\ell)$ then it needs to
have a very\ special propulsion system, with its rear accelerating faster than
the front. We do not see how such a rocket could be constructed.\footnote{Note
that the original Rindler reference frame $\boldsymbol{R}$ for which
$(0<z<\infty)$ is only supposed to be a theoretical construct, it obviously
cannot be realized by any material system.}
\end{remark}

\section{Bell `Paradox'}

In \cite{bell} it is proposed the following question:

\begin{quotation}
Three small spaceships, A, B, and C, drift freely in a region of spacetime
remote from other matter, without rotation and without relative motion, with B
and C equidistant from \ A (Fig.1).

%

%TCIMACRO{\FRAME{ftbpFU}{5.0436in}{2.0557in}{0pt}{\Qcb{Figure 1 in Bell
%\cite{bell}}}{}{Figure 2}{\special{ language "Scientific Word";
%type "GRAPHIC";  maintain-aspect-ratio TRUE;  display "USEDEF";
%valid_file "T";  width 5.0436in;  height 2.0557in;  depth 0pt;
%original-width 7.1849in;  original-height 2.9136in;  cropleft "0";
%tempfilename 'OK8WNW0G.bmp';tempfile-properties "XPR";}} }%
%BeginExpansion
\begin{figure}[htb]%
\centering
\includegraphics[width=9cm]{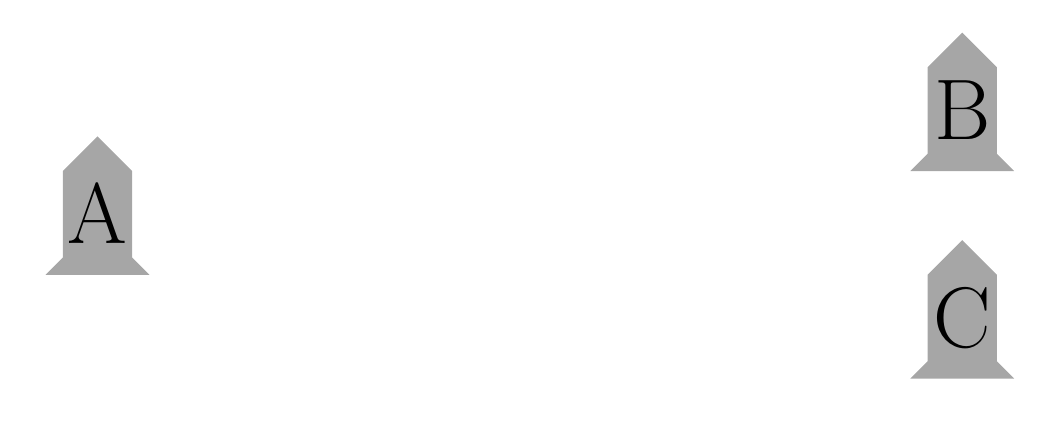}
\caption{Figure 1 in Bell \cite{bell} (adapted)}%
\end{figure}
%EndExpansion
%

On reception of a signal from A the motors of B and C are ignited and they
accelerate gently (Fig.2)

%TCIMACRO{\FRAME{ftbpFU}{4.8559in}{2.3419in}{0pt}{\Qcb{Figure 3: Figure 2 in
%Bell \cite{bell}}}{}{Figure 3}{\special{ language "Scientific Word";
%type "GRAPHIC";  maintain-aspect-ratio TRUE;  display "USEDEF";
%valid_file "T";  width 4.8559in;  height 2.3419in;  depth 0pt;
%original-width 7.1952in;  original-height 3.4566in;  cropleft "0";
%tempfilename 'OK8WNW0H.bmp';tempfile-properties "XPR";}} }%
%BeginExpansion
\begin{figure}[htb]%
\centering
\includegraphics[width=9cm]{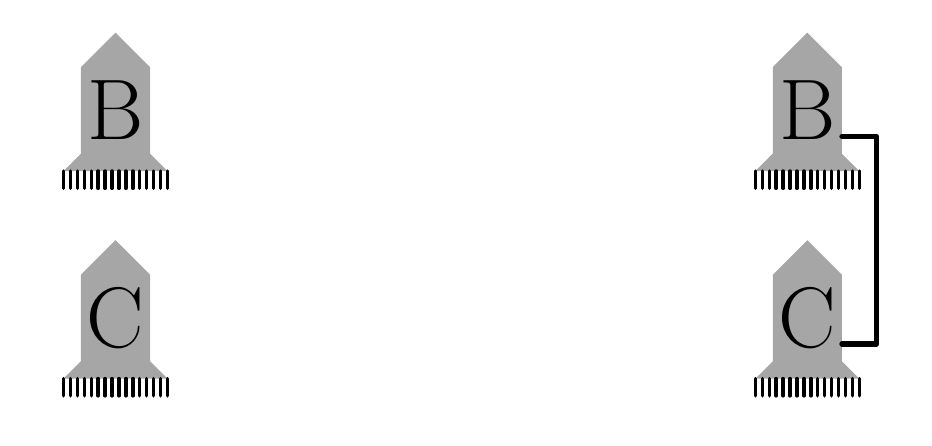}
\caption{Figure 3: Figure 2 in Bell \cite{bell} (adapted)}%
\end{figure}
%EndExpansion

Let ships B and C be identical, and have identical acceleration programmes.
Then (as reckoned by an observer at A) they will have at every moment the same
velocity, and so remain displaced one from the other by a fixed distance.
Suppose that a fragile thread is tied initially between projections form B to
C (Fig.3). If it is just long enough to span the required distance initially,
then as the rockets speed up, it will become to short, because of its need to
FitzGerald contract, and must finally break. It must break when at a
sufficiently high velocity the artificial prevention to the natural
contraction imposes intolerable stress.
\end{quotation}

Then Bell continues saying:

\begin{quotation}
Is this really so? This old problem came up for discussion once in the CERN
canteen. A distinguished experimental physicists refused to accept that the
thread would break, and regarded my assertion, that indeed it would, as a
personal misinterpretation of special relativity. We decided to appeal to the
CERN Theory Division for arbitration, and made a (not very systematic) canvas
of opinion in it. there emerged a clear consensus that the tread would
\textbf{not} break.

Of course many people who give this wrong answer at first get the right answer
on further reflection.
\end{quotation}

Recently Motl \cite{motl} wrote a note saying that Bell did not understand
Special Relativity since the correct answer to his question is the CERN
majority (first sight) view. Now, reading Motl's article one arrive at the
conclusion that he did not understand correctly the \emph{formulation} of
Bell's problem. Indeed, the problem that is correctly analyzed in \cite{motl}
was the one in each ships B\ and C are modelled as two distinct observers
following two different integral lines of the Rindler reference frame
$\boldsymbol{R}$ introduced in the previous section.

It is quite obvious to any one that read Section 1 that in this case (which is
\emph{not} the Bell's one)) B an C did not have the \emph{same} acceleration
programme as seem by observer A (represented by a particular integral line of
the inertial frame $\boldsymbol{I}=\partial/\partial t$ the $t$ axis in Figure
4).
%TCIMACRO{\FRAME{fhFU}{5.6066in}{3.8804in}{0pt}{\Qcb{Spacetime diagram for
%Bell's question with ships B (tick line on the left) and C (tick line on the
%right) having the same acceleration relative to the inertial observer A.}}%
%{}{Figure}{\special{ language "Scientific Word";  type "GRAPHIC";
%maintain-aspect-ratio TRUE;  display "USEDEF";  valid_file "T";
%width 5.6066in;  height 3.8804in;  depth 0pt;  original-width 7.2065in;
%original-height 4.9787in;  cropleft "0";  croptop "1";  cropright "1";
%cropbottom "0";  tempfilename 'OK8WNW0I.bmp';tempfile-properties "XPR";}} }%
%BeginExpansion
\begin{figure}[h]%
\centering
\includegraphics[width=11cm]{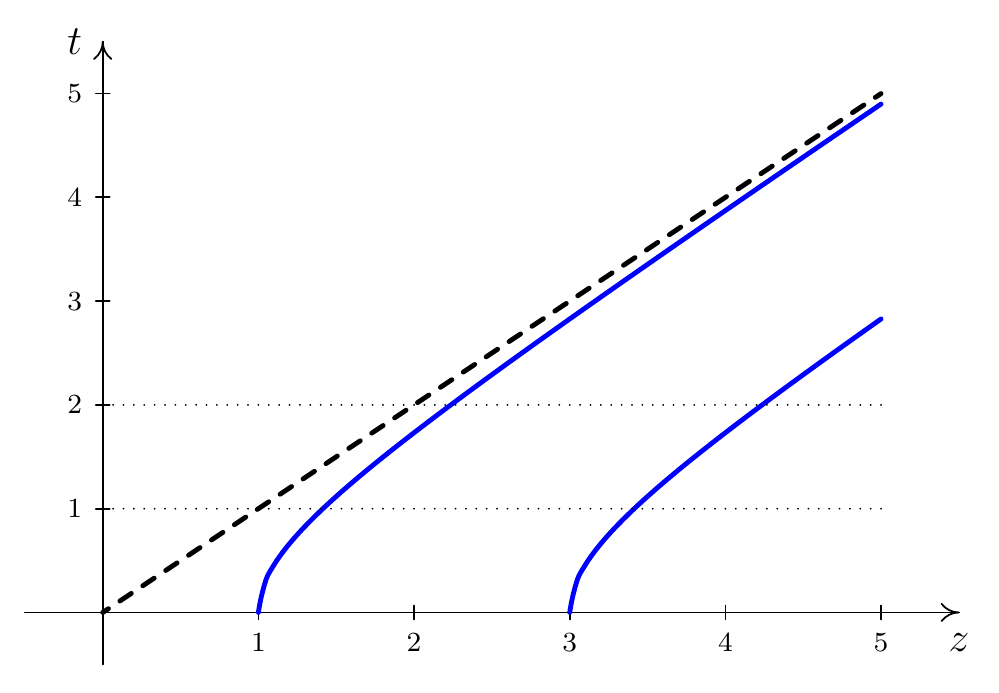}
\caption{Spacetime diagram for Bell's question with ships B (tick line on the
left) and C (tick line on the right) having the same acceleration relative to
the inertial observer A.}%
\end{figure}
%EndExpansion

In the case of Bell's question ships B and C are modelled (as a first
approximation) as observers, i.e., as the timelike curves%
\begin{align*}
t_{B}^{2}-x_{B}^{2} &  =-\frac{1}{a_{B}^{2}},\\
t_{C}^{2}-(x_{C}-d)^{2} &  =-\frac{1}{a_{C}^{2}}=-\frac{1}{a_{B}^{2}},
\end{align*}
where to illustrate the situation we draw Figure 4 with $a_{B}=1$ and $d=2$.
It is absolutely clear from Figure 4 that the distance between B and C any
instant $t>0$ as determined by the inertial observer is the same as it was at
$t=0$, when B and C start accelerating with the same accelerating programme.

A trivial calculation similar to the one in Subsection \ref{dist} above shows
that proper distance between B and C as determined by B (or C) is
\emph{increasing} with the coordinate time $\mathrm{t}$ used by these
observers which are modelled as integral lines of the Rindler reference frame
$\boldsymbol{R}$. As a consequence of this fact we arrive at the conclusion
that the thread cannot go during the acceleration period to its \emph{natural}
Lorentz deformed configuration and thus will break.\smallskip

Bell's problem illustrate that bodies subject to special acceleration programs
do not go to their Lorentz deformed configuration immediately. After the
acceleration programme ends the body will acquire adiabatically its Lorentz
deformed configuration. More on this issue is discussed in \cite{rs2001}.

\section{Does a Charge in Hyperbolic Motion Radiates?}

\subsection{The Answer Given by the Li\'{e}nard-Wiechert Potential}

It is usually assumed (see, e.g., \cite{jackson,
lyle1,lyle2,pano,parrott1,parrot} that the electromagnetic potential
$A=A_{\mu}(x)dx^{\mu}\in\sec T^{\ast}M$ generated by a charged particle in
hyperbolic motion with world line given by $\sigma:\mathbb{R\rightarrow}M$,
$s\mapsto\sigma(s)$, with parametric equations given by Eq.(\ref{3}) and
electric current given $J=\left.  J_{\mu}(x(s))dx^{\mu}\right\vert _{\sigma
}=\left.  eV_{\mu}(s))dx^{\mu}\right\vert _{\sigma}\sec T^{\ast}M$ where
\begin{align}
v^{\mu}(s) &  :=\frac{d}{ds}x^{\mu}\circ\sigma(s),~~~v:=(v^{0},\mathbf{v}%
)=\left(  \frac{1}{\sqrt{1-\mathbf{v}^{2}}},0,0,\frac{\mathbf{v}^{i}}%
{\sqrt{1-\mathbf{v}^{2}}}\right)  ,\label{R1}\\
J_{\mu}(x) &  =e\int dsv_{\mu}(s)\delta^{(4)}(x^{\prime}-x\circ\sigma
(s))\label{R2}%
\end{align}
is given by the solution of the differential equation%
\begin{equation}
\square A_{\mu}=J_{\mu}\label{R3}%
\end{equation}
through the well known formula
\begin{equation}
A_{\mu}(x)=e\int d^{4}xD_{r}(x-x^{\prime})J_{\mu}(x^{\prime})\label{R4}%
\end{equation}
where $D_{r}(x-x^{\prime})$ is the retarded Green function\footnote{I.e., a
solution of $\square D_{r}(x-x^{\prime})=\delta^{(4)}(x-x^{\prime})$.} given
by
\begin{align}
D_{r}(x-x^{\prime}) &  =\frac{1}{2\pi}\theta(x^{0}-x^{\prime0})\delta
^{(4)}[(x-x^{\prime})^{2}]\nonumber\\
&  =\frac{\theta(x^{0}-x^{\prime0})}{4\pi\mathrm{R}}\delta(x^{0}-x^{\prime
0}-\mathrm{R})\label{R5}%
\end{align}
with from the light cone constraint in Eq.(\ref{R5})
\begin{equation}
\mathrm{R}=\left\vert \mathbf{x}-\mathbf{x}(\sigma(s)\right\vert =\left\vert
x^{0}-x^{0}(s)\right\vert .\label{R6}%
\end{equation}
Thus using Eq.(\ref{R5}) in Eq.(\ref{R4}) gives the famous
Li\'{e}nard-Wiechert formula,i.e.,
\begin{equation}
A_{\mu}(x)=\frac{e}{4\pi}\left.  \frac{v_{\mu}(s)}{v\cdot\lbrack
x-x(\sigma(s))]}\right\vert _{s=s_{0}}\label{R7}%
\end{equation}
and putting $\gamma=1/\sqrt{1-\mathbf{v}^{2},}$ we have%
\begin{equation}
v\cdot\lbrack x-x(\sigma(s))]=\gamma\mathrm{R}(1-\mathbf{v}\bullet
\mathbf{n})\label{R9}%
\end{equation}
and thus%
\begin{equation}
A^{0}(t,x)=\frac{e}{4\pi}\left.  \frac{1}{(1-\mathbf{v\bullet n})\mathrm{R}%
}\right\vert _{\mathrm{ret}},~~\mathbf{A}(t,x)=\frac{e}{4\pi}\left.
\frac{\mathbf{v}}{(1-\mathbf{v\bullet n})\mathrm{R}}\right\vert _{\mathrm{ret}%
}\label{R10}%
\end{equation}
where \textrm{ret} means that that the value of the bracket must be calculated
at the instant $x^{0}(s_{0})=x^{0}-\mathrm{R}$.

We also have for the components of the field $F=dA\in\sec%
%TCIMACRO{\tbigwedge \nolimits^{2}}%
%BeginExpansion
{\textstyle\bigwedge\nolimits^{2}}
%EndExpansion
t^{\ast}M$%
\begin{equation}
F_{\mu\nu}(x)=\frac{e}{4\pi}\frac{1}{v\cdot\lbrack x-x(\sigma(s))]}\frac
{d}{ds}\left[  \frac{[x-x_{\sigma}(s)]_{\mu}v_{\nu}-[x-x_{\sigma}(s)]_{\nu
}v_{\mu}}{v\cdot\lbrack x-x(\sigma(s))]}\right]  _{\mathrm{ret}} \label{R11}%
\end{equation}
and taking into account that $[x-x_{\sigma}(s)]=(\mathrm{R},\mathrm{R}%
\mathbf{n}),~~~v_{\mu}=(\gamma,-\gamma\mathbf{v})$ and putting
$\overset{\bullet}{\mathbf{v}}=d\mathbf{v}/dt$ it is%
\begin{equation}
\frac{dv_{\mu}}{ds}=\gamma^{2}\left(  \gamma^{2}\mathbf{v\bullet\dot{v}%
},-(\mathbf{\dot{v}}+\gamma^{2}\mathbf{v}(\mathbf{v\bullet\dot{v}}))\right)
\label{R12}%
\end{equation}
and
\begin{equation}
\frac{d}{ds}[v\cdot(x-x(\sigma(s))]=-1+(x-x(\sigma(s)))_{\alpha}%
\frac{dv^{\alpha}}{ds} \label{R13}%
\end{equation}
and thus we get%

\begin{gather}
\mathbf{E}(t,\mathbf{x})=\frac{e}{4\pi}\left[  \frac{(\mathbf{n}-\mathbf{v}%
)}{\gamma^{2}(1-\mathbf{v\bullet n})^{3}\mathrm{R}^{2}}\right]  _{\mathrm{ret}%
}+\frac{e}{4\pi}\left[  \frac{\mathbf{n}\times\lbrack(\mathbf{n}%
-\mathbf{v})\times\mathbf{\dot{v}}}{\gamma^{2}(1-\mathbf{v\bullet n}%
)^{3}\mathrm{R}}\right]  _{\mathrm{ret}},\label{R14}\\
\mathbf{B}(t,x)=\mathbf{n}\times\mathbf{E}(t,\mathbf{x}).\label{R14a}%
\end{gather}
Since%
\begin{equation}
\mathbf{n}\times\lbrack(\mathbf{n}-\mathbf{v})\times\mathbf{\dot{v}%
}=(\mathbf{n\bullet\dot{v}})(\mathbf{n}-\mathbf{v})-\mathbf{n}\cdot
(\mathbf{n}-\mathbf{v})\mathbf{\dot{v}}\label{R15}%
\end{equation}
we see that for the hyperbolic motion where $\mathbf{v}$ is parallel to
$\mathbf{\dot{v}}$ and
\[
\mathbf{v}(t)=a_{\boldsymbol{\sigma}}\frac{t}{\sqrt{1+a_{\boldsymbol{\sigma}%
}^{2}t^{2}}}\mathbf{\hat{e}}_{3},~~~\mathbf{\dot{v}(}t\mathbf{)}%
=a_{\boldsymbol{\sigma}}\frac{1}{(1+a_{\boldsymbol{\sigma}}^{2}t^{2})^{3/2}%
}\mathbf{\hat{e}}_{3}%
\]
the Li\'{e}nard-Wiechert potential implies in a radiation field, i.e., a field
that goes in the infinity (radiation zone) as $1/R$.

In Jackson's book \cite{jackson} (page 667) one can read that when a charge is
accelerated in a reference frame where its speed is $\left\vert \mathbf{v}%
\right\vert \ll1$, the \ Poynting vector associated to the field given by
Eqs.(\ref{R14}) and (\ref{R14a}) is
\begin{equation}
\mathbf{S}=\mathbf{E}\times\mathbf{B}=\left\vert \mathbf{E}\right\vert
\mathbf{n} \label{R16}%
\end{equation}
and the power irradiated per solid angle is \cite{jackson}
\begin{equation}
\frac{dP}{d\Omega}=\frac{e^{2}}{(4\pi)^{2}}(\mathbf{n}\times\mathbf{\dot{v}})
\label{R17}%
\end{equation}
Thus the total instantaneous irradiated power (for a nonrelativistic
accelerated charge) is
\begin{equation}
P=\frac{2}{3}\frac{e^{2}}{4\pi}\left\vert \mathbf{\dot{v}}\right\vert ^{2},
\label{R18}%
\end{equation}
a result known as Larmor formula.

The correct formula valid for arbitrary speeds and with $P^{\mu}=mV^{\mu}$ (as
one can verify after some algebra) is%
\begin{align}
P  &  =-\frac{2}{3}\frac{1}{4\pi}\frac{e^{2}}{m^{2}}\left(  \frac{dP_{\mu}%
}{ds}\frac{dP^{\mu}}{ds}\right) \nonumber\\
&  =\frac{2}{3}\frac{1}{4\pi}e^{2}\gamma^{6}\left[  \left\vert \mathbf{\dot
{v}}\right\vert ^{2}-(\mathbf{v\times\dot{v})}^{2}\right]  . \label{r19}%
\end{align}

\begin{remark}
\emph{Eq.(\ref{R17})} show that the radiated power in a linear accelerator is,
of course, bigger for electrons than for, e.g., protons. However, as commented
by Jackson \emph{\cite{jackson}} even for electrons in a linear accelerator
with typical gain of 50 MeV/m the radiation loss is completely negligible In
the case of circular accelerators like synchrotrons \ since the momentum
$\mathbf{p}=\gamma m\mathbf{v}$ changes in direction rapidly we can show that
the radiated power \emph{(}predicted from the Li\'{e}nard-Wiechert
potential\emph{)} is
\begin{equation}
P=\frac{2}{3}\frac{1}{4\pi}\frac{e^{2}}{m^{2}}\gamma^{2}\omega^{2}\left\vert
\mathbf{p}\right\vert ^{2}\label{R20}%
\end{equation}
where $\omega$ is the angular momentum of the charged particle. This formula
fits well the experimental results.
\end{remark}

\subsection{Pauli's Answer}

In this section we use the same parametrization as before for the coordinates
of the charged particle in hyperbolic motion. Let $\mathfrak{e}$ (see Figure
5) be an arbitrary observation point with coordinates $x=(x_{{}}^{0}%
=t,x^{1},x^{2},x^{3}=z)$. In what follows for simplicity of writing we denote
the expression for the Lenard-Wiechert potential (Eq.(\ref{R7})) as
\begin{equation}
A_{\mu}(x)=\frac{e}{4\pi}\frac{v_{\mu}(s)}{v\cdot\lbrack x-x(\sigma
(s))]},\label{R22}%
\end{equation}
but we cannot forget that at the end of our calculations we must put $s=s_{0}%
$. We have, explicitly for the velocity of the particle (moving in the $x^{3}%
$-direction with $a_{\sigma}=1$)%
\begin{equation}
v^{0}(s)=\cosh s,~~~v^{3}(s)=\sinh s\label{r23}%
\end{equation}
and so%
\begin{align}
v\cdot\lbrack x-x(\sigma(s))] &  =x^{0}\cosh s-x^{3}\sinh s=\mathrm{x}%
^{3}\sinh(s-x^{0})\nonumber\\
&  =\mathrm{z}\sinh(s-t).\label{r24}%
\end{align}
Then, we have
\begin{equation}
A^{0}(x)=\frac{e}{4\pi}\frac{\cosh s}{\mathrm{z}\sinh(s-\mathrm{t})}%
,~~~A^{3}(x)=\frac{e}{4\pi}\frac{\sinh s}{\mathrm{z}\sinh(s-\mathrm{t}%
)}\label{r25}%
\end{equation}
which are Eqs (249) in Pauli's book \cite{pauli}.

Pauli's argument for saying that a charge in hyperbolic motion does not
radiate is the following:

\textbf{(i)} Consider the inertial reference frame $\boldsymbol{I}^{\prime}$
where the charge is momentarily at rest at the instant $(x_{\mathfrak{e}%
^{\prime}}^{0}-\mathrm{R})=t_{0}$. This is the time coordinate (in the
coordinates of the inertial frame $\boldsymbol{I}$) of the event
$\mathfrak{e}_{0}$ in Figure 5.%

%TCIMACRO{\FRAME{ftbpFU}{6.3373in}{4.0145in}{0pt}{\Qcb{Graphic for presenting
%Pauli's argument}}{}{Figure 5}{\special{ language "Scientific Word";
%type "GRAPHIC";  maintain-aspect-ratio TRUE;  display "USEDEF";
%valid_file "T";  width 6.3373in;  height 4.0145in;  depth 0pt;
%original-width 7.2385in;  original-height 4.5766in;  cropleft "0";
%tempfilename 'OK8WNW0J.bmp';tempfile-properties "XPR";}} }%
%BeginExpansion
\begin{figure}[ptb]%
\centering
\includegraphics[width=15cm]{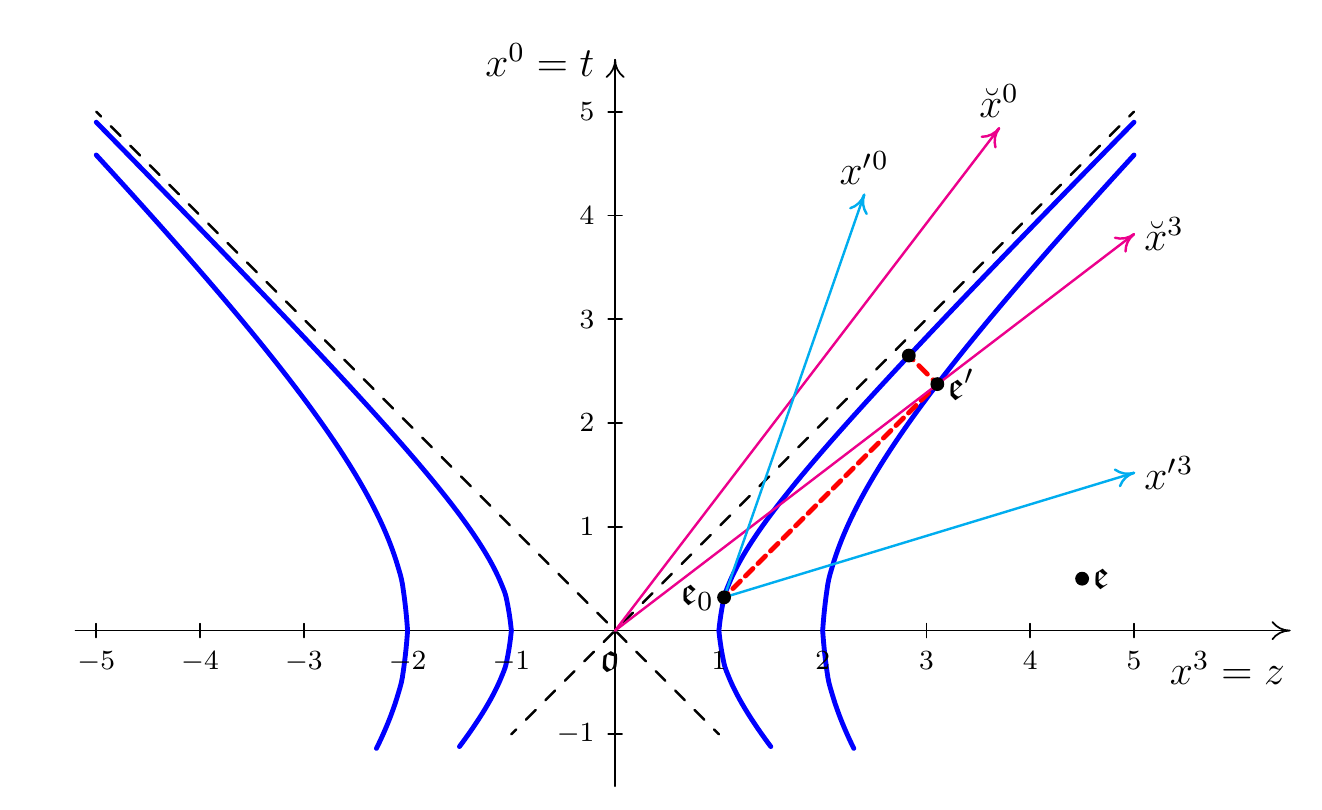}
\caption{Graphic for presenting Pauli's argument}%
\end{figure}
%EndExpansion

A naturally adapted coordinate system for the reference frame $\boldsymbol{I}%
^{\prime}$ is ($\mathfrak{v}\mathcal{=}\left\vert \mathbf{v}\right\vert $)%
\begin{align}
x^{\prime0}  &  =t_{0}+\gamma(x^{0}-\mathfrak{v}x^{3}),\nonumber\\
x^{\prime3}  &  =z_{0}+\gamma(x^{3}-\mathfrak{v}x^{0}),\nonumber\\
x^{\prime1}  &  =x^{1},~~~x^{\prime2}=x^{2}. \label{R26}%
\end{align}
and
\begin{align}
\frac{\partial x^{\prime0}}{\partial x^{0}}  &  =\gamma=\cosh s,~~~\frac
{\partial x^{\prime0}}{\partial x^{3}}=-\sinh s,\nonumber\\
\frac{\partial x^{\prime3}}{\partial x^{0}}  &  =-\gamma\mathfrak{v}=-\sinh
s,~~~\frac{\partial x^{\prime3}}{\partial x^{3}}=\cosh s. \label{r27}%
\end{align}
from where it follows that the components of the potential $A$ in the new
coordinates $\{x^{\prime\mu}\}$ are%
\begin{equation}
A^{\prime0}(x^{\prime})=\frac{e}{4\pi}\frac{1}{\mathrm{z}\sinh(s-\mathrm{t}%
)},~~~A^{\prime3}(x^{\prime})=0. \label{r28}%
\end{equation}

As a consequence of Eq.(\ref{r28}) it follows that the magnetic field
$\mathbf{B}^{\prime}\mathbf{\ }$as measured in the reference frame
$\boldsymbol{I}^{\prime}$ is null, thus the Poynting vector in this frame
$\mathbf{S}^{\prime}\mathbf{=\mathbf{E}^{\prime}\times B}^{\prime}=0$ and thus
(according to Pauli) an observer instantaneously at rest at event
$\mathfrak{e}_{0}$ with respect to the charge will detect no radiation$.$

\textbf{(ii)} To conclude his argument Pauli consider a second inertial
reference frame $\boldsymbol{\breve{I}}$ where the events $\mathfrak{o}$ and
$\mathfrak{e}^{\prime}$ are simultaneous and where $\mathfrak{e}^{\prime}$ is
an event on the world line of another observer at rest in the $\boldsymbol{R}$
frame which supposedly will receive ---if it exists---the radiation field
emitted by the charge at event $\mathfrak{e}_{0}$ (see Figure 5). A naturally
adapted coordinate system to $\boldsymbol{\breve{I}}$ is%
\begin{align}
\breve{x}^{0} &  =\breve{\gamma}(x^{0}-\mathfrak{\breve{v}}x^{3}),\nonumber\\
\breve{x}^{1} &  =x^{1},~~~\breve{x}^{2}=x^{2},\nonumber\\
\breve{x}^{3} &  =\breve{\gamma}(x^{3}-\mathfrak{\breve{v}}x^{0}),\label{r29}%
\end{align}
with%
\begin{equation}
\mathfrak{\breve{v}}=\sinh\mathrm{t/\cosh t},~~~\breve{\gamma}%
=(1-\mathfrak{\breve{v}}^{2})^{-1/2}=\cosh\mathrm{t.}\label{r30}%
\end{equation}

A trivial calculation gives%

\begin{equation}
\breve{A}^{0}(\breve{x})=\frac{e}{4\pi}\frac{\coth(s-\mathrm{t)}}%
{\sqrt{(\check{x}^{3})^{2}-((\check{x}^{0})^{2}}},~~~\breve{A}^{3}(\breve
{x})=\frac{e}{4\pi}\frac{1}{\sqrt{(\check{x}^{3})^{2}-(\check{x}^{0})^{2}}}.
\label{r31}%
\end{equation}
and since $\mathbf{\breve{B}=(}F_{32},F_{13},F_{21}\mathbf{)}=0$ it follows
that the Poynting vector $\mathbf{\breve{S}=\breve{E}\times\breve{B}}%
=0$\textbf{. }Thus an instantaneous observer $(\mathfrak{e}^{\prime
},\boldsymbol{\breve{I}}_{\mathfrak{e}^{\prime}})$ in the $\boldsymbol{\breve
{I}}$ frame momentously at rest relative to instantaneous observer
$(\mathfrak{e}^{\prime},\boldsymbol{R}_{\mathfrak{e}^{\prime}})$ observer \ in
the $\boldsymbol{R}$ frame at the considered event will also not detect any
radiation emitted from $\mathfrak{e}_{0}.$

\subsubsection{Calculation of Components of the Potentials in the
$\boldsymbol{R}$ Frame}

Using an obvious notation we write the components of the electromagnetic
potential in the in the $\boldsymbol{R}$ frame as $A(\boldsymbol{x}^{\prime
}(\mathfrak{e}))=(\mathrm{A}^{0}(\mathrm{t},\mathrm{z}),0,0,-\mathrm{A}%
^{3}(\mathrm{t},\mathrm{z}))$ and we have%

\begin{align}
\mathrm{A}_{0}  &  =\frac{\partial x^{0}}{\partial\mathrm{x}^{0}}A_{0}%
+\frac{\partial x^{3}}{\partial\mathrm{x}^{0}}A_{3}=\frac{e}{4\pi}\left.
\coth(\mathrm{t-}s)\right\vert _{s=s_{0}},\nonumber\\
\mathrm{A}_{3}  &  =\frac{\partial x^{0}}{\partial\mathrm{x}^{3}}A_{0}%
+\frac{\partial x^{3}}{\partial\mathrm{x}^{3}}A_{3}=-\frac{e}{4\pi\mathrm{z}%
}\left.  \tanh(s-\mathrm{t})\right\vert _{s=s_{0}}. \label{r32}%
\end{align}

So,%
\begin{align}
\text{ }\overrightarrow{\mathrm{E}}\mathrm{(t,z)} &  :=(0,0,\mathrm{F}%
_{03}(\mathrm{t,z})),~~~\overrightarrow{\mathrm{B}}\mathrm{(t,z)}%
=0,\label{33}\\
\mathrm{F}_{03}(\mathrm{t,z}) &  =\left.  \frac{\partial}{\partial\mathrm{t}%
}\mathrm{A}_{3}(\mathrm{t,z})\right\vert _{s=s_{0}}-\left.  \frac{\partial
}{\partial\mathrm{z}}\mathrm{A}_{0}(\mathrm{t,z})\right\vert _{s=s_{0}}%
\end{align}
and again the Poynting vector $\overrightarrow{\mathrm{E}}\times$
$\overrightarrow{\mathrm{B}}$ is null. So, by Paui's argument the observers at
rest in the $\boldsymbol{R}$ frame will detect no radiation.

\subsection{Is Pauli Argument Correct?}

In order to evaluate if Pauli's argument is correct we recall that the
Li\'{e}nard-Wiechert potential $A\in\sec%
%TCIMACRO{\tbigwedge \nolimits^{1}}%
%BeginExpansion
{\textstyle\bigwedge\nolimits^{1}}
%EndExpansion
T^{\ast}M$ by construction is in Lorenz gauge, i.e., $\delta A=0$ and
moreover\ it satisfy the homogeneous wave equation for all spacetime points
outside the worldline of the accelerated charge, i.e.,%
\begin{equation}
\Diamond A=-d\delta A-\delta dA=-\delta dA=0\label{34}%
\end{equation}
where $\Diamond$ is the Hodge Laplacian, and $\delta$ is the Hodge
coderivative. Since $F=dA\in\sec%
%TCIMACRO{\tbigwedge \nolimits^{2}}%
%BeginExpansion
{\textstyle\bigwedge\nolimits^{2}}
%EndExpansion
T^{\ast}M$ and%
\begin{equation}
\Diamond F=-d\delta dA-\delta ddA=-d\delta dA=0\label{35}%
\end{equation}
it follows that the electromagnetic field satisfies also a wave equation.

\begin{remark}
Well, it is common practice to call an electromagnetic field satisfying the
wave equation a electromagnetic wave. So, despite the fact that
$\overrightarrow{\mathrm{B}}$\textrm{ }$=0$ observers outside the worldline of
the accelerated charge \emph{(}and living in the same accelerated
laboratory\emph{)} will perceive a pure electric wave.
\end{remark}

In our case
\begin{equation}
F=\mathrm{F}_{03}d\mathrm{x}^{0}\wedge d\mathrm{x}^{3} \label{36}%
\end{equation}
and the energy momentum tensor of the electromagnetic field%
\begin{equation}
\mathbf{T}=\mathrm{T}_{\mu\nu}d\mathrm{x}^{\mu}\otimes d\mathrm{x}^{\nu}%
\in\sec T_{0}^{2}M \label{37}%
\end{equation}
in the coordinates $\{\mathrm{x}^{\mu}\}$ (naturally adapted to the Rindler
frame $\boldsymbol{R}$) has only the following non null component.%

\begin{equation}
\mathrm{T}^{00}(\mathrm{t},\mathrm{z})=\frac{1}{2}\left\vert \mathrm{F}%
_{03}(\mathrm{t},\mathrm{z})\right\vert ^{2}\label{38}%
\end{equation}
So an observer, following the worldline $\sigma^{\prime}$ with 
$\mathrm{z}=\mathrm{z}_{0}=~$constant ($\textrm{z} > 1$) will detected a
pseudo-energy density\ \textquotedblleft wave\textquotedblright\ passing
through the point where he is locate. Moreover, if this observer\textrm{
}carries with him an electric charge say $e^{\prime}$ he will certainly detect
that his charge is acted by the electromagnetic field with a ($1$-form) force%
\begin{equation}
\mathfrak{F}=e^{\prime}v_{\sigma^{\prime}}\lrcorner F=v_{\sigma^{\prime}}%
^{0}\mathrm{F}_{03}d\mathrm{x}^{3}\label{39}%
\end{equation}
and he certainly will need more pseudo energy or better more Minkowski energy
(fuel in his rocket) to maintain his charge (with mass $m^{\prime}$) at
constant acceleration than the energy that he would have to use to maintain at
a constant acceleration a particle with mass $m^{\prime}$ and null charge.

Also, since the energy arriving at the $\sigma^{\prime}$ worldline must be
coming from energy radiated by the charge following $\sigma$, an observer
maintaining the charge $e$ (of mass $m$) at constant acceleration will expend
more Minkowski energy than the one necessary for maintaining at a constant
acceleration a particle with mass $m$ and null charge.

\subsection{The Rindler (Pseudo) Energy}

It is a well known fact that outside the worldline $\sigma$ of the
accelerating charge the electromagnetic energy-momentum tensor has null
divergence, i.e.,satisfy
\begin{equation}
D\cdot\mathbf{T=0}\label{40}%
\end{equation}
where $D$ is the Levi-Civita connection of $\boldsymbol{g}$. Since
$\boldsymbol{K=}\frac{\partial}{\partial\mathrm{t}}$ is a Killing vector field
for the metric $\boldsymbol{g}$ as it is obvious looking at the representation
of $\boldsymbol{g}$ in terms of the coordinates $\{\mathrm{x}^{\mu}\}$ adapted
to the $\boldsymbol{R=}\frac{1}{\mathrm{z}}\boldsymbol{K}$ frame we have that
the current
\begin{equation}
\mathcal{J}_{R}=\mathrm{K}^{\nu}\mathrm{T}_{\nu\mu}d\mathrm{x}^{\mu}\label{41}%
\end{equation}
is conserved, i.e.,
\begin{equation}
\underset{\boldsymbol{g}}{\delta}\mathcal{J}_{R}=-\boldsymbol{\partial
\lrcorner}\mathcal{J}_{R}=-\frac{1}{\sqrt{-\det\boldsymbol{g}}}\frac{\partial
}{\partial\mathrm{x}^{\mu}}\left(  \sqrt{-\det\boldsymbol{g}}\mathrm{K}^{\nu
}\mathrm{T}_{\nu}^{\mu}\right)  =0.\label{42}%
\end{equation}

Then, of course, the \emph{scalar} quantity\footnote{If $N\subset M$ is the
region where $\mathcal{J}_{R}$ has support then $\partial N=\Xi+\Xi^{\prime
}+\digamma$ where $\Xi$ and $\Xi^{\prime}$ are spacelike surfaces and
$\mathcal{J}_{R}$ is null in $\digamma$ (spatial infinity).}%
\begin{equation}
\mathcal{E=}\int_{\Sigma^{^{\prime}}}\star\mathcal{J}_{R} \label{42a}%
\end{equation}
is a conserved one. However, take notice that differently of the case of the
similar current calculated with the Killing vector field $\partial/\partial t$
it does not qualify as the zero component of a momentum covector (\emph{not
}covector field). See details in \cite{rs2016}.

In our case we have
\begin{equation}
\frac{\partial}{\partial\mathrm{x}^{\mu}}\left(  \mathrm{zT}_{0}^{\mu}\right)
=0 \label{43}%
\end{equation}

Consider the accelerating charge following the $\sigma$ worldline (for which
$\mathrm{z}=1$ and $s=\mathrm{t}$) surrounded by a $2$-dimensional sphere
$\Sigma_{\mathrm{t}}$ of constant radius $\mathrm{r}=\mathfrak{R}$ at time
\textrm{t}. Now, from propertime $s_{1}=\mathrm{t}_{1}$ to propertime
$s_{2}=\mathrm{t}_{2}$ the surface $\Sigma_{\mathrm{t}}$ moves producing a
world tube in Minkowski spacetime.

Since
\begin{equation}
\frac{\partial}{\partial\mathrm{x}^{0}}\left(  \mathrm{zT}_{0}^{0}\right)
=-\frac{\partial}{\partial\mathrm{x}^{i}}\left(  \mathrm{zT}_{0}^{i}\right)
\label{44}%
\end{equation}
the quantity $\mathcal{E(}\mathrm{t}_{1}\mapsto\mathrm{t}_{2}\mathcal{)}$
given by%
\begin{align}
\mathcal{E(}\mathrm{t}_{1}  &  \mapsto\mathrm{t}_{2}\mathcal{)}=%
%TCIMACRO{\tint \nolimits_{\mathrm{t}_{1}}^{\mathrm{t}_{2}}}%
%BeginExpansion
{\textstyle\int\nolimits_{\mathrm{t}_{1}}^{\mathrm{t}_{2}}}
%EndExpansion
d\mathrm{t}%
%TCIMACRO{\tiiint }%
%BeginExpansion
{\textstyle\iiint}
%EndExpansion
\mathrm{r}^{2}\sin\theta d\mathrm{r}d\mathrm{\theta}d\mathrm{\varphi}%
\frac{\partial}{\partial\mathrm{t}}\left(  z\mathrm{T}_{0}^{0}\right)  =-%
%TCIMACRO{\tint \nolimits_{\mathrm{t}_{1}}^{\mathrm{t}_{2}}}%
%BeginExpansion
{\textstyle\int\nolimits_{\mathrm{t}_{1}}^{\mathrm{t}_{2}}}
%EndExpansion
d\mathrm{t}%
%TCIMACRO{\tiiint }%
%BeginExpansion
{\textstyle\iiint}
%EndExpansion
\mathrm{r}^{2}d\mathrm{r}d\Omega\frac{\partial}{\partial\mathrm{x}^{i}}\left(
\mathrm{zT}_{0}^{i}\right) \nonumber\\
&  =-%
%TCIMACRO{\tint \nolimits_{\mathrm{t}_{1}}^{\mathrm{t}_{2}}}%
%BeginExpansion
{\textstyle\int\nolimits_{\mathrm{t}_{1}}^{\mathrm{t}_{2}}}
%EndExpansion
d\mathrm{t}%
%TCIMACRO{\tiint }%
%BeginExpansion
{\textstyle\iint}
%EndExpansion
\left(  \mathrm{zT}_{0}^{i}\right)  \mathrm{n}_{i}\text{ }\mathfrak{R}%
^{2}d\Omega\label{45a}%
\end{align}
(where $\{\mathrm{r,\theta,\varphi}\}$ are polar coordinates associated to
$\{\mathrm{x}^{1},\mathrm{x}^{2},\mathrm{x}^{3}\}$ and $\mathrm{n}_{i}$ are
the components of the normal vector to $\Sigma_{\mathrm{t}}$ ) is null since
$\mathrm{T}_{0}^{i}=0.$

Thus if the observer following $\sigma$ (of course, at rest relative to the
accelerating charge) decide to call $\mathcal{E(\mathrm{t}}_{1}\mapsto
\mathcal{\mathrm{t}}_{2}\mathcal{)}$ the energy radiated by the charge\ he
will arrive at the conclusion that he did not see any radiated energy.

But of course, $\mathcal{E(\mathrm{t}}_{1}\mapsto\mathcal{\mathrm{t}}%
_{2}\mathcal{)}$ is not the extra Minkowski energy (calculated above)
necessary for the observer to maintain the charge at constant acceleration.
Parrott \cite{parrot} quite appropriately nominate $\mathcal{E(\mathrm{t}}%
_{1}\mapsto\mathcal{\mathrm{t}}_{2}\mathcal{)}$ the \emph{pseudo-energy},
other people as authors of \cite{chm} call it Rindler energy.

\begin{conclusion}
What seems clear at least to us is that whereas any one can buy Minkowski
energy \emph{(}e.g., in the form of fuel\emph{)} for his rocket no one can buy
the \textquotedblleft magical\textquotedblright\ Rindler energy.
\end{conclusion}

\subsection{The Turakulov Solution}

In a paper published in the \emph{Journal of Geometry and Physics} \cite{tura}
Turakulov presented a solution for the problem of finding the electromagnetic
field of a charge in uniformly accelerate motion by direct solving the wave
equation for the potential $A\in\sec%
%TCIMACRO{\tbigwedge \nolimits^{1}}%
%BeginExpansion
{\textstyle\bigwedge\nolimits^{1}}
%EndExpansion
T^{\ast}M$ using a separation of variables method instead of using the
Li\'{e}nard-Wiechert potential used in the previous discussion. Since this
solution is not well known we recall and analyze it here with some details.

Turakulov started his analysis with the coordinates $(\mathrm{t,x,y,z})$
introduced in Section 2 and proceeds as follows. In the $\mathrm{t}=$ constant
Euclidean semi-spaces he introduced \footnote{Toroidal coordinates (also caled
bishperical coordinates) in discussed in Section 10.3 in volume II\ of the
classical book by Morse and Feshbach \cite{mf}.} \emph{toroidal} coordinates
$(u,v,\varphi)$ by%

\begin{align}
\mathrm{z}  &  =\frac{a\sinh u}{\cosh u+\cos v},~~~\rho=\frac{a\sin v}{\cosh
u+\cos v},\nonumber\\
u  &  =\tanh^{-1}\left(  \frac{2a\mathrm{z}}{\mathrm{z}^{2}+\rho^{2}+a^{2}%
}\right)  ,~~~v=\tanh^{-1}\left(  \frac{2a\mathrm{z}}{\mathrm{z}^{2}+\rho
^{2}-a^{2}}\right)  . \label{t1}%
\end{align}
(where $\rho=+\sqrt{x^{2}+y^{2}}$) and also introduce their pseudo Euclidean
generalizations for the other domains, i.e.,%

\begin{align}
\mathrm{z}  &  =\frac{a\sin u}{\cos u+\cos v},~~~\rho=\frac{a\sin v}{\cos
u+\cos v},\nonumber\\
u  &  =\tan^{-1}\left(  \frac{2a\mathrm{z}}{-\mathrm{z}^{2}+\rho^{2}+a^{2}%
}\right)  ,~~~v=\tan^{-1}\left(  \frac{2a\mathrm{z}}{-\mathrm{z}^{2}+\rho
^{2}-a^{2}}\right)  . \label{t2}%
\end{align}
Let $\sigma$ be the world line an uniformly accelerate charge, as we know it
corresponds to $\mathrm{z}=$constant and thus the surfaces $u=$ constant forms
a family of spheres defined by the equation%
\begin{equation}
(z-a\coth u_{0})+\rho^{2}=a\sinh^{-1}u \label{t3}%
\end{equation}
involving the charge. The Minkowski metric metric in region \textrm{I }and
\textrm{II }using the coordinates $(\mathrm{t,}u,v,\rho)$ reads%
\begin{equation}
\boldsymbol{g=}\left(  \frac{a}{\cosh u+\cos v}\right)  ^{2}\left(  \sinh
^{2}u~d\mathrm{t}\otimes d\mathrm{t}-du\otimes du-dv\otimes dv-\sin
^{2}vd\varphi\otimes d\varphi\right)  \label{t4}%
\end{equation}
and for regions F and P it is%

\begin{equation}
\boldsymbol{g=}\left(  \frac{a}{\cosh u+\cos v}\right)  ^{2}\left(  -\sin
^{2}u~d\mathrm{t}\otimes d\mathrm{t}+du\otimes du-dv\otimes dv-\sin
^{2}vd\varphi\otimes d\varphi\right)  . \label{t5}%
\end{equation}

As we know the potential $A^{T}$ in the Lorenz gauge $\delta A^{T}=0$
satisfies the wave equation $\delta dA^{T}=0$ \ Then supposing (as usual) that
the potential is tangent to the the integral lines of $\boldsymbol{R}$ we can
write\footnote{Here the value of the charge is $e/4\pi=1$.}
\begin{equation}
A^{T}=\Theta(u,v)d\mathrm{t}\label{t6}%
\end{equation}
and the general solution of the wave equation is
\begin{equation}
\Theta(u,v)=\alpha_{0}(\cosh u-1)+%
%TCIMACRO{\tsum \nolimits_{n=1}^{\infty}}%
%BeginExpansion
{\textstyle\sum\nolimits_{n=1}^{\infty}}
%EndExpansion
\alpha_{n}\sinh u\frac{d}{du}P_{n}(\cosh u)P_{n}(\cos\nu),\label{t7}%
\end{equation}
where $P_{n}$ are Legendre polynomials and $\alpha_{0,}\alpha_{n}$ are
constants. The field of a charge is simply specified only by the first term
with $\alpha_{0}=e$ the value of the charge generating the field. Thus, if the
charge is at $u=\infty$ we have for regions \textrm{I} and \textrm{II }and P
and F%
\begin{equation}
A_{\mathrm{I,II}}^{T}=e(\cosh u-1)d\mathrm{t,~~~}A_{\text{P,F}}^{T}=e(\cos
u-1)d\mathrm{t}.\label{t8}%
\end{equation}
In terms of the coordinates $(t,x,y,z)$, writing $A^{T}=A_{\mu}^{T}dx^{\mu}$
we have the following solution valid for all regions\footnote{We have verified
using the Mathematica software that indeed $A_{0}$ and $A_{3}$ satisfy the
wave equation. Note that ther is are signal misprints in the formulas for
$A_{0}$ and $A_{3}$ in \cite{tura}\ and the modulus $\sqrt{\left\vert
z^{2}-t^{2}\right\vert }$ in those formulas are not necessary.}:%
\begin{gather}
A_{0}^{T}=-\frac{z}{z^{2}-t^{2}}\left(  \frac{t^{2}-\rho^{2}+z^{2}-a^{2}%
}{\Lambda_{+}\Lambda_{-}}-1\right)  ,\nonumber\\
A_{3}^{T}=\frac{t}{z^{2}-t^{2}}\left(  \frac{t^{2}-\rho^{2}+z^{2}-a^{2}%
}{\Lambda_{+}\Lambda_{-}}-1\right)  ,\nonumber\\
A_{1}^{T}=A_{2}^{T}=0,\nonumber\\
\Lambda_{\pm}(t,x,y,z)=\sqrt{(\sqrt{z^{2}-t^{2}}\pm a)^{2}+x^{2}+y^{2}%
}.\label{19}%
\end{gather}

From these formulas we infer that%

\begin{equation}
F^{T}=F_{\mathrm{t}u}d\mathrm{t}\wedge du=-e\sinh ud\mathrm{t}\wedge du
\label{t10}%
\end{equation}
and thus an observer comoving with the charge will see only an
\textquotedblleft electric field\textquotedblright\ which for him is in the
$u$-direction and the \emph{pseudo} energy evaluated beyond a given sphere
$u=u_{0}$ of radius $\boldsymbol{r}$ is%
\begin{equation}
\mathcal{E}=\frac{e^{2}}{2\boldsymbol{r}}. \label{t11}%
\end{equation}
Thus, Turakulov concludes as did Pauli did that there is no radiation. But is
his conclusion correct?

\subsubsection{Does the Turakulov Solution Implies that a Charge in Hyperbolic
Motion does not Radiate?}

Recall that in subsection 4.3 we showed that supposing that the
Li\'{e}nard-Wiechert solution is the correct one then Pauli's argument is
incorrect since an observer following another integral line of $\boldsymbol{R}%
$ will see an electric \textquotedblleft wave\textquotedblright\ (recall
Eq.(\ref{38})) \ We now makes the same analysis as the one we did in the case
of the Turakulov solution in order to find the correct answer to our question.
We first explicitly calculate the electric and magnetic fields in the inertial
frame $\boldsymbol{I}=\partial/\partial t$. We have%

\begin{gather}
E_{x}=\frac{8a^{2}xz}{\Lambda_{+}^{3}\Lambda_{-}^{3}},~~~~~E_{y}=\frac
{8a^{2}yz}{\Lambda_{+}^{3}\Lambda_{-}^{3}},~~~~~E_{z}=\frac{-4a^{2}%
[x^{2}+y^{2}+a^{2}-z^{2}+t^{2}]}{\Lambda_{+}^{3}\Lambda_{-}^{3}}%
,\nonumber\\[0.06in]
B_{x}=\frac{8a^{2}yt}{\Lambda_{+}^{3}\Lambda_{-}^{3}},~~~~B_{y}=\frac
{-8a^{2}xt}{\Lambda_{+}^{3}\Lambda_{-}^{3}},~~~B_{z}=0.\label{J000}%
\end{gather}

The Poincar\'{e} invariants of the Turakulov solution $I_{1}:=\mathbf{E}%
^{2}-\mathbf{B}^{2}$ and $I_{2}:=\mathbf{E\bullet B}~$ are
\begin{equation}
I_{1}=\frac{16a^{4}}{\Lambda_{+}^{6}\Lambda_{-}^{6}}[(x^{2}+y^{2}-z^{2}%
+t^{2})^{2}+4(x^{2}+y^{2})(z^{2}+t^{2})],~~I_{2}=0.\label{J00}%
\end{equation}
This shows that an inertial observer at rest at $(x,y,z)$ will detect a
\emph{time dependent }electromagnetic field configuration passing though his
observation point. Of course, it is \emph{not} a null field, but it certainly
qualify as an\emph{ }electromagnetic wave. And what is important for our
analysis is that the field carries energy and momentum from the accelerating
charge to the point $(x,y,z)$.

Indeed, consider a charge $q$ at rest in the Rindler frame following an
integral line $\sigma^{\prime}$ of $\boldsymbol{R}$ with constant Rindler
coordinates $(\mathrm{t},\mathrm{x}=\mathrm{x}_{0},\mathrm{y}=\mathrm{y}_{0}$
$\mathrm{z}=\mathrm{z}_{0})$ and thus with inertial coordinates $(t,\mathrm{x}%
_{0},y=\mathrm{y}_{0},$ $z=\sqrt{\mathrm{z}_{0}^{2}+t^{2}})$.

As determined by the inertial observer the density of \emph{real} energy and
the Poynting vector arriving from the uniformly accelerated charge moving
along the $z$-axis of the inertial frame to where the charge $q$ is locate
are:%
\begin{gather}
\frac{1}{2}(\mathbf{E}^{2}+\mathbf{B}^{2})\nonumber\\
=\frac{1}{2}\mathring{\Lambda}_{+}^{-6}\mathring{\Lambda}_{-}^{-6}%
(128(\mathrm{x}_{0}^{2}+\mathrm{y}_{0}^{2})t^{2}+64a^{4}(\mathrm{x}_{0}%
^{2}+\mathrm{y}_{0}^{2})\mathrm{z}_{0}^{2}+16a^{4}(\mathrm{x}_{0}%
^{2}+\mathrm{y}_{0}^{2}+a^{2}-\mathrm{z}_{0}^{2})^{2}),\nonumber\\
\mathbf{S}=\mathbf{i}\frac{32a^{4}\mathrm{x}_{0}}{\mathring{\Lambda}_{+}%
^{6}\mathring{\Lambda}_{-}^{6}}(\mathrm{x}_{0}^{2}+\mathrm{y}_{0}^{2}%
+a^{2}-\mathrm{z}_{0}^{2})t+\mathbf{j}\frac{-32a^{4}\mathrm{y}_{0}}%
{\mathring{\Lambda}_{+}^{6}\mathring{\Lambda}_{-}^{6}}(\mathrm{x}_{0}%
^{2}+\mathrm{y}_{0}^{2}+a^{2}-\mathrm{z}_{0}^{2})t\nonumber\\
+\mathbf{k}\frac{64a^{4}\sqrt{z_{0}^{2}+t^{2}}}{\mathring{\Lambda}_{+}%
^{6}\mathring{\Lambda}_{-}^{6}}\mathbf{(}\mathrm{x}_{0}^{2}+\mathrm{y}_{0}%
^{2})t\mathbf{,}\nonumber\\
\mathring{\Lambda}_{\pm}=\sqrt{(\mathrm{z}_{0}\pm a)^{2}+\mathrm{x}_{0}%
^{2}+\mathrm{y}_{0}^{2}}\label{J2}%
\end{gather}

Thus, we see that indeed there is a flux of\ \emph{real} energy and momentum
arriving at the charge $q$ located at $(t,\mathrm{x}_{0},y=\mathrm{y}_{0},$
$z=\sqrt{\mathrm{z}_{0}^{2}+t^{2}}).$

Moreover, the Lorentz force $\mathbf{F}_{L}$ acting on the charge $q$
(according to the inertial observer) is%
\begin{equation}
\mathbf{F}_{L}=q\mathbf{E+qv}_{\sigma^{\prime}}\times\mathbf{B} \label{J1a}%
\end{equation}
depends on $t$ and is doing work on the charge $q$. So, an observer comoving
with the charge $q$ will need to expend more \emph{real} energy to carry this
charge than to carry a particle with zero charge.\smallskip

More important: since the energy arriving at the charge $q$ is the one
produced by the charge $e$ generating the field we arrive at the conclusion,
as in the case of the Pauli solution that an observer carrying the charge $e$
will speed more energy (fuel of its rocket) than when it carries a particle
with zero charge.

\begin{remark}
We already observed in \emph{\cite{mr}} that the use of the retarded Green's
function may result in non sequitur solutions in some cases. Most important is
the fact that in \emph{\cite{tura2}} it is observed that the Green's function
for a massless scalar field is the integral $(\omega=k_{0})$
\begin{equation}
G(x,x^{\prime})=\frac{1}{(2\pi)^{4}}\int d^{3}\mathbf{k}\int d\omega
\frac{e^{-i(\omega(t-t^{\prime})-\mathbf{k}\cdot(\mathbf{x-x}^{\prime}))}%
}{\mathbf{k}^{2}-\omega^{2}}\label{t12}%
\end{equation}
and the evaluation of the integral is done in all classical presentations in
the complex $\omega$-plane and thus its result depends, as is well known from
the path of integration chosen. But, contrary to what is commonly accepted
this is not necessary for the integrand is not singular. This can be shown as
follows. Recalling that $G$ depends only on
\[
\tau^{2}-\mathfrak{r}^{2}=(t-t^{\prime})^{2}-(\mathbf{x-x}^{\prime})^{2}%
\]
we can choose a coordinate system where $(\mathbf{x-x}^{\prime})^{2}=0$ for
the point under consideration, Then, introducing the coordinates
\begin{gather}
\varkappa=\omega^{2}-\mathbf{k}^{2},~~~\xi=\tanh^{-1}(\left\vert
\mathbf{k}\right\vert /\omega),\nonumber\\
\omega(t-t^{\prime})-\mathbf{k}\cdot(\mathbf{x-x}^{\prime})=\varkappa
\varsigma\cosh\xi\label{t13}%
\end{gather}
the \emph{Eq.( \ref{t13}) }becomes after some algebra%
\begin{equation}
G(\tau,\mathfrak{r})=\frac{1}{4\pi^{3}}\int d\varkappa\int d\xi\int
d\theta\int d\varphi\sin\theta\sinh^{3}\xi\varkappa^{2}e^{i\varkappa
\varsigma\cosh\xi}.\label{t14}%
\end{equation}
This important result obtained in\emph{ \cite{tura2} }shows explicitly that it
is possible to evaluate the Green's function without introducing the
\textquotedblleft famous" $i\varepsilon$ prescription! Turakulov also observed
that putting $\ \mathfrak{\lambda}=\varkappa\varsigma$ the \emph{Eq.(\ref{t14}%
)} gives
\begin{equation}
G(\tau,\mathfrak{r})=\frac{\pi^{2}}{\varsigma^{2}}\int d\lambda\lambda\int
d\xi\sinh^{2}\xi e^{i\lambda\cosh\xi}.\label{t15}%
\end{equation}

The conclusion is thus that integration only predetermines the factor $1/$
$\varsigma^{2}$ and it is now possible to select any path of integration in
the complex plane, which means that the retarded Green's function is create by
inserting a non-existence singularity into the integrand!

Moreover, in it is shown in \emph{\cite{tura2}} that the use of the retarded
Green's function produces problems with energy-conservation when, e.g., a
charge is accelerated in an external potential. Finally we observe that in
\emph{\cite{tura1}} it is shown that\ when there are infinitesimally small
changes of the acceleration\ there is emission of radiation.
\end{remark}

\section{The Equivalence Principle}

Consider first the statements (a) and (b):

(a) an observer (say Mary) living in a small constantly accelerated reference
frame (e.g., a `small' world tube, with non transparent walls of the reference
frame $\boldsymbol{R}$) following an integral line $\sigma$ of the
$\boldsymbol{R}$ frame and for which $\left.  D_{\boldsymbol{R}}%
\boldsymbol{R}\right\vert _{\sigma}=\boldsymbol{a}_{|\sigma}$;

(b) an observer (say John) living in a `small' reference frame, (e.g., a
`small' world tube, with non transparent walls of the reference frame
$\boldsymbol{Z}$ in a Lorentzian spacetime structure $(M,\mathbf{g}%
,\mathbf{D,\tau}_{\mathbf{g}},\uparrow)$ modelling a gravitational field
(generated by some energy-momentum distribution) in General Relativity theory
and such that $\left.  \mathbf{D}_{\boldsymbol{Z}}\boldsymbol{Z}\right\vert
_{\lambda}=\boldsymbol{a}_{|\lambda}=\boldsymbol{a}_{|\sigma}$.

Then a common formulation of the \emph{Equivalence Principle\footnote{A
thoughtful dicussion of the Equivalence Principle and the so-called Principle
of Local Lorentz Invariance is given in \cite{rs2001}}} says that Mary or John
cannot with \emph{local}\footnote{Of course, by local mathematicians means an
($4$-dimensional) open set $U$ of the appropriate spacetime manifold. So, by
doing experiments in $U$ observers will detect using a gradiometer tidal force
fields (proportional to the Riemann curvature tensor) if at rest in
$\boldsymbol{Z}$ in a real gravitational field and will not detect any tidal
force field if living in $\boldsymbol{R}$ in Minkowski spacetime. For more
details see, e.g., \cite{ohru,rs2001}.} experiments determine if she(he) lives
in an uniformly accelerated frame in Minkowski spacetime or in the
gravitational field modelled by $(M,\mathbf{g},\mathbf{D,\tau}_{\mathbf{g}%
},\uparrow)$.

Now, as well known (since long ago) and as proved rigorously (under well
determined conditions) in \cite{parrot} a charge in a static gravitational
field in General Relativity theory does not radiated if it follows an integral
line of a reference frame like $\boldsymbol{Z}$ in (b). An observer commoving
with the charge will see only an electric field and thus will see no radiation
since the Poynting vector is null.

Does this implies that the Equivalence Principle holds for local experiments
with charged matter?\smallskip

Well, if we accept that the Li\'{e}nard-Wiechert solution the correct one,
then the answer from the analysis given in the previous section is \emph{no}
(see also, \cite{lyle1,lyle2,parrot}. In particular Parrot's argument is the
following: since there is no radiation in the true gravitational field an
observer at rest in the Schwarzschild spacetime following a worldline
$\lambda$ will spend the same amount of \textquotedblleft
energy\textquotedblright\ to maintain at constant acceleration $\boldsymbol{a}%
_{|\lambda}=\boldsymbol{a}_{|\sigma}$ a particle with mass $m$ and null charge
and one with mass $m$ and charge $e\neq0$.

Since we already know that in the $\boldsymbol{R}$ frame it is clear that an
observer $\sigma$ will spend \textit{different} amounts (of Minkowski) energy
to maintain at constant acceleration $\boldsymbol{a}_{|\lambda}=\boldsymbol{a}%
_{|\sigma}$ a particle with mass $m$ and null charge and one with mass $m$ and
charge $e\neq0$.

Of course, even supposing that the Li\'{e}nard-Wiechert solution is the
correct one many people does not agree with this conclusion and some of the
arguments of the opposition \ is discussed in \cite{parrot}.

\begin{remark}
From our point of view we think necessary to comment that Parrot' s argument
would be a really strong one only if the concept of energy \emph{(}and
momentum\emph{)} would be well defined in General Relativity, which is
definitively not the case \emph{\cite{rc2016,rs12016,rs2016}}. However, take
notice that the quantity defined as \textquotedblleft energy\textquotedblright%
\ by Parrot \emph{(}the zero component of current of the form given by
\emph{Eq.(\ref{41})}, were in this case $\boldsymbol{K}$ is a timelike Killing
vector field for the Schwarzschild metric is not the componet of any
energy-momentum covector field, it looks more as the concept of energy in
Newtonian physics. . Anyway, the quantity of\ the pseudo \textquotedblleft
energy\textquotedblright\ \ necessary to carry a particle in uniformily
accelerated motion will certainly be different in the two cases of a charged
and a non charged particle. In our opinion what is necessary is to construct
an analysis of the problem charge in a gravitational theory where
energy-momentum of a system can be defined and is a conserved quantity
\emph{\cite{rc2016,rs12016}}.
\end{remark}

On the other hand if we accept that Turakulov solution as the correct one than
again the Equivalence Principle is violated and for the same reason than in
the case of the Li\'{e}nard-Wiechert solution as discussed in Section
4.5.1.\smallskip

So, which solution, Li\'{e}nard-Wiechert or Turakulov is the correct one?

An answer can be given to the above question only with a clever experiment and
for the best of our knowledge no such experiment has been done yet.

\section{Some Comments on the Unhru Effect}

\subsection{Minkowski and Fulling-Unruh Quantization of the Klein-Gordon
Field}

(\textbf{u1}) To discuss the Unruh effect it is useful to introduce
coordinates such that the solution of the Klein-Gordon equation in these
variables becomes as simple as possible. A standard choice is to take
$(\mathfrak{t,x,y,z})$ and \textrm{ }$(\mathfrak{t}^{\prime}\mathfrak{,x}%
^{\prime}\mathfrak{,y}^{\prime}\mathfrak{,z}^{\prime})$ for regions \textrm{I
}and\textrm{ II }defined by\footnote{Note that $(\mathfrak{t},\mathfrak{z})$
differs form the coodinates $(\mathrm{t},\mathrm{z})$ introduced in Section
2.}%

\begin{gather}
\mathfrak{t}=\frac{1}{a}\tanh^{-1}(\frac{t}{z}),~~~\mathfrak{z}=\frac{1}%
{2a}\ln[a(z^{2}-t^{2})],~~~\mathfrak{x}=x,~~~\mathfrak{y}=y\nonumber\\
t=\frac{1}{a}\exp(a\mathfrak{z})\sinh(a\mathfrak{t),~~~}z=\frac{1}{a}%
\exp(a\mathfrak{z})\cosh(a\mathfrak{t),~~}\left\vert z\right\vert \geq
t,~~z>0,\nonumber\\
\mathfrak{t}^{\prime}=\frac{1}{a}\tanh^{-1}(\frac{t}{z}),~~~\mathfrak{z}%
^{\prime}=\frac{1}{2a}\ln[a^{2}(z^{2}-t^{2})],~~~\mathfrak{x}^{\prime
}=x,~~~\mathfrak{y}^{\prime}=y.,\nonumber\\
t=\frac{1}{a}\exp(a\mathfrak{z}^{\prime})\sinh(a\mathfrak{t}^{\prime
}\mathfrak{),~~~}z=-\frac{1}{a}\exp(a\mathfrak{z}^{\prime})\cosh
(a\mathfrak{t}^{\prime}\mathfrak{),~~}\left\vert z\right\vert \geq
t,~~z<0,\nonumber\\
\mathfrak{t,z\in(-\infty,\infty),~~~~~}a\in\mathbb{R}^{+}. \label{F1}%
\end{gather}

Take notice that in regions \textrm{I} and \textrm{II }the coordinates
$\mathfrak{t}$ and $\mathfrak{z}$ are respectively timelike and spacelike and
in region \textrm{II }the decreasing of $\mathfrak{t}$ corresponds to the
increase of $t$.

The Minkowski metric in these coordinates (and in the regions \textrm{I
}and\textrm{ II) }reads
\begin{gather}
\boldsymbol{g}=\exp(2a\mathfrak{z})d\mathfrak{t\otimes}d\mathfrak{t-}%
d\mathfrak{x\otimes}d\mathfrak{x-}d\mathfrak{y\otimes}d\mathfrak{y-}%
\exp(2a\mathfrak{z})d\mathfrak{z\otimes}d\mathfrak{z}=\eta_{\mathbf{ab}%
}\mathfrak{g}^{\mathbf{a}}\otimes\mathfrak{g}^{\mathbf{b}},\nonumber\\
\mathfrak{g}^{\mathbf{0}}=\exp(a\mathfrak{z})d\mathfrak{t,~~g}^{\mathbf{1}%
}=d\mathfrak{x,~~g}^{\mathbf{2}}=d\mathfrak{y,~~g}^{\mathbf{3}}=\exp
(a\mathfrak{z})d\mathfrak{z.} \label{f2}%
\end{gather}

\textbf{(u2}) The right and left Rindler reference frames are represented by
\begin{align}
\boldsymbol{R}  &  =\frac{1}{\exp(a\mathfrak{z})}\mathfrak{\partial/}%
\partial\mathfrak{t,~~~}t\in\mathfrak{(-\infty,\infty)}\text{,~~~}\left\vert
z\right\vert \geq t,~~z>0,\nonumber\\
\boldsymbol{L}  &  =\frac{1}{\exp(a\mathfrak{z})}\mathfrak{\partial/}%
\partial\mathfrak{t,~~~}t\in\mathfrak{(-\infty,\infty)}\text{,~~~}\left\vert
z\right\vert \geq t,~~z<0. \label{f3}%
\end{align}
and they are \emph{not} Killing vector fields.\footnote{This can easily be
verifed taking into account that $\mathcal{L}_{\boldsymbol{R}}\boldsymbol{g}%
=2\eta_{\mathbf{ab}}\mathcal{L}_{\boldsymbol{R}}\mathfrak{g}^{\mathbf{a}%
}\otimes\mathfrak{g}^{\mathbf{b}}$ and recalling that if $R=\boldsymbol{g}%
(\boldsymbol{R},\mathfrak{~})=\mathfrak{g}^{\mathbf{0}}$ we may evalute
\cite{rc2016} as $\mathcal{L}_{\boldsymbol{R}}\mathfrak{g}^{\mathbf{a}%
}=d(\mathfrak{g}^{\mathbf{0}}\cdot\mathfrak{g}^{\mathbf{a}})+\mathfrak{g}%
^{\mathbf{0}}\lrcorner d\mathfrak{g}^{\mathbf{a}}.$}

Consider the integral line, say $\sigma$ of $\boldsymbol{R}$ given by
$\ \mathfrak{x,y=}$ constant and $\mathfrak{z=z}_{0}=$ constant. We
immediately find that its proper acceleration is%
\begin{equation}
a_{\sigma}=1/\sqrt{g_{00}(\mathfrak{z}_{0})}. \label{f3a}%
\end{equation}

(\textbf{u3}) However, the vector fields
\begin{gather}
\boldsymbol{I}=\mathbf{\partial/\partial}t,\nonumber\\
\boldsymbol{Z}_{\mathrm{I}}=\mathbf{\partial/\partial}\mathfrak{t},\text{with
}t\in(-\infty,\infty),\left\vert z\right\vert \geq t\text{ and }%
z>0,\nonumber\\
\boldsymbol{Z}_{\mathrm{II}}=\mathbf{\partial/\partial}\mathfrak{t},\text{with
}t\in(-\infty,\infty),\left\vert z\right\vert \geq t\text{ and }z<0,
\label{16x}%
\end{gather}
are Killing vector fields, i.e., $\mathcal{L}_{\partial/\partial
t}\boldsymbol{g}=\mathcal{L}_{\boldsymbol{Z}_{\mathrm{I}}}\boldsymbol{g}%
=\mathcal{L}_{\boldsymbol{Z}_{\mathrm{II}}}\boldsymbol{g}=0$. The inertial
reference frame $\boldsymbol{I}$ besides being locally synchronizable is also
propertime synchronizable, i.e., $\boldsymbol{g}(\boldsymbol{I,~})=dt$ and the
fields $\boldsymbol{Z}_{\mathrm{I}}$ and $\boldsymbol{Z}_{\mathrm{II}}$
although does not qualify as reference frames (according to our definition)
play an important role for our considerations of the Unruh effect. The reason
is that both fields in the regions where they have support are such that
\begin{align}
Z_{\mathrm{I}}  &  =\boldsymbol{g}(\boldsymbol{Z}_{\mathrm{I}},\boldsymbol{~}%
)=\exp(2a\mathfrak{z})d\mathfrak{t},\text{with }t\in(-\infty,\infty
),\left\vert z\right\vert \geq t\text{ and }z>0,\nonumber\\
Z_{\mathrm{II}}  &  =\boldsymbol{g}(\boldsymbol{Z}_{\mathrm{II}}%
,\boldsymbol{~})=\exp(2a\mathfrak{z})d\mathfrak{t},\text{with }t\in
(-\infty,\infty),\left\vert z\right\vert \geq t\text{ and }z>0. \label{16XX}%
\end{align}

Thus the field $\boldsymbol{I}$ can be used to foliate all $M$ as $M=\cup
_{t}(\mathbb{R\times}\Sigma\mathbb{(}t\mathbb{))}$ where $\Sigma
\mathbb{(}t\mathbb{)\simeq R}^{3}$ is a Cauchy surface. Moreover, the field
$Z_{\mathrm{I}}$ \emph{(}respectively $Z_{\mathrm{II}}$) can be used to
foliate region $\mathrm{I}$, (respectively region \textrm{II}) as
$\mathrm{I}=\cup_{\mathrm{t}}(\mathbb{R\times}\Sigma_{\mathrm{I}}%
\mathbb{(}\mathrm{t}\mathbb{))}$ \emph{(}respectively $\mathrm{II}%
=\cup_{\mathrm{t}}(\mathbb{R\times}\Sigma_{\mathbb{\mathrm{II}}}%
\mathbb{(}\mathfrak{t}\mathbb{))}$\emph{)} where $\Sigma_{\mathrm{I}%
}\mathbb{(}\mathfrak{t}\mathbb{)}\simeq\Sigma_{\mathrm{I}}$ and $\Sigma
_{\mathrm{II}}\mathbb{(}\mathrm{t}\mathbb{)}\simeq\Sigma_{\mathrm{II}}$ are
Cauchy surfaces.\smallskip

We now briefly describe how the Unruh effect for a complex Klein-Gordon field
is presented in almost all texts\footnote{E.g., in
\cite{chm,davies,gmm,socolovski,suli,unruh,wald2}. The presentations
eventually differ in the use of other coordinate systems.} dealing with the
issue. \smallskip

(\textbf{u4)} Let $\phi$ $\in\sec(\mathbb{C\otimes}%
%TCIMACRO{\tbigwedge \nolimits^{0}}%
%BeginExpansion
{\textstyle\bigwedge\nolimits^{0}}
%EndExpansion
T^{\ast}M)$. Our departure point is to first solve the Klein-Gordon equation%
\begin{equation}
-\delta d\phi+\mu^{2}\phi=0\label{f4}%
\end{equation}
valid for all $M$, in the global naturally adapted coordinates (in ELP gauge)
to $\boldsymbol{I}$ and next to solve it in regions $\mathrm{I}$ and
$\mathrm{II}$ using the coordinates defined in Eq.(\ref{F1}) (and then extend
this new solution for all $M$). In the first case we use the $t=0$ as Cauchy
surface to given initial data. In the second case we use the $\mathfrak{t}=0$
Cauchy surface to give initial data (see below).

The positive energy solutions will be called \emph{Minkowski modes} for the
first case and \emph{Fulling-Unruh modes} for the second case\ (i.e., the
solutions in regions $\mathrm{I}$ and $\mathrm{II}$). In order to simplify the
writing of the formulas that follows we introduce the notations%
\begin{gather}
\phi_{M}(x)=\phi_{M}(t,x,y,z),~~\phi_{\mathrm{I}}(\mathfrak{l})=\phi
_{\mathrm{I}}(\mathfrak{t,x,y,z}),~~\phi_{\mathrm{II}}(\mathfrak{l}^{\prime
})=\phi_{\mathrm{II}}(\mathfrak{t,x,y,z}),,\nonumber\\
k\cdot x=k_{\alpha}x^{\alpha},~~\omega_{\mathbf{k}}=k_{0}=+\sqrt
{\mathbf{k}^{2}+\mu^{2}},~~k\cdot k=(k_{0})^{2}-\mathbf{k}^{2}=\mu
^{2},~~\mathbf{k}^{2}=\mathbf{k}\bullet\mathbf{k},\nonumber\\
\mathbf{q=(}k_{1},k_{2}\mathbf{),~~r=(}x^{1},x^{2}\mathbf{)=(}x,y\mathbf{)}%
\text{ and }\mathbf{q\bullet r=}k_{1}x^{1}+k_{2}x^{2},~~\nu=+\sqrt
{\mathbf{q}^{2}+\mu^{2}}. \label{f5}%
\end{gather}

Observing that in region \textrm{II} the timelike coordinate $\mathfrak{t}%
^{\prime}$ decreases when $t$ increases we have that the elementary modes (of
positive energy) which are solutions of the Klein-Gordon equation in the three
regions:%
\begin{align}
\phi_{M\mathbf{k}}(x)  &  =[(2\pi)^{3}2\omega_{\mathbf{k}})]^{-1/2}e^{-ik\cdot
x},\nonumber\\
\phi_{\mathrm{I\nu}\mathbf{q}}(\mathfrak{l})  &  =[(2\pi)^{2}2\nu
)]^{-1/2}F_{\mathrm{I}\nu\mathbf{q}}(\mathfrak{z})e^{-i(\nu\mathfrak{t}%
-\mathbf{q\bullet r})},\nonumber\\
\phi_{\mathrm{I\nu}\mathbf{q}}(\mathfrak{l}^{\prime})  &  =[(2\pi)^{2}%
2\nu)]^{-1/2}F_{\mathrm{II}\nu\mathbf{q}}(\mathfrak{z})e^{+i(\nu
\mathfrak{t}^{\prime}+\mathbf{q\bullet r})}, \label{f6}%
\end{align}
with
\begin{align}
F_{\mathrm{I}\nu\mathbf{q}}(\mathfrak{z})  &  =(2\pi^{-1})^{1/2}%
C_{\mathrm{I}\mathbf{q}}\frac{1}{\Gamma(i\nu)}(\frac{\nu}{2a})^{i\nu}K_{i\nu
}(\nu\mathfrak{z}),\nonumber\\
F_{\mathrm{II}\nu\mathbf{q}}(\mathfrak{z}^{\prime})  &  =(2\pi^{-1}%
)^{1/2}C_{\mathrm{II}\mathbf{q}}(a)\frac{1}{\Gamma(i\nu)}(\frac{\nu}%
{2a})^{i\nu}K_{i\nu}(\nu\mathfrak{z}^{\prime}), \label{f7}%
\end{align}
where $C_{\mathrm{I}\mathbf{q}}$ are arbitrary \textquotedblleft phase
factor\textquotedblright, $\Gamma$ is the gamma function and $K_{i\nu}$ are
the modified Bessel functions of second kind.

\begin{remark}
Before we continue it is important to emphasize that the concept of energy
defined in regions $\mathrm{I}$ and $\mathrm{II}$ are indeed the pseudo-energy
concept that we discussed in previous section.\smallskip
\end{remark}

(\textbf{u5}) We use the positive frequencies in standard way in order
construct Hilbert spaces $\mathcal{H}$, $\mathcal{H}_{\mathrm{I}}$ and
$\mathcal{H}_{\mathrm{II}}$ by defining the well known scalar products for the
spaces of positive energy-solutions. This is done by introducing the spaces of
square integrable functions $\mathcal{K}_{M},\mathcal{K}_{\mathrm{I}}$ and
$\mathcal{K}_{\mathrm{II}}$ respectively of the forms%
\begin{align}
\Phi_{M}(x)  &  =\int d^{3}\mathbf{k[}a(\mathbf{k)}\phi_{M\mathbf{k}}%
(x)+\bar{a}^{\ast}(\mathbf{k)}\phi_{M\mathbf{k}}^{\ast}(x)]\nonumber\\
\Phi_{\mathrm{I}}(\mathfrak{l)}  &  =\int_{0}^{\infty}d\nu\int d^{2}%
\mathbf{q[}b_{\mathrm{I}\nu}(\mathbf{q)}\phi_{\mathrm{I\nu}\mathbf{q}%
}(\mathfrak{l})+\bar{b}_{\mathrm{I}\nu}^{\ast}(\mathbf{q)}\phi_{\mathrm{I\nu
}\mathbf{q}}^{\ast}(\mathfrak{l})]\nonumber\\
\Phi_{\mathrm{II}}(\mathfrak{l}^{\prime}\mathfrak{)}  &  =\int_{0}^{\infty
}d\nu\int d^{2}\mathbf{q[}b_{\mathrm{II}\nu}(\mathbf{q)}\phi_{\mathrm{II\nu
}\mathbf{q}}(\mathfrak{l}^{\prime})+\bar{b}_{\mathrm{II}\nu}^{\ast
}(\mathbf{q)}\phi_{\mathrm{II\nu}\mathbf{q}}^{\ast}(\mathfrak{l}^{\prime})]
\label{f18}%
\end{align}
where $a,b_{\mathrm{I}\nu},b_{\mathrm{II}\nu},\bar{a},\bar{b}_{\mathrm{I}\nu
},\bar{b}_{\mathrm{II}\nu}$ are arbitrary square integrable functions
(elements of $\mathfrak{L(\mathbb{R}}^{3}\mathfrak{)}$).

Take notice that $\hat{\phi}_{\mathrm{I}}+\hat{\phi}_{\mathrm{II}}$ can be
extended to all $M$ by extending $\phi_{\mathrm{I\nu}\mathbf{q}}$ and
$\phi_{\mathrm{II\nu}\mathbf{q}}(\mathfrak{l})$ to all $M.$

Now, we construct in the space of these functions the usual inner products
($J=M,\mathrm{I},\mathrm{II}$)
\begin{equation}
\langle\Phi_{J},\Psi_{J}\rangle_{J}=i\int_{\Sigma}d\Sigma n^{a}(\Phi_{J}%
^{\ast}\frac{\partial}{\partial x_{J}^{a}}\Psi_{J}-\Phi_{J}\frac{\partial
}{\partial x_{J}^{a}}\Psi_{J}^{\ast}) \label{f19}%
\end{equation}
where $J=M,\mathrm{I},\mathrm{II}$ and $x_{J}^{a}$ denotes the appropriate
variables for each domain and finally we construct as usual the Hilbert spaces
$\mathcal{H},\mathcal{H}_{\mathrm{I}}$ and $\mathcal{H}_{\mathrm{II}}$ by
completion of the respective $\mathcal{K}$ spaces and $n^{a}$ are the
components of the normal to the spacelike surface $\Sigma$.\smallskip

In particular, choosing $\Sigma$ to be hypersurface $t=0$ for the Minkowski
modes and $\mathfrak{t}=0$ for the Rindler modes we have
\begin{gather}
\langle\phi_{M\mathbf{k}},\phi_{M\mathbf{k}^{\prime}}\rangle_{M}%
=\delta(\mathbf{k-k}^{\prime}),~~~~\langle\phi_{M\mathbf{k}}^{\ast}%
,\phi_{M\mathbf{k}^{\prime}}^{\ast}\rangle_{M}=-\delta(\mathbf{k-k}^{\prime
}),\nonumber\\
\langle\phi_{\mathrm{I\nu}\mathbf{q}},\phi_{\mathrm{I\nu}^{\prime}%
\mathbf{q}^{\prime}}\rangle_{\mathrm{I}}=\delta(\nu-\nu^{\prime}%
)\delta(\mathbf{q-q}^{\prime}),~~~~\langle\phi_{\mathrm{I\nu}\mathbf{q}}%
,\phi_{\mathrm{I\nu}^{\prime}\mathbf{q}^{\prime}}\rangle_{\mathrm{I}}%
=-\delta(\nu-\nu^{\prime})\delta(\mathbf{q-q}^{\prime}),\nonumber\\
\langle\phi_{\mathrm{II\nu}\mathbf{q}},\phi_{\mathrm{I\nu}^{\prime}%
\mathbf{q}^{\prime}}\rangle_{\mathrm{II}}=\delta(\nu-\nu^{\prime}%
)\delta(\mathbf{q-q}^{\prime}),~~~~\langle\phi_{\mathrm{II\nu}\mathbf{q}}%
,\phi_{\mathrm{II\nu}^{\prime}\mathbf{q}^{\prime}}\rangle_{\mathrm{II}%
}=-\delta(\nu-\nu^{\prime})\delta(\mathbf{q-q}^{\prime}),\nonumber\\
\langle\phi_{M\mathbf{k}},\phi_{M\mathbf{k}^{\prime}}^{\ast}\rangle
_{M}=0,~~~\langle\phi_{\mathrm{I\nu}\mathbf{q}},\phi_{\mathrm{I\nu}^{\prime
}\mathbf{q}^{\prime}}^{\ast}\rangle_{\mathrm{I}}=0,~~~\langle\phi
_{\mathrm{II\nu}\mathbf{q}},\phi_{\mathrm{II\nu}^{\prime}\mathbf{q}^{\prime}%
}^{\ast}\rangle_{\mathrm{II}}=0. \label{F19N1}%
\end{gather}

(\textbf{u6}) From $\mathcal{H},\mathcal{H}_{\mathrm{I}}$ and $\mathcal{H}%
_{\mathrm{II}}$ we construct the Fock-Hilbert space $\mathcal{F}%
(\mathcal{H}),$ $\mathcal{F}(\mathcal{H}_{\mathrm{I}})$ and $\mathcal{F}%
(\mathcal{H}_{\mathrm{II}})$ which describe all possible physical states of
the quantum fields
\begin{subequations}
\label{uu20}%
\begin{gather}
\hat{\phi}_{M}(x)=%
%TCIMACRO{\tint }%
%BeginExpansion
{\textstyle\int}
%EndExpansion
d^{3}\mathbf{k}\left[  \mathbf{a}\left(  \mathbf{k}\right)  \phi_{M\mathbf{k}%
}+\mathbf{\bar{a}}^{\dagger}\left(  \mathbf{k}\right)  \phi_{M\mathbf{k}%
}^{\ast}\right]  ,\label{a}\\
\hat{\phi}_{\mathrm{I}}(\mathfrak{l})=\int_{0}^{\infty}d\nu%
%TCIMACRO{\tint }%
%BeginExpansion
{\textstyle\int}
%EndExpansion
d^{2}\mathbf{q}\left[  \mathbf{b}_{\mathrm{I}\nu}\left(  \mathbf{q}\right)
\phi_{\mathrm{I}_{\nu}\mathbf{q}}(\mathfrak{l})+\mathbf{\bar{b}}%
_{\mathrm{I}\nu}^{\dagger}\left(  \mathbf{q}\right)  \phi_{_{\mathrm{I}\nu
}\mathbf{q}}^{\ast}(\mathfrak{l})\right]  ,\label{b}\\
\hat{\phi}_{\mathrm{II}}(\mathfrak{l}^{\prime})=\int_{0}^{\infty}d\nu%
%TCIMACRO{\tint }%
%BeginExpansion
{\textstyle\int}
%EndExpansion
d^{2}\mathbf{q}\left[  \mathbf{b}_{\mathrm{II}\nu}\left(  \mathbf{q}\right)
\phi_{\mathrm{II}_{\nu}\mathbf{q}}(\mathfrak{l}^{\prime})+\mathbf{\bar{b}%
}_{\mathrm{II}\nu}^{\dagger}\left(  \mathbf{q}\right)  \phi_{_{\mathrm{I}\nu
}\mathbf{q}}^{\ast}(\mathfrak{l}^{\prime})\right]  ,\label{c}%
\end{gather}
which are operator valued distributions acting respectively on $\mathcal{F}%
(\mathcal{H}),\mathcal{F}(\mathcal{H}_{\mathrm{II}})$ $\mathcal{F}%
(\mathcal{H})$ and where the $\mathbf{a}_{,}\mathbf{a}^{\dagger}$,
$\mathbf{b}_{\mathrm{I}\nu},\mathbf{b}_{\mathrm{I}\nu}^{\dagger}$ and
$\mathbf{b}_{\mathrm{II}\nu},\mathbf{b}_{\mathrm{II}\nu}^{\dagger}$
(respectively $\mathbf{\bar{a}}_{,}\mathbf{\bar{a}}^{\dagger}$, $\mathbf{\bar
{b}}_{\mathrm{I}\nu},\mathbf{\bar{b}}_{\mathrm{I}\nu}^{\dagger}$ and
$\mathbf{\bar{b}}_{\mathrm{II}\nu},\mathbf{\bar{b}}_{\mathrm{II}\nu}^{\dagger
}$) are destruction and creation operators for \emph{positive} (respectively
negative) charged particles. We have for the \emph{non null} commutators:%
\end{subequations}
\begin{gather}
\lbrack\mathbf{\bar{a}}\left(  \mathbf{k}\right)  ,\mathbf{\bar{a}}^{\dagger
}\left(  \mathbf{k}^{\prime}\right)  ]=[\mathbf{a}\left(  \mathbf{k}\right)
,\mathbf{a}^{\dagger}\left(  \mathbf{k}^{\prime}\right)  ]=\delta
(\mathbf{k-k}^{\prime}),\nonumber\\
\lbrack\mathbf{\bar{b}}_{\mathrm{I}\nu}\left(  \mathbf{q}\right)
,\mathbf{\bar{b}}_{\mathrm{I}\nu^{\prime}}\left(  \mathbf{q}^{\prime}\right)
]=[\mathbf{b}_{\mathrm{I}\nu}\left(  \mathbf{q}\right)  ,\mathbf{b}%
_{\mathrm{I}\nu^{\prime}}\left(  \mathbf{q}^{\prime}\right)  ]=\delta(\nu
-\nu^{\prime})\delta(\mathbf{q-q}^{\prime}),\nonumber\\
\lbrack\mathbf{\bar{b}}_{\mathrm{II}\nu}\left(  \mathbf{q}\right)
,\mathbf{\bar{b}}_{\mathrm{II}\nu^{\prime}}\left(  \mathbf{q}^{\prime}\right)
]=[\mathbf{b}_{\mathrm{II}\nu}\left(  \mathbf{q}\right)  ,\mathbf{b}%
_{\mathrm{II}\nu^{\prime}}\left(  \mathbf{q}^{\prime}\right)  ]=\delta(\nu
-\nu^{\prime})\delta(\mathbf{q-q}^{\prime}).\label{f19n2}%
\end{gather}
We suppose that we have a second quantum field construction for all Minkowski
spacetime (with eigenfunctions properly extended for all domains) once we
choose as the one-particle Hilbert space $\mathcal{H}_{\mathrm{II}}%
\oplus\mathcal{H}_{\mathrm{I}}$. Now, take notice that \cite{wald2}
\begin{equation}
\mathcal{F}(\mathcal{H}_{\mathrm{II}}\oplus\mathcal{H}_{\mathrm{I}}%
)\simeq\mathcal{F}(\mathcal{H}_{\mathrm{II}})\otimes\mathcal{F}(\mathcal{H}%
_{\mathrm{I}}).\label{ffock}%
\end{equation}
\smallskip

(\textbf{u7}) The Minkowski vacuum and the vacua for regions $\mathrm{I}%
,\mathrm{II}$ are defined respectively by the states $|0\mathbf{\rangle}%
_{M}\in\mathcal{F}(\mathcal{H}),|0\mathbf{\rangle}_{\mathrm{I}}\in
\mathcal{F}(\mathcal{H}_{\mathrm{I}}),|0\mathbf{\rangle}_{\mathrm{II}}%
\in\mathcal{F}(\mathcal{H}_{\mathrm{II}})$ such that
\begin{gather}
\mathbf{a}\left(  \mathbf{k}\right)  |0\mathbf{\rangle}_{M}=\mathbf{\bar{a}%
}\left(  \mathbf{k}\right)  |0\mathbf{\rangle}_{M}=0~\forall\mathbf{k,}%
\nonumber\\
\mathbf{b}_{\mathrm{I}\nu}(\mathbf{q)}|0\mathbf{\rangle}_{\mathrm{I}%
}=\mathbf{\bar{b}}_{\mathrm{I}\nu}(\mathbf{q)}|0\mathbf{\rangle}_{\mathrm{I}%
}=0,\text{and }\mathbf{b}_{\mathrm{II}\nu}\left(  \mathbf{q}\right)
|0\mathbf{\rangle}_{\mathrm{II}}=\mathbf{\bar{b}}_{\mathrm{II}\nu}\left(
\mathbf{q}\right)  |0\mathbf{\rangle}_{\mathrm{II}}=0,\forall\mathbf{q,}\nu.
\label{ffock1}%
\end{gather}
The respective particle number operators for modes $\mathbf{k}$,
$\mathrm{I}\nu$\textbf{ }and $\mathrm{II}\nu$ are $N_{\mathbf{k}}%
=\mathbf{a}^{\dagger}\left(  \mathbf{k}\right)  \mathbf{a}\left(
\mathbf{k}\right)  ,~~\bar{N}_{\mathbf{k}}=\mathbf{a}^{\dagger}\left(
\mathbf{k}\right)  \mathbf{a}\left(  \mathbf{k}\right)  ,~~N_{\mathrm{I}%
\nu\mathbf{q}}=\mathbf{b}_{\mathrm{I}\nu}^{\dagger}\left(  \mathbf{q}\right)
\mathbf{b}_{\mathrm{I}\nu}\left(  \mathbf{q}\right)  ,~~\bar{N}_{\mathrm{I}%
\nu\mathbf{q}}=\mathbf{b}_{\mathrm{I}\nu}^{\dagger}\left(  \mathbf{q}\right)
\mathbf{b}_{\mathrm{I}\nu}\left(  \mathbf{q}\right)  $ and $N_{\mathrm{II}%
\nu\mathbf{q}}=\mathbf{b}_{\mathrm{II}\nu}^{\dagger}\left(  \mathbf{q}\right)
\mathbf{b}_{\mathrm{II}\nu}\left(  \mathbf{q}\right)  ,~~\bar{N}%
_{\mathrm{II}\nu\mathbf{q}}=\mathbf{b}_{\mathrm{II}\nu}^{\dagger}\left(
\mathbf{q}\right)  \mathbf{b}_{\mathrm{II}\nu}\left(  \mathbf{q}\right)  .$ Of
course,
\begin{align}
_{M}\langle0|N_{\mathbf{k}}|0\mathbf{\rangle}_{M}  &  =0,~_{\mathrm{I}}%
\langle0|N_{\mathrm{I}\nu\mathbf{q}}|0\mathbf{\rangle}_{\mathrm{I}%
}=0,~_{\mathrm{II}}\langle0|N_{\mathrm{II}\nu\mathbf{q}}|0\mathbf{\rangle
}_{\mathrm{II}}=0,\nonumber\\
_{M}\langle0|\bar{N}_{\mathbf{k}}|0\mathbf{\rangle}_{M}  &  =0,~_{\mathrm{I}%
}\langle0|\bar{N}_{\mathrm{I}\nu\mathbf{q}}|0\mathbf{\rangle}_{\mathrm{I}%
}=0,~_{\mathrm{II}}\langle0|\bar{N}_{\mathrm{II}\nu\mathbf{q}}%
|0\mathbf{\rangle}_{\mathrm{II}}=0. \label{21}%
\end{align}

(\textbf{u8}) In some presentations it is supposed that the quantum field in
regions $\mathrm{I+II}$ obtained through the above quantization procedures can
be described by
\begin{equation}
\hat{\phi}_{\mathrm{I}}+\hat{\phi}_{\mathrm{II}}\label{f21a}%
\end{equation}
acting on $\mathcal{F}(\mathcal{H}_{\mathrm{II}}\oplus\mathcal{H}_{\mathrm{I}%
})$. However, here we suppose that the quantum field $\hat{\phi}^{\prime}$ in
regions $\mathrm{I+II}$ is described by an \textquotedblleft entangled
field\textquotedblright\ made from $\hat{\phi}_{\mathrm{I}}(x)$ and $\hat
{\phi}_{\mathrm{II}}(x)$ acting on $\mathcal{F}(\mathcal{H}_{\mathrm{II}%
})\otimes\mathcal{F}(\mathcal{H}_{\mathrm{I}})$, i.e., described by%
\begin{equation}
\hat{\phi}^{\prime}=\mathbf{1}_{\mathrm{II}}\otimes\hat{\phi}_{\mathrm{I}%
}+\hat{\phi}_{\mathrm{II}}\otimes\mathbf{1}_{\mathrm{I}}%
\end{equation}
acting (see Eq.(\ref{ffock})) on the Fock-Hilbert space $\mathcal{F}%
(\mathcal{H}_{\mathrm{1I}})\otimes\mathcal{F}(\mathcal{H}_{\mathrm{1}})$

Moreover, it is taken as obvious that (see e.g., \cite{unruh}) that it is not
necessary to analyze what happens in regions \textrm{F} and \textrm{P. }

\subsection{\textquotedblleft Deduction\textquotedblright\ of the Unruh
Effect}

(\textbf{u9}) As it is well known the delta functions in Eqs (\ref{F19N1}) and
(\ref{f19n2}) leads to problems and so to continue the analysis it is usual to
introduce in the Hilbert spaces\footnote{Note that $\mathcal{H},\mathcal{H}%
_{\mathrm{I}}$ and $\mathcal{H}_{\mathrm{II}}$ are isomorphic to
$\mathfrak{L}^{2}\mathfrak{(\mathbb{R}}^{3}\mathfrak{)}$.}\ $\mathcal{H}%
,\mathcal{H}_{\mathrm{I}}$ and $\mathcal{H}_{\mathrm{II}}$ countable basis,
which we denote in Fourier space by%
\begin{equation}
f_{\mathbf{m},\mathbf{l},\varrho}(\mathbf{k})=\varrho^{-\frac{3}{2}}%
\exp\left(  -\frac{2\pi i\mathbf{k\bullet l}}{\varrho}\right)  \mathbf{\chi
}_{\left[  (\left\vert \mathbf{m}\right\vert -1/2)\varrho,(\left\vert
\mathbf{m}\right\vert +1/2)\varrho\right]  }(\mathbf{k}),\label{f19n3}%
\end{equation}
where $\varrho\in\mathbb{R}^{+}$ (has inverse length dimension) and
$\mathbf{\chi}_{S}$ is the characteristic function of the set $S$\footnote{For
each $\mathbf{m}=(m_{1},m_{2},m_{3})$ it is $S=\{(x^{1},x^{2},x^{3}%
)~~|~~(m_{i}-1/2)\varrho<x^{i}<(m_{i}+1/2)\varrho,~~x^{i}\in\mathbb{R}%
,i=1,2,3\}$.}. The functions $f_{\mathbf{m},\mathbf{l},\varrho}(\mathbf{k})$
are localized in Fourier space around\footnote{The $m_{i},\ell_{i}%
\in\mathbb{Z}$, $i=1,2,3$.} $\mathbf{m}=(m_{1},m_{2},m_{3})$ and have wave
number vector $\mathbf{l}=(\ell_{1},\ell_{2},\ell_{3})$, and thus in
$\mathfrak{\mathbb{R}}^{3}$ they are localized around $\mathbf{l}$ with wave
number vector $\mathbf{m}$. We immediately have that\footnote{Take notice that
in the term $\exp\left(  -\frac{2\pi ik_{i}(\ell_{i}-\ell_{i}^{\prime}%
)}{\varrho}\right)  $ in Eq.(\ref{f19n4}) $k_{i}\ell_{i}$ does not means that
we are summing in the indice $i$.}
\begin{gather}
\int d\mathbf{k}f_{\mathbf{m},\mathbf{l},\varrho}^{\ast}(\mathbf{k}%
)f_{\mathbf{m}^{\prime},\mathbf{l}^{\prime},\varrho}(\mathbf{k})\nonumber\\
:=\frac{1}{\varrho^{3}}\delta_{\mathbf{mm}^{\prime}}\prod\limits_{i}%
\int_{(m_{i}-1/2)\varrho}^{(m_{i}+1/2)\varrho}dk_{i}\exp\left(  -\frac{2\pi
ik_{i}(\ell_{i}-\ell_{i}^{\prime})}{\varrho}\right)  =\delta_{\mathbf{mm}%
^{\prime}}\delta_{\mathbf{\ell\ell}^{\prime}}.\label{f19n4}%
\end{gather}
and
\begin{gather}
\sum\nolimits_{\mathbf{l}\in\mathbb{Z}^{3}}f_{\mathbf{m},\mathbf{l},\varrho
}(\mathbf{k})f_{\mathbf{m},\mathbf{l},\varrho}(\mathbf{k}^{\prime
})=\mathbf{\chi}_{\left[  (\left\vert \mathbf{m}\right\vert -1/2)\varrho
,(\left\vert \mathbf{m}\right\vert +1/2)\varrho\right]  }(\mathbf{k}%
)\delta(\mathbf{k-k}^{\prime}),\nonumber\\
\sum\nolimits_{\mathbf{l},\mathbf{m}\in\mathbb{Z}^{3}}f_{\mathbf{m}%
,\mathbf{l},\varrho}(\mathbf{k})f_{\mathbf{m},\mathbf{l},\varrho}%
(\mathbf{k}^{\prime})=\delta(\mathbf{k-k}^{\prime}).\label{f19n5}%
\end{gather}

(\textbf{u10}) Now, in the Hilbert spaces $\mathcal{H},$ $\mathcal{H}%
_{\mathrm{I}}$ and $\mathcal{H}_{\mathrm{II}}$ we construct the\emph{ positive
frequencies} solutions of the Klein-Gordon equation, i.e.,
\begin{align}
\Phi_{M,_{\mathbf{m},\mathbf{l},\varrho}}(x) &  =\int d^{3}\mathbf{k}%
f_{\mathbf{m},\mathbf{l},\varrho}(\mathbf{k})\phi_{M\mathbf{k}}(x),\nonumber\\
\Phi_{\mathrm{I}},_{\mathbf{m},\mathbf{l},\varrho}(\mathfrak{l)} &  =\int%
_{0}^{\infty}d\nu\int d^{2}\mathbf{q}f_{\mathbf{m},\mathbf{l},\varrho
}(\mathbf{k})\phi_{\mathrm{I}\nu\mathbf{q}}(\mathfrak{l}),\nonumber\\
\Phi_{\mathrm{II}},_{\mathbf{m},\mathbf{l},\varrho}(\mathfrak{l}^{\prime
}\mathfrak{)} &  =\int_{0}^{\infty}d\nu\int d^{2}\mathbf{q}f_{\mathbf{m}%
,\mathbf{l},\varrho}(\mathbf{k})\phi_{\mathrm{II}\nu\mathbf{q}}(\mathfrak{l}%
^{\prime}).\label{f19n6}%
\end{align}
We have
\begin{gather}
\langle\Phi_{M,_{\mathbf{m},\mathbf{l},\varrho}},\Phi_{M,_{\mathbf{n}%
,\mathbf{l}^{\prime},\varrho}}\rangle_{M}=\delta_{\mathbf{mm}^{\prime}}%
\delta_{\ell\ell^{\prime}},~~\langle\Phi_{M,_{\mathbf{m},\mathbf{l},\varrho}%
}^{\ast},\Phi_{M,_{\mathbf{n},\mathbf{l}^{\prime},\varrho}}^{\ast}\rangle
_{M}=-\delta_{\mathbf{mm}^{\prime}}\delta_{\ell\ell^{\prime}},\nonumber\\
\langle\Phi_{\mathrm{I},_{\mathbf{m},\mathbf{l},\varrho}},\Phi_{\mathrm{I}%
,_{\mathbf{m}^{\prime},\mathbf{l}^{\prime},\varrho}}\rangle_{\mathrm{I}%
}=\delta_{\mathbf{mm}^{\prime}}\delta_{\ell\ell^{\prime}},~~\langle
\Phi_{\mathrm{I},_{\mathbf{m},\mathbf{l},\varrho}}^{\ast},\Phi_{\mathrm{I}%
,_{\mathbf{m}^{\prime},\mathbf{l}^{\prime},\varrho}}^{\ast}\rangle
_{\mathrm{I}}=-\delta_{\mathbf{mm}^{\prime}}\delta_{\ell\ell^{\prime}%
},\nonumber\\
\langle\Phi_{\mathrm{II},_{\mathbf{m},\mathbf{l},\varrho}},\Phi_{\mathrm{II}%
,_{\mathbf{m}^{\prime},\mathbf{l}^{\prime},\varrho}}\rangle_{\mathrm{II}%
}=\delta_{\mathbf{mm}^{\prime}}\delta_{\ell\ell^{\prime}},~~\langle
\Phi_{\mathrm{II},_{\mathbf{m},\mathbf{l},\varrho}}^{\ast},\Phi_{\mathrm{II}%
,_{\mathbf{m}^{\prime},\mathbf{l}^{\prime},\varrho}}^{\ast}\rangle
_{\mathrm{II}}=-\delta_{\mathbf{mm}^{\prime}}\delta_{\ell\ell^{\prime}%
},\nonumber\\
\langle\Phi_{M,_{\mathbf{m},\mathbf{l},\varrho}},\Phi_{M,_{\mathbf{n}%
,\mathbf{l}^{\prime},\varrho}}^{\ast}\rangle_{M}=0,~~\langle\Phi
_{\mathrm{I},_{\mathbf{m},\mathbf{l},\varrho}},\Phi_{\mathrm{I},_{\mathbf{m}%
^{\prime},\mathbf{l}^{\prime},\varrho}}^{\ast}\rangle_{\mathrm{I}}%
=0,~~\langle\Phi_{\mathrm{II},_{\mathbf{m},\mathbf{l},\varrho}},\Phi
_{\mathrm{II},_{\mathbf{m}^{\prime},\mathbf{l}^{\prime},\varrho}}^{\ast
}\rangle_{\mathrm{II}}=0\label{F19N7}%
\end{gather}
and so%
\begin{align}
\phi_{M\mathbf{k}}(x) &  =\sum\nolimits_{\mathbf{l},\mathbf{m}\in
\mathbb{Z}^{3}}f_{\mathbf{m},\mathbf{l},\varrho}(\mathbf{k})\Phi
_{M,_{\mathbf{m},\mathbf{l},\varrho}}(x),\nonumber\\
\phi_{\mathrm{I}\nu\mathbf{q}}(\mathfrak{l}) &  =\sum\nolimits_{\mathbf{l}%
,\mathbf{m}\in\mathbb{Z}^{3}}f_{\mathbf{m},\mathbf{l},\varrho}(\mathbf{k}%
)\Phi_{\mathrm{I}},_{\mathbf{m},\mathbf{l},\varrho}(\mathfrak{l),}\nonumber\\
\phi_{\mathrm{II}\nu\mathbf{q}}(\mathfrak{l}^{\prime}) &  =\sum
\nolimits_{\mathbf{l},\mathbf{m}\in\mathbb{Z}^{3}}f_{\mathbf{m},\mathbf{l}%
,\varrho}(\mathbf{k})\Phi_{\mathrm{II}},_{\mathbf{m},\mathbf{l},\varrho
}(\mathfrak{l}^{\prime}\mathfrak{).}\label{F19N8}%
\end{align}

The field operators are then written as
\begin{subequations}
\label{uuu20}%
\begin{gather}
\hat{\phi}_{M}(x)=\sum\nolimits_{\mathbf{l},\mathbf{m}\in\mathbb{Z}^{3}%
}\left[  \mathbf{a}_{\mathbf{m},\mathbf{l}_{,\varrho}}\phi_{M,_{\mathbf{m}%
,\mathbf{l},\varrho}}(x)+\mathbf{\bar{a}}^{\dagger}\phi_{M,_{\mathbf{m}%
,\mathbf{l},\varrho}}^{\ast}(x)\right]  ,\label{a'}\\
\hat{\phi}_{\mathrm{I}}(\mathfrak{l})=\sum\nolimits_{\mathbf{l},\mathbf{m}%
\in\mathbb{Z}^{3}}\left[  \mathbf{b}_{\mathrm{I}\mathbf{m},\mathbf{l}%
_{,\varrho}}\phi_{\mathrm{I}_{\nu}\mathbf{q}}(\mathfrak{l})+\mathbf{\bar{b}%
}_{\mathrm{I}\mathbf{m},\mathbf{l}_{,\varrho}}^{\dagger}\phi_{_{\mathrm{I}\nu
}\mathbf{q}}^{\ast}(\mathfrak{l})\right]  ,\label{b'}\\
\hat{\phi}_{\mathrm{II}}(\mathfrak{l}^{\prime})=\sum\nolimits_{\mathbf{l}%
,\mathbf{m}\in\mathbb{Z}^{3}}\left[  \mathbf{b}_{\mathrm{II}\nu}%
\phi_{\mathrm{II}_{\nu}\mathbf{q}}(\mathfrak{l})+\mathbf{\bar{b}}%
_{\mathrm{II}\nu}^{\dagger}\phi_{_{\mathrm{I}\nu}\mathbf{q}}^{\ast
}(\mathfrak{l})\right]  ,\label{c'}%
\end{gather}
with%
\end{subequations}
\begin{gather}
\mathbf{a}_{\mathbf{m},\mathbf{l}_{,\varrho}}=\int d^{3}\mathbf{k}%
f_{\mathbf{m},\mathbf{l},\varrho}^{\ast}(\mathbf{k})\mathbf{a}\left(
\mathbf{k}\right)  ,\nonumber\\
\mathbf{b}_{\mathrm{I}\mathbf{m},\mathbf{l}_{,\varrho}}=\int_{0}^{\infty}%
d\nu\int d^{2}\mathbf{q}f_{\mathbf{m},\mathbf{l},\varrho}^{\ast}%
(\mathbf{k})\mathbf{b}_{\mathrm{I\nu}}(\mathbf{q}),~~\mathbf{b}_{\mathrm{II}%
\mathbf{m},\mathbf{l}_{,\varrho}}=\int_{0}^{\infty}d\nu\int d^{2}%
\mathbf{q}f_{\mathbf{m},\mathbf{l},\varrho}^{\ast}(\mathbf{k})\mathbf{b}%
_{\mathrm{II\nu}}(\mathbf{q})\label{f19n9}%
\end{gather}
and analogus equations for the operators $\mathbf{\bar{a}}_{\mathbf{m}%
,\mathbf{l}_{,\varrho}},\mathbf{\bar{b}}_{\mathrm{I}\mathbf{m},\mathbf{l}%
_{,\varrho}}$ and $\mathbf{\bar{b}}_{\mathrm{II}\mathbf{m},\mathbf{l}%
_{,\varrho}}$. The non null commutators are
\begin{gather}
\lbrack\mathbf{a}_{\mathbf{m},\mathbf{l}_{,\varrho}},\mathbf{a}_{\mathbf{m}%
^{\prime},\mathbf{l}_{,\varrho}^{\prime}}^{\dagger}]=\delta_{\mathbf{mm}%
^{\prime}}\delta_{\mathbf{ll}^{\prime}},[\mathbf{b}_{\mathrm{J}\mathbf{m}%
,\mathbf{l}_{,\varrho}},\mathbf{b}_{\mathrm{J}^{\prime}\mathbf{m}^{\prime
},\mathbf{l}_{,\varrho}^{\prime}}]=\delta_{\mathrm{JJ}^{\prime}}%
\delta_{\mathbf{mm}^{\prime}}\delta_{\mathbf{ll}^{\prime}},\nonumber\\
~~[\mathbf{b}_{\mathrm{J}\mathbf{m},\mathbf{l}_{,\varrho}},\mathbf{b}%
_{\mathrm{J}\mathbf{m}^{\prime},\mathbf{l}_{,\varrho}^{\prime}}]=\delta
_{\mathrm{JJ}^{\prime}}\delta_{\mathbf{mm}^{\prime}}\delta_{\mathbf{ll}%
^{\prime}}\label{F19N10}%
\end{gather}
with $\mathrm{J=I,II}$ (and analogous equations involving the operators
$\mathbf{\bar{a}}_{\mathbf{m},\mathbf{l}_{,\varrho}},\mathbf{\bar{b}%
}_{\mathrm{I}\mathbf{m},\mathbf{l}_{,\varrho}}$ and $\mathbf{\bar{b}%
}_{\mathrm{II}\mathbf{m},\mathbf{l}_{,\varrho}}$). Of course,%
\begin{equation}
_{M}\langle0|\mathbf{a}_{\mathbf{m},\mathbf{l}_{,\varrho}}\mathbf{a}%
_{\mathbf{m}^{\prime},\mathbf{l}_{,\varrho}^{\prime}}^{\dagger}|0\rangle
_{M}=1,~_{\mathrm{I}}\langle0|\mathbf{b}_{\mathbf{m},\mathbf{l}_{,\varrho}%
}\mathbf{b}_{\mathbf{m}^{\prime},\mathbf{l}_{,\varrho}^{\prime}}^{\dagger
}|0\rangle_{\mathrm{I}}=1,~_{\mathrm{II}}\langle0|\mathbf{b}_{\mathbf{m}%
,\mathbf{l}_{,\varrho}}\mathbf{b}_{\mathbf{m}^{\prime},\mathbf{l}_{,\varrho
}^{\prime}}^{\dagger}|0\rangle_{\mathrm{II}}=1\label{f1911}%
\end{equation}
and analogous equations involving the operators $\mathbf{\bar{a}}%
_{\mathbf{m},\mathbf{l}_{,\varrho}},\mathbf{\bar{b}}_{\mathrm{I}%
\mathbf{m},\mathbf{l}_{,\varrho}}$ and $\mathbf{\bar{b}}_{\mathrm{II}%
\mathbf{m},\mathbf{l}_{,\varrho}}$.

(\textbf{u11}) The Fulling-Rindler vacuum\textrm{ }$|0\mathbf{\rangle}%
_{F}:=|0\mathbf{\rangle}_{\mathrm{II}}\otimes|0\mathbf{\rangle}_{\mathrm{I}%
}\in\mathcal{F}(\mathcal{H}^{\prime})$ is then defined by
\begin{equation}
\mathbf{1}_{\mathrm{II}}\otimes\mathbf{b}_{\mathrm{I}\mathbf{m},\mathbf{l}%
_{,\varrho}}|0\mathbf{\rangle}_{F}=\mathbf{1}_{\mathrm{II}}\otimes
\mathbf{\bar{b}}_{\mathrm{I}\mathbf{m},\mathbf{l}_{,\varrho}}|0\mathbf{\rangle
}_{F}=0\text{,~ }\mathbf{b}_{\mathrm{II}\mathbf{m},\mathbf{l}_{,\varrho}%
}\otimes\mathbf{1}_{\mathrm{I}}|0\mathbf{\rangle}_{F}=\mathbf{\bar{b}%
}_{\mathrm{II}\mathbf{m},\mathbf{l}_{,\varrho}}\otimes\mathbf{1}_{\mathrm{I}%
}|0\mathbf{\rangle}_{F}=0. \label{u2c}%
\end{equation}

(\textbf{u12}) Let $\hat{\phi}_{M,\mathrm{I+II}}$ be the representation in
$\mathcal{F}(\mathcal{H}_{\mathrm{II}})\otimes\mathcal{F}(\mathcal{H}%
_{\mathrm{I}})$ of the restriction of the field $\hat{\phi}_{M}$ given by
Eq.(\ref{a}) to regions $\mathrm{I+II}$. It is a well known fact
\cite{fulling} that the Minkowski quantization\ of the Klein-Gordon field and
the Unruh quantization producing\ $\hat{\phi}^{\prime}$ are \emph{not} unitary
equivalent\footnote{See Appendix B to know how this reult is obtained in the
algebraic approach to quantum theory..}.

Anyhow, it is supposed that we can identify
\begin{equation}
\left.  \mathcal{F}(\mathcal{H})\right\vert _{\mathcal{H}^{\prime}%
}=\mathcal{F}(\mathcal{H}^{\prime})=\mathcal{F}(\mathcal{H}_{\mathrm{1}%
})\otimes\mathcal{F}(\mathcal{H}_{\mathrm{1I}})\label{u3n}%
\end{equation}
and writing
\[
\hat{\phi}_{M,\mathrm{I+II}}=\mathbf{1}_{\mathrm{II}}\otimes\hat{\phi
}_{M,\mathrm{I}}+\hat{\phi}_{M,\mathrm{II}}\otimes\mathbf{1}_{\mathrm{I}}%
\]
we thus put%
\begin{equation}
\hat{\phi}_{M,\mathrm{I+II}}=\hat{\phi}^{\prime}.\label{u3a}%
\end{equation}
(\textbf{u13}) Under these conditions the relation between those
representations is supposed to be given by the well known Bogolubov
transformations which express the operators $\mathbf{b,b}^{\dagger}$ as
functions of the operators $\mathbf{a,a}^{\mathbf{\dagger}}$. We have
$(\mathrm{J=I,II})$%
\begin{align}
\mathbf{b}_{\mathrm{J}\mathbf{m},\mathbf{l}_{,\varrho}} &  =\sum
\nolimits_{\mathbf{l},\mathbf{m}\in\mathbb{Z}^{3}}\mathbf{a}_{\mathbf{m}%
,\mathbf{l}_{,\varrho}}\Xi_{\mathrm{J}\mathbf{m,l,m}^{\prime},\mathbf{l}%
^{\prime},\varrho}+\mathbf{\bar{a}}_{\mathbf{m},\mathbf{l}_{,\varrho}%
}^{\dagger}\Upsilon_{\mathrm{J}\mathbf{m,l,m}^{\prime},\mathbf{l}^{\prime
},\varrho},\nonumber\\
\mathbf{\bar{b}}_{\mathrm{J}\mathbf{m},\mathbf{l}_{,\varrho}} &
=\sum\nolimits_{\mathbf{l},\mathbf{m}\in\mathbb{Z}^{3}}\mathbf{a\dagger
}_{\mathbf{m},\mathbf{l}_{,\varrho}}\Upsilon_{\mathrm{J}\mathbf{m,l,m}%
^{\prime},\mathbf{l}^{\prime},\varrho}+\mathbf{\bar{a}}_{\mathbf{m}%
,\mathbf{l}_{,\varrho}}\Xi_{\mathrm{J}\mathbf{m,l,m}^{\prime},\mathbf{l}%
^{\prime},\varrho}.\label{f19n12}%
\end{align}
The explicit calculation of the operators $\mathbf{b}_{\mathrm{J}%
\mathbf{m},\mathbf{l}_{,\varrho}}$ and $\mathbf{\bar{b}}_{\mathrm{J}%
\mathbf{m},\mathbf{l}_{,\varrho}}$ is done by first evaluating $\Xi
_{\mathrm{J}\mathbf{m,l,m}^{\prime},\mathbf{l}^{\prime},\varrho}$ and
$\Upsilon_{\mathrm{J}\mathbf{m,l,m}^{\prime},\mathbf{l}^{\prime},\varrho}$.
The well known result is \cite{takagi}%
\begin{align}
\Xi_{\mathrm{J}\mathbf{m,l,m}^{\prime},\mathbf{l}^{\prime},\varrho} &
=\int_{0}^{\infty}d\nu\int_{-\infty}^{\infty}dp_{1}\int\int\int\int
dk_{1}dk_{2}dp_{2}dp_{3}[f_{m_{1},\mathbf{\ell}_{1},\varrho}^{\ast}%
(\nu)f_{m_{1}^{\prime},\mathbf{\ell}_{1}^{\prime},\varrho}(p_{1})\nonumber\\
&  \times f_{m_{2},\mathbf{\ell}_{2},\varrho}^{\ast}(k_{1})f_{m_{3}%
,\mathbf{\ell}_{3},\varrho}^{\ast}(k_{2})f_{m_{2},\mathbf{\ell}_{2},\varrho
}(p_{2})f_{m_{3},\mathbf{\ell}_{3},\varrho}(p_{3})\Xi_{\mathrm{J\nu
},\mathbf{pk}}\label{F1913}%
\end{align}
(with analogous expression for $\Upsilon_{\mathrm{J}\mathbf{m,l,m}^{\prime
},\mathbf{l}^{\prime},\varrho}$ where $\Xi_{\mathrm{J\nu},\mathbf{pk}}$ is
substituted by $\Upsilon_{\mathrm{I\nu},\mathbf{pk}}$ ) with%
\begin{align}
\Xi_{\mathrm{I\nu},\mathbf{pk}} &  =\frac{1}{2\pi}\delta(p_{1}-k_{1}%
)\delta(k_{2}-p_{2})e^{\frac{\pi\nu}{2}}\left\vert \Gamma(i\nu)\right\vert
\left(  \frac{\nu}{\omega_{\mathbf{k}}}\right)  ^{\frac{1}{2}}\left(
\frac{\omega_{\mathbf{k}}+p_{3}}{\omega_{\mathbf{k}}-p_{3}}\right)
^{\frac{i\nu}{2}},\nonumber\\
\Upsilon_{\mathrm{I\nu},\mathbf{pk}} &  =\frac{1}{2\pi}\delta(p_{1}%
-k_{1})\delta(k_{1}-p_{1})e^{-\frac{\pi\nu}{2}}\left\vert \Gamma
(i\nu)\right\vert \left(  \frac{\nu}{\omega_{\mathbf{k}}}\right)  ^{\frac
{1}{2}}\left(  \frac{\omega_{\mathbf{k}}+p_{3}}{\omega_{\mathbf{k}}-p_{3}%
}\right)  ^{\frac{i\nu}{2}}\label{f19n14}%
\end{align}
Next $\mathbf{b}_{\mathrm{J}\mathbf{m},\mathbf{l}_{,\varrho}}$\ and
$\mathbf{\bar{b}}_{\mathrm{J}\mathbf{m},\mathbf{l}_{,\varrho}}$ are
approximated for the case where $\varrho$ is very small and such that $\varrho
m_{3}\approx1$ by the corresponding $\mathbf{b}_{\mathrm{J}\nu}\left(
\mathbf{q}\right)  $. We have that
\begin{equation}
\nu\mapsto\nu_{m_{_{3}}}:m_{3}\varrho,~~~\omega_{\mathbf{k}}\mapsto
\omega_{\mathbf{m}^{\prime}}:=\sqrt{\varrho^{2}\sum_{i}(m_{i}^{\prime}%
)^{2}+\mu^{2}}\label{f19n15}%
\end{equation}
and thus using this approximation we write
\begin{align}
\Xi_{\mathrm{I}\mathbf{m,l,m}^{\prime},\mathbf{l}^{\prime},\varrho} &
=\frac{\varrho}{\sqrt{2\pi}}\Theta(m_{3}+\frac{1}{2})\delta_{m_{1}%
,m_{1}^{\prime}}\delta_{\ell_{1}^{\prime},0}\delta_{m_{2},m_{2}^{\prime}%
}\delta_{m_{3}^{\prime},0}\delta_{\ell_{2},\ell_{2}^{\prime}}\delta_{\ell
_{3},\ell_{3}^{\prime}}\nonumber\\
&  \times\frac{1}{\sqrt{\omega_{\mathbf{n}}}}\frac{1}{\sqrt{1-e^{-2\pi
\nu_{m_{3}}}}}\left(  \frac{\omega_{\mathbf{m}^{\prime}}+m_{3}^{\prime}%
\varrho}{\omega_{\mathbf{m}^{\prime}}-m_{3}^{\prime}\varrho}\right)
^{\frac{i\nu_{_{m_{3}}}}{2}},\nonumber\\
\Upsilon_{\mathrm{I}\mathbf{m,l,m}^{\prime},\mathbf{l}^{\prime},\varrho} &
=\frac{\varrho}{\sqrt{2\pi}}\Theta(m_{3}+\frac{1}{2})\delta_{m_{1}%
,m_{1}^{\prime}}\delta_{\ell_{1}^{\prime},0}\delta_{m_{2},-m_{2}^{\prime}%
}\delta_{m_{3}^{^{\prime}},0}\delta_{\ell_{2},-\ell_{2}^{\prime}}\delta
_{\ell_{3},-\ell_{3}^{\prime}}\nonumber\\
&  \times\frac{1}{\sqrt{\omega_{\mathbf{n}}}}\frac{1}{\sqrt{1-e^{-2\pi
\nu_{m_{3}}}}}\left(  \frac{\omega_{\mathbf{m}^{\prime}}+m_{3}^{\prime}%
\varrho}{\omega_{\mathbf{m}^{\prime}}-m_{3}^{\prime}\varrho}\right)
^{\frac{i\nu_{_{m_{3}}}}{2}}.\label{f19n16}%
\end{align}
where the errors $\triangle\Xi_{\mathrm{I}\mathbf{m,l,m}^{\prime}%
,\mathbf{l}^{\prime},\varrho}$ and $\triangle\Upsilon_{\mathrm{I}%
\mathbf{m,l,m}^{\prime},\mathbf{l}^{\prime},\varrho}$ are estimated to be
of\ order $\varrho$.

Denoting by $|0,\mathrm{II,I}\mathbf{\rangle}_{M}$ the restriction of the
Minkowski vacuum state $|0\mathbf{\rangle}_{M}$ to the region $\mathrm{II+I}$
we have putting $\nu_{m_{3}}=\nu_{j}/a$ that, e.g., the expectation value of
particles of type $\mathbf{b}_{\mathrm{I}\mathbf{m},\mathbf{l}_{,\varrho}%
}^{\dagger}$ in the state
%TCIMACRO{\TEXTsymbol{\vert}}%
%BeginExpansion
$\vert$%
%EndExpansion
$0,\mathrm{II,I}\mathbf{\rangle}_{M}$ is:
\begin{align}
&  _{M}\langle0,\mathrm{II,I}|\mathbf{1}_{\mathrm{II}}\otimes\mathbf{b}%
_{\mathrm{I}\mathbf{m},\mathbf{l}_{,\varrho}}^{\dagger}\mathbf{b}%
_{\mathrm{I}\mathbf{m},\mathbf{l}_{,\varrho}}|0,\mathrm{II,I}\mathbf{\rangle
}_{M}\nonumber\\
&  =\frac{\varrho^{2}}{2\pi}\delta_{\ell_{1}0~M}\langle0,\mathrm{II,I}%
||0,\mathrm{II,I}\mathbf{\rangle}_{M}\frac{1}{e^{2\pi\nu_{j}/a}-1}%
\sum\nolimits_{j\in\mathbb{Z}}\frac{1}{\omega_{j}}\label{U45n}%
\end{align}

Eq.(\ref{U45}) shows that even if we suppose that $_{M}\langle0,\mathrm{I,II}%
||0,\mathrm{II,I}\mathbf{\rangle}_{M}=~_{M}\langle0|0\mathbf{\rangle}_{M}=1$,
the vector $\mathbf{b}_{\mathrm{I}\mathbf{m},\mathbf{l}_{,\varrho}%
}|0,\mathrm{II,I}\mathbf{\rangle}_{M}\in\mathcal{F}(\mathcal{H}_{\mathrm{1}%
})\otimes\mathcal{F}(\mathcal{H}_{\mathrm{1I}})$ has not a finite norm, thus
showing that the procedure we have been using until now is not a mathematical
legitimate one.\smallskip

(\textbf{u14}) Nevertheless, taking the\ above approximation for the Bogolubov
transformation as a good one for at least a region where $\varrho m_{3}%
\approx1$ , the state $|0,\mathrm{II,I}\mathbf{\rangle}_{M}$ is written%

\begin{gather}
|0,\mathrm{II,I}\mathbf{\rangle}_{M}\nonumber\\
=\Omega^{-1}\exp\left\{
%TCIMACRO{\tsum \nolimits_{j,m_{1}}}%
%BeginExpansion
{\textstyle\sum\nolimits_{j,m_{1}}}
%EndExpansion
e^{-2\pi\nu_{m_{1}}}\left(  (\mathbf{b}_{\mathrm{II}\mathbf{m},\mathbf{l}%
_{,\varrho}}^{+})^{n_{j}}\otimes\mathbf{1}_{\mathrm{I}}+\mathbf{1}%
_{\mathrm{II}}\otimes(\mathbf{b}_{\mathrm{I}\mathbf{m},\mathbf{l}_{,\varrho}%
}^{+})^{n_{j}}\right)  \right\}  |0\mathbf{\rangle}_{\mathrm{II}}%
\otimes|0\mathbf{\rangle}_{\mathrm{I}}\nonumber\\
=\Omega^{-1}%
%TCIMACRO{\tprod \limits_{j}}%
%BeginExpansion
{\textstyle\prod\limits_{j}}
%EndExpansion%
%TCIMACRO{\tsum \nolimits_{n}}%
%BeginExpansion
{\textstyle\sum\nolimits_{n}}
%EndExpansion
e^{-\pi n\nu_{j}/a}|\check{n}_{j}\rangle_{\mathrm{II}}\otimes|\check{n}%
_{j}\rangle_{\mathrm{I}}, \label{U46}%
\end{gather}
where $\Omega$ is a normalization constant and $|\check{n}_{j}\rangle
_{\mathrm{J}}=|\check{n}_{j}\rangle_{\mathrm{J}}+|0\rangle_{\mathrm{J}}$,
$\mathrm{J}=\mathrm{I,II}$.

(\textbf{u15}) Using the fact that regions \textrm{I} and \textrm{II }are
causally disconnected, i.e., observers following integral lines of
$\boldsymbol{R}$, can only detect \emph{right} Rindler particles it is
supposed that these observers can only describe (according to standard quantum
mechanics prescription) the state of the Minkowski quantum vacuum by a mixed
state \cite{wald2}, i.e., a density matrix obtained by tracing over the states
of the region \textrm{II} the pure state density matrix $\hat{\rho
}=|0,\mathrm{I,II}\mathbf{\rangle}_{M}\langle0,\mathrm{I,II}\mathbf{|}_{M}$.
The result is
\begin{equation}
\hat{\rho}_{\mathrm{I}}=\mathrm{tr}_{\mathrm{I}}(\hat{\rho})=\Omega^{-1}%
%TCIMACRO{\tprod \limits_{j}}%
%BeginExpansion
{\textstyle\prod\limits_{j}}
%EndExpansion%
%TCIMACRO{\tsum \nolimits_{n}}%
%BeginExpansion
{\textstyle\sum\nolimits_{n}}
%EndExpansion
e^{-2\pi n\nu_{j}/a}|n_{j}\rangle_{\mathrm{I}}\otimes\text{~}_{\mathrm{I}%
}\langle n_{j}|,\label{U45}%
\end{equation}
which looks like a thermal spectrum with temperature parameter $a/2\pi$.

\begin{remark}
Take notice that\ for an observer following the worldline $\sigma$ with
$\mathfrak{z}=$constant in region \textrm{I} the local temperature of the
thermal radiation is \emph{\cite{unruh}}%
\begin{equation}
T(\mathfrak{z})=\frac{1}{\sqrt{g_{00}(\mathfrak{z})}}\frac{a}{2\pi} \label{u6}%
\end{equation}
and thus $T(\mathfrak{z})\sqrt{g_{00}(\mathfrak{z})}$ is a constant. This is
extremely important for otherwise thermodynamical equilibrium \emph{(}%
according to Tolman's version \emph{\cite{tolman})} would not be possible in
the $\boldsymbol{R}$ frame.\smallskip
\end{remark}

(\textbf{u16}) Given Eq.(\ref{U45}) since $n\nu_{j}$ is the value of the
\emph{pseudo energy} in the \ $|n_{j}\rangle_{\mathrm{I}}$ state and since
$\hat{\rho}_{\mathrm{I}}$ looks like a thermal density matrix $\rho
_{T}=e^{-H/T}$ it is claimed that:

T\emph{he Minkowski vacuum in region I is seem by observers living there as a
thermal bath at temperature }$a/2\pi$\emph{ of the so-called Rindler
particles, which can excite well designed detectors. }%
\cite{ginsburg,gmm,unruh,unwald,sciama,suli,wald2}\emph{ \ Even more, it is
claimed,}(\emph{e.g., in }\cite{suli})\emph{ that the Rindler particles are
irradiated from the boundary of the region I }(\emph{which is supposed to be
\textquotedblleft analogous\textquotedblright\ to the horizon of a blackhole
which is supposed to radiate due to the so-called Hawking effect}%
)\emph{.}\smallskip

(\textbf{u17}) The fact is that a rigorous mathematical analysis of the
problem, based on the algebraic approach to field theory\footnote{First
applied to the Unruh effect problem in \cite{kay}.} (which for completeness,
we recall in Appendix B), it is possible to show that the hypothesis given by
Eq.(\ref{u3n}) and thus Eq.(\ref{U45n}) are \emph{not }correct. Indeed, there
we recall that strictly speaking the density matrix $\hat{\rho}$ and thus
$\hat{\rho}_{\mathrm{I}}$ are meaningless. Also, many people has serious
doubts if Fulling-Rindler vacuum $|0\mathbf{\rangle}_{F}:=|0\mathbf{\rangle
}_{\mathrm{II}}\otimes|0\mathbf{\rangle}_{\mathrm{I}}$.can be physically
realizable. These arguments are, in our opinion) stronger ones and the reader
is invited to at least give a look in Appendix B (where the main references on
original papers dealing with the issue of the algebraic approach to the Unruh
effect may be found) in order to have an idea of the truth of what has just
been stated.\smallskip

(\textbf{u18}) As it is the case\ of the problem of the electromagnetic field
generated by a charge in hyperbolic motion, there are several researchers that
are convinced that the Unruh effect does \emph{not} exist.

Besides the inconsistencies recalled in Appendix B several others are
discussed, e.g., in \cite{colosi,fnmbk1999,aer,buver} The most important one
in our opinion, has been realized in \cite{fnmbk1999} where it is shown that
both in the conventional approach as well as in the algebraic approach to
quantum field theory it is impossible to perform the quantization of Unruh
modes in Minkowski spacetime. Authors claim (and we agree with them) that
Unruh quantization in a Rindler frame implies setting a \emph{boundary
condition} for the quantum field operator which changes the topological
properties and symmetry group of the spacetime (where the Rindler reference
frame has support) and leads to a field theory in the two disconnected regions
\textrm{I} and \textrm{II}. They concluded that the Rindler effect does not
exist.\smallskip

(\textbf{u19}) Despite this fact, in a recent publication \cite{clmv} authors
that pertain to the majority view (i.e., those that believe in the existence
of the thermal radiation) state:

\begin{quotation}
\textquotedblleft Then, instead of waiting for experimentalists to perform the
experiment, we use standard classical electrodynamics to anticipate its output
and show that it reveals the presence of a thermal bath with temperature
$T_{U}$ in the accelerated frame. Unless one is willing to question the
validity of classical electrodynamics, this must be seen as a virtual
observation of the Unruh effect\textquotedblright.
\end{quotation}

Well, authors of \cite{clmv} also believe that a charge in hyperbolic motion
radiates, and that the correct solution to the problem is the one given by the
Li\'{e}nard-Wiechert potential. But what will be of the statement that we
cannot doubt classical electrodynamics if turns out that the Turakulov
solution is the correct one (i.e., experimentally confirmed)? \smallskip

Another important question is the following one: does a detector following an
integral line of $\boldsymbol{R}$ get excited?\smallskip

(\textbf{u20}) Several thoughtful analysis of the problem done from the point
of view of an inertial reference frame shows that the detector get excited.
This is discussed in \cite{chm} and a very simple model of a detector showing
that the statement is correct may be found in \cite{noth}. But, of course, it
is necessary to leave clear that this excitation energy can only come from the
source that maintains the detector accelerated and it is not an excitation due
to fluctuations of the zero point of the field as claimed, e.g. in \cite{aer}.

\section{Conclusions}

There are some problems in Relativity Theory that are source of controversies
since a long time. One of them has to do with the question if a charge in
uniformly accelerated motion radiates. This problem is important, in
particular, in its connection with one of the forms of the Equivalence
Principle. In this paper we recalled that there are two different solutions
for the electromagnetic field generated by a charge in hyperbolic motion, the
Li\'{e}nard-Wiechert (LW) one (obtained by the retarded Green function) and
the less known one discovered by Turakulov in 1994 (and which we have verified
to be correct, in particular using the software Mathematica). According to the
LW solution the charge radiates and claims that an observer comoving with the
charge does not detect any radiation is shown to be wrong. This is done by
analyzing the different concepts of energy used by people that claims that no
radiation is detected. Turakulov claims in \cite{tura} \ that his solution
implies that there are no radiation. However, we have proved that he is also
wrong, the reason being essentially the same as in the case of the
Li\'{e}nard-Wiechert solution. On the other hand we recalled that a charge at
rest in Schwarzschild spacetime does not radiate. Thus, if the LW or the
Turakulov solution is the correct one, then experiment with charges may show
that the Equivalence Principle is false.

Another problem which we investigate is the so-called Bell's \textquotedblleft
paradox\textquotedblright. We discussed it in details since it is, as yet, a
source of \ misunderstandings.

Finally, we briefly recall how the so-called Unruh effect is obtained in
almost all texts using some well ideas of quantum field theory. We comment
that this standard approach seems to imply that an observer in hyperbolic
motion is immersed in a thermal bath with temperature proportional to its
proper acceleration. Acceptance that this is indeed the case is almost the
majority view among physicists. However, fact is that the standard approach
does not resist a rigorous mathematical analysis, in particular when one use
the algebraic approach to quantum field theory. Thus as it is the case with
the problem of determining the electromagnetic field of a charge in hyperbolic
motion there are dissidents of the majority view. Having studied the arguments
of several papers we presently agree with \cite{buver,fnmbk1999} that there is
no Unruh effect. However, it is not hard to show that a detector in hyperbolic
motion on the Minkowski vacuum gets excited, but the energy producing such
excitation, contrary to some claims (as, e.g., in \cite{aer}) does not come
from the fluctuations of the zero point field, but comes from the source
pushing the charge.

\appendix{}

\section{Some Notations and Definitions}

(\textbf{a1}) Let $M$ be a four dimensional, real, connected, paracompact and
non-compact manifold. We recall that a \emph{Lorentzian manifold} as a pair
$(M,\mathbf{g})$, where $\mathbf{g}\in\sec T_{2}^{0}M$ is a Lorentzian metric
of signature $(1,3)$, i.e., $\forall\mathfrak{e}\in M,T_{x}M\simeq
T_{\mathfrak{e}}^{\ast}M\simeq\mathbb{R}^{4}$. Moreover, $\forall x\in
M,(T_{x}M,\mathbf{g}_{x})\simeq\mathbb{R}^{1,3}$, where $\mathbb{R}^{1,3}$ is
the Minkowski \emph{vector }space We define a \emph{Lorentzian spacetime} $M$
as pentuple $(M,\mathbf{g},\boldsymbol{D},\tau_{\mathbf{g}},\uparrow)$, where
$(M,\mathbf{g},\tau_{\mathbf{g}},\uparrow)$) is an \emph{oriented} Lorentzian
manifold (oriented by $\tau_{\boldsymbol{g}}$) and \emph{time}
oriented\footnote{Please, consult, e.g., \cite{rc2016}.} by $\uparrow$, and
$D$ is the Levi-Civita connection of $\boldsymbol{g}$. Let $U\subseteq M$ be
an open set covered, say, by coordinates $(y^{0},y^{1},y^{2},y^{3})$. . Let
$U\subseteq M$ be an open set covered by coordinates $\{x^{\mu}\}$. Let
$\{e_{\mu}=\partial_{\mu}\}$ be a coordinate basis of $T\mathcal{U}$ and
$\{\boldsymbol{\vartheta}^{\mu}=dx^{\mu}\}$ the dual basis on $T^{\ast
}\mathcal{U}$, i.e., $\boldsymbol{\vartheta}^{\mu}(\partial_{\nu})=\delta
_{\nu}^{\mu}$. If $\mathbf{g}=g_{\mu\nu}\boldsymbol{\vartheta}^{\mu}%
\otimes\boldsymbol{\vartheta}^{\nu}$ is the metric on $T\mathcal{U}$ we denote
by $\mathtt{g}=g^{\mu\nu}\boldsymbol{\partial}_{\mu}\otimes
\boldsymbol{\partial}_{\nu}$ the metric of $T^{\ast}\mathcal{U}$, such that
$g^{\mu\rho}g_{\rho\nu}=\delta_{\nu}^{\mu}$. We introduce also
$\{\boldsymbol{\partial}^{\mu}\}$ and $\{\boldsymbol{\vartheta}_{\mu}\}$,
respectively, as the reciprocal bases of $\{e_{\mu}\}$ and
$\{\boldsymbol{\vartheta}_{\mu}\}$, i.e., we have
\begin{equation}
\mathbf{g}(\boldsymbol{\partial}_{\nu},\boldsymbol{\partial}^{\mu}%
)=\delta_{\nu}^{\mu},~~~g(\boldsymbol{\vartheta}^{\mu},\boldsymbol{\vartheta
}_{\nu})=\delta_{\nu}^{\mu}. \label{a1}%
\end{equation}

(\textbf{a2}) Call $(M\simeq\mathbb{R}^{4},\boldsymbol{g},D,\tau
_{\boldsymbol{g}},\uparrow)$ the \emph{Minkowski spacetime structure}. When
$M\simeq\mathbb{R}^{4}$ there is (infinitely) global charts. Call
$(x^{0},x^{1},x^{2},x^{3})$ the coordinates of one of those charts. These
coordinates are said to be in \emph{Einstein-Lorentz-Poincar\'{e}} \ (ELP)
gauge. In these coordinates
\begin{equation}
\boldsymbol{g}=\eta_{\mu\nu}dx^{\mu}\otimes dx^{\nu}\text{and }\mathtt{g}%
=\eta^{\mu\nu}\frac{\partial}{\partial x^{\mu}}\otimes\frac{\partial}{\partial
x^{\nu}} \label{a2}%
\end{equation}
where the matrix with entries $\eta_{\mu\nu}$ and also the matrix with entries
$\eta^{\mu\nu}$ are diagonal matrices $\mathrm{diag}(1,-1,-1,-1)$.\smallskip

(\textbf{a3})\ In a general Lorentzian structure if $\mathbf{Q}\in\sec
TU\subset\sec TM$ is a time-like vector field such that $\boldsymbol{g}%
(\mathbf{Q},\mathbf{Q})=1$, then there exist, in a coordinate neighborhood
$U$, three space-like vector fields\texttt{ }$\boldsymbol{e}_{\mathbf{i}}$
which together with $\mathbf{Q}$ form an orthogonal moving frame for $x\in U$
\cite{choquet,rc2016}.\smallskip

(\textbf{a4}) A \emph{moving frame} at $x\in M$ is a basis for the tangent
space $T_{x}M$. An orthonormal (moving) frame at $x\in M$ is a basis of
orthonormal vectors for $T_{x}M$.\smallskip

(\textbf{a5}) An \emph{observer} in a general Lorentzian spacetime is a future
pointing time-like curve $\sigma:\mathbb{R}\supset I\rightarrow M$ such
that\texttt{ }$\boldsymbol{g}(\sigma_{\ast},\sigma_{\ast})=1$. The timelike
curve $\sigma$ is said to be the worldline of the observer.\smallskip

(\textbf{a6}) An \emph{instantaneous observer} is an element of $TM$, i.e., a
pair $(x,\mathcal{Q}),$ where $x\in M$, and $\mathcal{Q}\in T_{x}M$ is a
future pointing unit timelike vector. $\mathrm{Span}\mathcal{Q}\subset T_{x}M$
is the \emph{local time axis} of the observer and $\mathcal{Q}^{\perp}$ is the
\emph{observer rest space.}

(\textbf{a7}) Of course, $T_{x}M=\mathrm{Span}\mathcal{Q}\oplus\mathcal{Q}%
^{\perp}$, and we denote in what follows $\mathrm{Span}\mathcal{Q=}T$ and
$\mathcal{Q}^{\perp}=H$, which is called the\textit{ rest space} of the
instantaneous observer. If $\sigma:\mathbb{R}\supset I\rightarrow M$ is an
observer, then $(\sigma u,\sigma_{\ast}u)$ is said to be the local observer at
$u$ and write $T_{\sigma u}M=T_{u}\oplus H_{u}\ ,\ u\in I$.\smallskip

(\textbf{a8}) The \emph{orthogonal projections} are the mappings%
\begin{equation}
\mathbf{p}_{u}=T_{\sigma u}M\rightarrow H_{u}\ ,\ \mathbf{q}_{u}:T_{\sigma
u}M\rightarrow T_{u}. \label{proj0}%
\end{equation}
Then if $\mathbf{Y}$\textbf{ }is a vector field over $\sigma$ then
$\mathbf{pY}$ and $\mathbf{qY}$ are vector fields over $\sigma$ given by
\begin{equation}
(\mathbf{pY})_{u}=\mathbf{p}_{u}(\mathbf{Y}_{u}),\text{ }(\mathbf{qY}%
)_{u}=\mathbf{q}_{u}(\mathbf{Y}_{u}). \label{proj00}%
\end{equation}

(\textbf{a9}) Let $(x,\mathcal{Q})$ be a instantaneous observer and
$\mathbf{p}_{x}:T_{x}M\rightarrow H$ the orthogonal projection. The
\emph{projection tensor} is the symmetric bilinear mapping \ $\mathbf{h}%
:\sec($ $TM\times TM)\rightarrow\mathbb{R}$ such that for any $\mathbf{U,W}\in
T_{x}M$ we have:%
\begin{equation}
\mathbf{h}_{x}(\mathbf{U,W})=\boldsymbol{g}_{x}(\mathbf{pU},\mathbf{pW)}
\label{proj1}%
\end{equation}

Let $\{x^{\mu}\}$ be coordinates of a chart covering $U\subset M$, $x\in U$
and $\alpha_{\mathcal{Q}}=\boldsymbol{g}_{x}(\mathcal{Q},~\mathcal{)}$.\ We
have the properties:%

\begin{equation}%
\begin{tabular}
[c]{|c|c|}\hline
(a) & $\mathbf{h}_{X}=\boldsymbol{g}_{X}-\alpha_{\mathcal{Q}}\otimes
\alpha_{\mathcal{Q}}$\\\hline
(b) & $\left.  \mathbf{h}\right\vert _{\mathcal{Q}^{\perp}}=\left.
\boldsymbol{g}_{x}\right\vert _{\mathcal{Q}^{\perp}}$\\\hline
(c) & $\mathbf{h}(\mathcal{Q},)=0$\\\hline
(d) & $\mathbf{h}(\mathbf{U,)}=\boldsymbol{g}(\mathbf{U,)\Leftrightarrow}$
\texttt{ }$\boldsymbol{g}(\mathbf{U,}\mathcal{Q})=0$\\\hline
(e) & $\mathbf{p}=h_{\nu}^{\mu}\left.  \frac{\partial}{\partial x^{\mu}%
}\right\vert _{x}\otimes\left.  dx^{\nu}\right\vert _{x}$\\\hline
(f) & \textrm{trace(}$h_{\nu}^{\mu}\left.  \frac{\partial}{\partial x^{\mu}%
}\right\vert _{x}\otimes\left.  dx^{\nu}\right\vert _{x})=-3$\\\hline
\end{tabular}
\ \ \ \ \ \ \ \ \ \label{proj2}%
\end{equation}

The result quote in (\textbf{a3}) together with the above definitions suggest
to introduce the following notions:\smallskip

(\textbf{a10}) A \emph{reference frame} for $U\subseteq M$ in a spacetime
structure $(M,\boldsymbol{g},D,\tau_{\boldsymbol{g}},\uparrow)$ is a time-like
vector field which is a section of $TU$ such that each one of its integral
lines is an observer.\smallskip

(\textbf{a11}) Let $\mathbf{Q}\in\sec TM,$ be a reference frame. A chart in
$U\subseteq M$ of an oriented atlas of $M$ with coordinate functions
$(\boldsymbol{y}^{\mu})$ and coordinates $(\boldsymbol{y}^{0}(\mathfrak{e}%
)=y^{0},\boldsymbol{y}^{1}(\mathfrak{e})=y^{1},\boldsymbol{y}^{2}%
(\mathfrak{e)=}y^{2},\boldsymbol{y}^{3}(\mathfrak{e})=y^{3})$ such that
$\partial/\partial y^{0}\in\sec TU$ is a timelike vector field and the
$\partial/\partial y^{i}\in\sec TU$ ($i=1,2,3$) are spacelike vector fields is
said to be a possible naturally adapted coordinate chart to the frame
$\mathbf{Q}$\textbf{\ }\emph{(}denoted $\emph{(nacs-}\mathbf{Q}\emph{)}$ in
what follows\emph{)} if the space-like components of $\mathbf{Q}$ are null in
the natural coordinate basis $\{\partial/\partial x^{\mu}\}$ of $TU$
associated with the chart. We also say that $(y^{0},y^{1},y^{2},y^{3})$ \ are
naturally adapted coordinates to the frame $\mathbf{Q}$.\smallskip

\begin{remark}
It is crucial, in order to avoid misunderstandings, to have in mind that most
of the reference frames used in the formulation of physical theories are
theoretical objects, i.e., a reference frame does not need to have material
support in the region were it has mathematical support.
\end{remark}

(\textbf{a12}) References frames in Lorentzian spacetimes can be
\emph{classified} according to the decomposition of $D\mathbf{Q\ }$and
according to their \emph{synchronizability}. Details may be found in
\cite{rc2016}. We analyze in detail the nature of the right Rindler reference
frame in Section 2. Here we only recall that $\mathbf{Q}$ is locally
synchronizable if it rotation tensor $\omega$ (coming form the decomposition
of $Q=\boldsymbol{g}(\mathbf{Q},~)$ and we can show $\omega
=0\Longleftrightarrow Q\wedge dQ=0$. Also, $\mathbf{Q}$ is synchronizable if
besides being irrotational also there exists a function $H$ \ on $U$ and a
timelike coordinate, say $u$ (part of a naturally adapted coordinate system to
$\mathbf{Q}$) such that $Q=Hdu$. Finally, $\mathbf{Q}$ is said to be
propertime synchronizable if $Q=du$.

(\textbf{a13}) We also used in the main text the following conventions:%

\begin{align}
\boldsymbol{g}(\boldsymbol{A}\mathbf{,}\boldsymbol{B}) &  =\boldsymbol{A}%
\mathbf{\cdot}\boldsymbol{B},~~~\mathrm{g}(C,D)=C\cdot D,\nonumber\\
\boldsymbol{A}\mathbf{,}\boldsymbol{B} &  \in\sec TM,~~~~C,D\in\sec%
%TCIMACRO{\tbigwedge \nolimits^{1}}%
%BeginExpansion
{\textstyle\bigwedge\nolimits^{1}}
%EndExpansion
T^{\ast}M.\label{a7}%
\end{align}
and the scalar product of Euclidean vector fields is denoted by $\bullet
$.\smallskip

(\textbf{a14}) Moreover, $d$ and $\delta$ denotes the differential and Hodge
codifferential operators acting on sections of $%
%TCIMACRO{\tbigwedge }%
%BeginExpansion
{\textstyle\bigwedge}
%EndExpansion
T^{\ast}M$ and $\lrcorner$ denotes the left contraction operator of form
fields \cite{rc2016}.

\section{$C^{\star}$ Algebras and the Unruh \textquotedblleft
Effect\textquotedblright}

The reason for including this Appendix in this paper is for the interested
reader to have an idea of how much he can \emph{trust} the standard approach
recalled in the main text which result in the claim that Rindler observers
live in a thermal bath. The algebraic approach to quantum field theory is
based on $C^{\star}$-algebras\footnote{For a susccint presentation of
$C^{\ast}$-algebras, enough for the understanding of the following see, e.g.,
\cite{david}. There the reader will find \ there the main references on the
algebraic (and axiomatic) approach to quantum field theory. Also, the reader
who wants to know all the details \ concerning the algebraic approach to the
Unruh effect must study the texts quoted below which has been heavily used in
the writing of this Appendix B.} which are now briefly recalled.

(\textbf{b1}) Let then be $\mathcal{A}$ a $C^{\ast}$-algebra over $\mathbb{C}$
whose some of its elements may be associated to the observables\footnote{I.e.,
the self-adjoints elements of $\mathcal{A}$} (associated to the quantum field
$\hat{\phi}$). We recall that a \emph{representation} of a $C^{\ast}$-algebra
is a linear mapping%
\begin{equation}
f:\mathcal{A\rightarrow B}(\mathfrak{H}),~~A\mapsto f(A),~~f(A^{\star
})=f(A)^{\dagger}. \label{al0}%
\end{equation}
where $\mathcal{B}(\mathfrak{H})$ is an algebra of bounded linear operators on
a Hilbert space $\mathfrak{H}$. The observables are associated with elements
$A=A^{\ast}$, where $^{\star}$ denotes the \emph{involution} operation in
$\mathcal{A}$, i.e., $\mathcal{AA}^{\star}=1$ and $^{\dagger}$ denotes the
Hermitian conjugate in $\mathcal{B}(\mathfrak{H})$

(\textbf{b2}) A representation $(f,\mathfrak{H})$ of $\mathcal{A}$ is said
\emph{faithful} if $f(A)=0$ if $A=0$ and $(f,\mathcal{H})$ is irreducible if
the only closed subspaces of $\mathfrak{H}$ invariant under $f$ are $\{0\}$
and $\mathfrak{H}$.

(\textbf{b3}) Let $\mathfrak{L}\subset\mathfrak{H}$ be a non zero closed
subspace of invariant under $f$. Let $\mathbf{\hat{P}}_{\mathcal{L}}$ be the
\emph{orthogonal projection} operator on $\mathcal{L}$. A
\emph{subrepresentation} of $f_{\mathcal{L}}$ is the mapping%
\begin{equation}
f_{\mathcal{L}}:\mathcal{A\rightarrow B}(\mathfrak{H}),~~~A\mapsto
f(A)\mathbf{\hat{P}}_{\mathcal{L}}. \label{al10a}%
\end{equation}

(\textbf{b3}) Two representations, say $(f_{1},\mathfrak{H}_{1})$ and
$(f_{2},\mathfrak{H}_{2})$ of $\mathcal{A}$ are said to be unitarily
equivalent is there exists an isomorphism $\mathbf{U:}\mathfrak{H}%
_{1}\rightarrow\mathfrak{H}_{2},$ such that
\begin{equation}
\mathbf{U}f_{1}(\mathcal{A})\mathbf{U}^{-1}=f_{2}(\mathcal{A}). \label{al3}%
\end{equation}

(\textbf{b4}) A \emph{state} on $\mathcal{A}$ is a mapping
\begin{align}
\omega &  :\mathcal{A\rightarrow}\mathbb{R},\nonumber\\
\omega(1)  &  =1,~~~\omega(A^{\star}A)\geq0,\forall A\in\mathcal{A}.
\label{al4}%
\end{align}

(\textbf{b5}) A \emph{pure state }$\omega$\emph{ }on $\mathcal{A}$ is one that
cannot be written as a non-trivial convex linear combination other states. On
the other hand a \emph{state }$\omega$\emph{ }on $\mathcal{A}$ is said to be
\emph{mixed }if it can be written as a non-trivial convex linear combination
other states.\smallskip

(\textbf{b6}) It is important to recall that a result (theorem) due to
Gel'fand, Naimark and Segal (GNS) \cite{gn,segal} establishes that for any
$\omega$\emph{ }on $\mathcal{A}$ there always exists a representation
$(f_{\omega},\mathfrak{H}_{\omega})$ of $\mathcal{A}$ and\emph{ }$\Phi
_{\omega}\in\mathfrak{H}_{\omega}$ (usually called a \emph{cyclic vector})
such that $f_{\omega}(\mathcal{A})\Phi_{\omega}$ is dense in $\mathfrak{H}%
_{\omega}$ and
\begin{equation}
\omega(A)=\langle\Phi_{\omega}|f_{\omega}(A)|\Phi_{\omega}\rangle. \label{al5}%
\end{equation}
Moreover the\ GNS result warrants that up to unitary equivalence, $(f_{\omega
},\mathfrak{H}_{\omega})$ is the unique \emph{cyclic }representation of
$\mathcal{A}$.\smallskip

(\textbf{b7}) The\emph{ folium }$\mathfrak{F(}\omega\mathfrak{)}$ of $\omega
$\emph{ }on $\mathcal{A}$ is the set of all abstract states that can be
expressed as density matrices on the Hilbert space of the GNS representation
determined by $\mathfrak{H}_{\omega}$.\smallskip

(\textbf{b8}) Given states $\omega_{1},\omega_{2}$ on $\mathcal{A}$ they are
said \emph{quasi-equivalent} if and only if $\mathfrak{F(}\omega
_{1}\mathfrak{)=F(}\omega_{2}\mathfrak{)}$. The states $\omega_{1},\omega_{2}$
on $\mathcal{A}$ are said to be \emph{disjoint }if $\mathfrak{F(}\omega
_{1})\cap\mathfrak{F(}\omega_{2}\mathfrak{)=\varnothing}$.\smallskip

\textbf{(b9}) It is possible to show that:

(i) \emph{Any irreducible representation have no proper subrepresentations and
in this case if }$\omega_{1}$\emph{ and }$\omega_{2}$\emph{ are pure states,
quasi-equivalence reduces to unitary equivalence and disjointness reduces to
non-unitary equivalences;}

(ii) \emph{When }$\omega_{1}$\emph{ and }$\omega_{2}$\emph{ are mixed states
they in general are not quasi equivalent or disjoint.\smallskip}

This happens when, e.g., $\omega_{1}$ has disjoint representations and one of
then is unitarily equivalent to $\omega_{2}$.\smallskip

(\textbf{b10}) For our considerations it is important to recall the following
result \cite{br2}:

\emph{The states }$\omega_{1}$\emph{ and }$\omega_{2}$\emph{ are disjoint if
and only if the GNS representation of }$f_{\omega_{1}+\omega_{2}}$\emph{
determined by }$\omega_{1}+\omega_{2}$\emph{ satisfies}%
\begin{equation}
(f_{\omega_{1}+\omega_{2}},\mathfrak{H}_{\omega_{1}+\omega_{2}})=(f_{\omega
_{1}}\oplus f_{\omega_{2}},\mathfrak{H}_{\omega_{1}}\oplus\mathfrak{H}%
_{\omega_{2}}), \label{ai6}%
\end{equation}
\emph{ \ i.e., the direct sum of the representations }$f_{\omega_{1}}$\emph{
and }$f_{\omega_{2}}$\emph{. Elements of }$\mathfrak{H}_{\omega_{1}+\omega
_{2}}$ are\emph{ denoted by}%
\begin{equation}
|\Phi_{\omega_{1}+\omega_{2}}\rangle=|\Phi_{\omega_{1}}\rangle\oplus
|\Phi_{\omega_{2}}\rangle\label{AI7}%
\end{equation}

(\textbf{b11}) To continue the presentation it is necessary to use a
particular $C^{\ast}$-algebra, namely the Weyl algebra\footnote{Also called
Symplectic Clifford Algebra \cite{cru,Z}.} $\mathcal{A}_{W}(M)$ which encodes
(see, e.g., \cite{ch}), in particular an exponential version of the canonical
commutation relations for the Klein-Gordon field used in the analysis of the
Unruh effect in this paper. Use of the Weyl algebras is opportune because in a
version appearing in \cite{kw} it leads to a net of algebras $\{\mathcal{A}%
(U)\}$ where if $U\subset M$\ is an open set of compact closure which
qualifies as a globally hyperbolic spacetime structure $(U,\left.
\boldsymbol{g}\right\vert _{U},\left.  D\right\vert _{U},\left.
\tau_{\boldsymbol{g}}\right\vert _{U},\uparrow)$ then if $U\subset U^{\prime
}\subset M$ it is $\mathcal{A}(U)\subset\mathcal{A}(U^{\prime})$.\smallskip

(\textbf{b12}) It is also necessary to know the following result
\cite{bkr,br1,br2}:

\emph{Let }$\boldsymbol{Z}\in\sec TU$\emph{ \ where\ }$U$\emph{ qualifies as a
globally hyperbolic spacetime which is foliated with Cauchy
surfaces\footnote{$u$ is a parameter indexing the foliation.} }$\Sigma
(u).$\emph{ Let }$n$\emph{\ }$\in\sec TM$\emph{ be the unit normal to }%
$\Sigma$\emph{, a member of the foliation. Only if for some }$\varepsilon\in
R$\emph{, }$Z$\emph{ satisfies}%
\begin{equation}
\boldsymbol{Z}\cdot\boldsymbol{Z}\geq\varepsilon\boldsymbol{Z}\cdot
\boldsymbol{n}\geq\varepsilon^{2} \label{al8}%
\end{equation}
\emph{there exists a procedure that associates with }$\Sigma$\emph{ a
so-called quasi-free\ state }$\omega_{\Sigma}$\emph{ on }$A(M)$\emph{.}

(\textbf{b13}) \emph{Quasi-free states} are the ones for which the $n$-point
functions of quantum field theory are determined by the two point functions
and their importance here lies in the fact that it can be shown that the GNS
representation of $\omega_{\Sigma}$ has a natural Fock-Hilbert space structure
$\mathcal{F(}\Sigma\mathcal{)}$ where $\omega_{\Sigma}$ is represented by the
vacuum state $|0\rangle_{\Sigma}\in\mathcal{F(}\Sigma\mathcal{)}$. Thus,
$\omega_{\Sigma}$ qualifies as a candidate for the vacuum state.

\begin{remark}
Note that if we take $\boldsymbol{Z}$ equal to $\boldsymbol{I}$ since it is
irrotational \emph{(}and a Killing vector field\emph{)}, \ it can be used to
foliate $M$ and for $\boldsymbol{I}$ \ \emph{Eq.(\ref{al8})} is satisfied.
Then we naturally can construct $\omega_{M}$ on $\mathcal{A}$ representing the
state $|0\rangle_{M}\in\mathcal{F(H)}$. Also, if we take $\boldsymbol{Z=}$
$\boldsymbol{Z}_{\mathrm{I}}$ \emph{or} $\boldsymbol{Z=}$ $\boldsymbol{Z}%
_{\mathrm{II}}$ \emph{(}as defined in \emph{Eqs.(\ref{16x})) }since these
fields besides being Killing vector fields are also irrotational, they can be
used to foliate regions \textrm{I} and \textrm{II }where the respective Cauchy
surfaces are of course, spacelike surfaces orthogonal respectively to
$\boldsymbol{Z}_{\mathrm{I}}$ and $\boldsymbol{Z}_{\mathrm{Ii}}$. In these
cases, \emph{Eq.(\ref{al8})} is violated near the \textquotedblleft
horizon\textquotedblright.and it is not possible to
construct\emph{\footnote{The states $\omega_{\mathrm{I}}$ on $\mathcal{A}%
(\mathrm{I})$ and $\omega_{\mathrm{II}}$ on $\mathcal{A}(\mathrm{II})$ are
called Boulware vacuum states\cite{bou}.}} $\omega_{\mathrm{I}}$ on
$\mathcal{A}(\mathrm{I})$ and $\omega_{\mathrm{II}}$ on $\mathcal{A}%
(\mathrm{II})$.These states are the ones associate with the vacuum
states\ $|0\rangle_{\mathrm{I}}$ and $|0\rangle_{\mathrm{II}}$ described
above. \smallskip
\end{remark}

(\textbf{b14)} We have now the fundamental result:

\emph{The states }$\left.  \omega_{M}\right\vert _{\mathcal{A}(\mathrm{I)}}%
$\emph{ }(\emph{respectively }$\left.  \omega_{M}\right\vert _{\mathcal{A}%
(\mathrm{II)}}$)\emph{ and }$\omega_{\mathrm{I}}$\emph{ }(\emph{respectively
}$\omega_{\mathrm{II}}$)\emph{ are disjoint.\smallskip}

\textbf{(b15) }To understand what is the meaning of this statement it is
necessary to recall the definition of a \emph{von Neumann algebra}
\cite{vN}.(denoted $W^{\ast}$-algebra). It is a special type of a $C^{\ast}%
$-algebra of bounded operators on a Hilbert space that is closed in the weak
operator topology and contains the identity operator.\smallskip

(\textbf{b16}) What is important for us here is that if $\mathcal{A}$ is a
$C^{\ast}$-algebra identified with the space of bound operators $\mathfrak{B}%
(\mathcal{H})$ of an appropriate Hilbert space\ then $\mathcal{A}$ is a
$W^{\ast}$-algebra if and only if
\begin{equation}
\mathcal{A=A}^{\prime\prime}, \label{a19}%
\end{equation}
where $\mathcal{A}^{\prime}$ denotes the so called \emph{commutant }of
$\mathcal{A}$, i.e., the set of operators that commute with all elements of
$\mathcal{A}$. Of course, $\mathcal{A}^{\prime\prime}$ denotes the commutant
of the commutant and is called bicommutant.\smallskip

(\textbf{b17}) Given a representation $(f,\mathfrak{H})$ of $\mathcal{A}$ we
denote $f^{\prime\prime}(\mathcal{A})$ the so-called \emph{double commutant
}of $f(\mathcal{A}).$ It is called the von Neumann algebra and denoted
$W_{f}(\mathcal{A})$. If the commutant $f^{\prime}(\mathcal{A})$ is an Abelian
algebra $W_{f}(\mathcal{A})$ is called type $\mathbf{I}$ and it is the case
given von Neumann theorem that if $\ \omega$ is an state on $\mathcal{A}$ then
$W_{f}(\mathcal{A})$ can be identified with $\mathfrak{B}(\mathcal{H}_{\omega
})$ for a GNS representation $(f_{\omega},\mathcal{H}_{\omega}).\smallskip$

(\textbf{b18}) A \emph{factorial state} $\omega$ on $\mathcal{A}$ (and their
GNS representation $\Phi_{\omega}$ $\in\mathcal{H}_{\omega}$) is one for which
the only\ multiples of the identity are elements of $W_{f\omega}%
(\mathcal{A})\cap W_{f\omega}(\mathcal{A})^{\prime}$.\smallskip

(\textbf{b19}) A\emph{ normal state} $\omega$ on $\mathcal{A}$ (and their GNS
representation $\Phi_{\omega}$ $\in\mathcal{H}_{\omega}$) is one whose
canonical extension to a state $\breve{\omega}\in W_{f\omega}(\mathcal{A})$ is
countably additive.\smallskip

(\textbf{b20}) Von Neumann algebras can also be of types \cite{bla}
$\mathbf{II}$ and $\mathbf{III}$. Type $\mathbf{III}$ are important for the
sequel and it is one where factors are factors that do not contain any nonzero
finite projections at all.\smallskip

(\textbf{b21}) Given these definitions it is possible to show the following
results concerning $C^{\ast}$-algebras:\smallskip

(\textbf{b21a}) If $f$ and $f^{\prime}$ are non degenerate representations of
$\mathcal{A}$, then they are quasi-equivalent if and only if there is a
$^{\ast}$-isomorphism
\begin{gather}
\boldsymbol{i}:W_{f}(\mathcal{A})\rightarrow W_{f^{\prime}}(\mathcal{A}%
),\nonumber\\
\boldsymbol{i}(f(A))=f^{\prime}(A) \label{al10}%
\end{gather}

(\textbf{b21b}) The representations $f$ and $f^{\prime}$ are quasi equivalent
if\ an only if $f$ has no subrepresentation disjoint from $f^{\prime}$ and vice-versa.

(\textbf{b21c}) A representation of a $\mathcal{A}$ is factorial if and only
if every subrepresentation of $f$ is quasi equivalent to $f^{\prime}%
$.\smallskip

From (\textbf{b21a}) it follows (see, e.g., \cite{ch}) that $f_{\omega
_{\mathrm{I}}}$ (respectively $f_{\omega_{\mathrm{II}}}$) and \ $f_{\left.
\omega_{M}\right\vert _{\mathcal{A(\mathrm{I})}}}$ (respectively $f_{\left.
\omega_{M}\right\vert _{\mathcal{A(\mathrm{II})}}}$) are not \ isomorphic
since $W_{f_{\omega\mathrm{I}}}(\mathcal{A})$ (respectively $W_{f_{\omega
\mathrm{I}}}(\mathcal{A})$) is a von Neumann algebra of type $\mathbf{I}%
$\textbf{ }whereas $W_{\ f_{\left.  \omega_{M}\right\vert
_{\mathcal{A(\mathrm{I})}}}}(\mathcal{A})$ ( respectively $W_{\ f_{\left.
\omega_{M}\right\vert _{\mathcal{A(\mathrm{II})}}}}(\mathcal{A})$) is a
von\ Neumann algebra of type $\mathbf{III}$ \cite{araki}.\smallskip

(\textbf{b22}) It is the case that in general not to be quasi equivalent does
not implies being disjoint., but in our particular case $\omega_{\mathrm{I}}$
(respectively $\omega_{\mathrm{II}}$) is a pure state which is irreducible and
as such has no no trivial representation. Also, $\left.  \omega_{M}\right\vert
_{\mathcal{A(\mathrm{I})}}$ (respectively $\left.  \omega_{M}\right\vert
_{\mathcal{A(\mathrm{II})}}$) is factorial and (c) implies that it is
equivalent to each one of its subrepresentation. Finally, from (\textbf{a}) it
follows that \ $f_{\omega_{\mathrm{I}}}$ (respectively $f_{\omega
_{\mathrm{II}}}$) \ and \ $f_{\left.  \omega_{M}\right\vert
_{\mathcal{A(\mathrm{I})}}}$ (respectively $f_{\left.  \omega_{M}\right\vert
_{\mathcal{A(\mathrm{II})}}}$) is disjoint if and only if they are not quasi equivalent.

\emph{Now, what does it means that }$f_{\omega_{\mathrm{I}}}$\emph{
}(\emph{respectively }$f_{\omega_{\mathrm{II}}}$)\emph{ \ and \ }$f_{\left.
\omega_{M}\right\vert _{\mathcal{A(\mathrm{I})}}}$\emph{ }(\emph{respectively
}$f_{\left.  \omega_{M}\right\vert _{\mathcal{A(\mathrm{II})}}}$)\emph{ is
disjoint?\smallskip}

(\textbf{b23}) Recall, e.g., that what $\omega_{M}$ has to say about region
\textrm{I }is given by\textrm{ }$\left.  \omega_{M}\right\vert
_{\mathcal{A(\mathrm{I})}}$ and from what we already recalled above cannot be
represented by a density matrix in the representation $f_{\omega_{\mathrm{I}}%
}$, in particular for any representation on $\mathcal{A(}\mathrm{I).}$This
happens because it is impossible to write $\mathcal{A(}M)$ as a tensor product
$\mathcal{A}^{\prime}\mathcal{\otimes A(}\mathrm{I})$ for some $\mathcal{A}%
^{\prime}$. This result is called \emph{expressive incompleteness}%
.\emph{\smallskip}

(\textbf{b24}) Despite expressive incompleteness we have the following result
by Verch \cite{verch}:

On\emph{ }$U\subset$\emph{ }$I\subset M$\emph{ }(which is open and of compact
closure)\emph{ let }$f_{\ \omega_{_{M}}}|$ $\mathcal{A(}U)$\ be the GNS
representation constructed from $\omega_{_{M}}$ restrict to the image
$\omega_{_{M}}|\mathcal{A(}U)$ under \emph{ }$f_{\ \omega_{_{M}}}\ $of
$\mathcal{A(}U)$ (and completing in the natural topology of $\mathcal{H}%
_{\omega_{_{M}}}$) and analogous construct\footnote{Please, do not confuse
$\omega_{\mathrm{I}}|\mathcal{A(}U\mathcal{)}$ with $\left.  \omega
_{\mathrm{I}}\right\vert _{\mathcal{A(}U\mathcal{)}}.$} $\omega_{\mathrm{I}%
}|\mathcal{A(}U\mathcal{)}$ the image of $\omega_{_{\mathrm{I}}}$ under\emph{
}$f_{\ \omega_{_{\mathrm{I}}}}|\mathcal{A(}U\mathcal{)}$\emph{\footnote{The
states $\omega_{_{M}}|\mathcal{A(}U)$ and $\omega_{_{\mathrm{I}}}%
|\mathcal{A(}U)$ are quasi free Hadamard states, i.e., states for which}.
Then,}$f_{\ \omega_{_{M}}}|$ $\mathcal{A(}U)\ $and $f_{\ \omega_{_{\mathrm{I}%
}}}|\mathcal{A(}U\mathcal{)}$ \emph{ are quasi equivalent}.\smallskip

(\textbf{b25}) The result presented in (\textbf{b24}) is the only one that
would permit legitimately to physicists to talk about $\omega_{M}$ and
$\omega_{\mathrm{I}}$ as being quasi equivalents, for indeed as already
recalled $f_{\omega_{M}}$ and $f_{\omega_{\mathrm{I}}}$ are indeed disjoint
representations of the algebra of observables $\mathcal{A}$ and thus not
unitarily equivalents.\smallskip

(\textbf{b26}) Anyway, the above result implies that only if we do
measurements on observables of the algebra $\mathcal{A}$ in regions of non
compact closure can distinguish \ the representations $f_{\omega_{M}}$ and
$f_{\omega_{\mathrm{I}}}.$

(\textbf{b27}) Finally one can ask the question: is $f_{\left.  \omega
_{M}\right\vert _{\mathcal{A(}U\mathcal{)}}}$ and $f_{\left.  \omega
_{M}\right\vert _{\mathcal{A(}U\mathcal{)}}}$ \ where again $U\subset$\emph{
}$I\subset M$\emph{ }(open and of compact closure) quasi equivalent?

The answer to this question is (for the best of our knowledge) \emph{not}
known and this is another hindrance that makes one to affirm that no
convincing theoretical proof that the Unruh effect is a real effect
exists.\smallskip

(\textbf{b28}) In the standard \textquotedblleft deduction\textquotedblright%
\ (Section 6.1) of the Unruh effect it is claimed that the uniformly
accelerated observer detects a thermal bath. Supports that the effect is a
real one try to endorse their claim by using the notion of KMS
states\footnote{Recall that a KMS state is an algebraic state $(\zeta
_{u},\beta)$ on $\mathcal{A}$ where $\zeta_{u}:\mathcal{A\rightarrow A}$ one
parameter group of automorphisms and $0\leq\beta<\infty$ such that the
condition $\omega(A\zeta_{u\beta}B)=\omega(BA)$. It is a basic result that a
state satisfying the \ KMS condition at t act as a thermal reservoir, in the
sense that any finite system coupled to it reaches thermal equilibrium at
\textquotedblleft temperature\textquotedblright\ $T=\beta^{-1}$.} (which as
well known generalizes the notion of equilibrium state)
\cite{k,ms,bkr,br1,br2}. In fact, Sewell \cite{sew} argues that the
restriction of the Minkowski vacuum $\omega_{M}$ to region \textrm{I, }i.e.,
$\left.  \omega_{M}\right\vert _{\mathcal{A(}\mathrm{I)}}$\textrm{ (=}$\left.
\omega_{M}\right\vert _{\mathrm{I}}$\textrm{) }can be formulated as an
algebraic state on $\mathcal{A}_{\mathrm{I}}$ which satisfies the KMS
condition at temperature $\beta^{-1}=a/2\pi$ relative to the notion of time
translation defined by vector field $\boldsymbol{Z}_{\mathrm{I}}%
=\partial/\partial\mathfrak{t}$ (which then generates the one-parameter group
\ of automorphism $a_{u=\mathfrak{t}}$). However,\ it is necessary to have in
mind that the proof that $\left.  \omega_{M}\right\vert _{\mathrm{I}}$ is a
KMS state does not imply that it is a thermal bath of Rindler particles. The
assumption that it is is only a suggestive one. The reason for that statement
is that as commented in the main text a detector can indeed be excited when in
uniform accelerated motion, but the excitation energy does \emph{not} come
from the \emph{pseudo energy} of any hypothetical thermal bath, but from the
\emph{real} energy (as inferred from an inertial reference frame) of the
source accelerating the device.

\end{document}